\documentclass{aa}

\usepackage{url}
\usepackage{natbib}
\usepackage{array}
\usepackage[bf]{subfigure}
\usepackage[varg]{txfonts}
\usepackage{graphicx}

\newcommand{\beq}{\begin{equation}}
\newcommand{\eeq}{\end{equation}}
\newcommand{\bdi}{\begin{displaymath}}
\newcommand{\edi}{\end{displaymath}}

\newcommand{\herschel}{\textit{Herschel}}
\newcommand{\getsources}{\textsl{getsources}}
\newcommand{\getfilaments}{\textsl{getfilaments}}

\newcommand{\degree}{$^{\circ}$}
\newcommand{\micron}{$\mu$m}
\newcommand{\nh}{$N_{\mathrm{H}_2}$}

\newcommand{\mline}{\textit{M}$_{\mathrm{line}}$}

\newcommand{\mtot}{\textit{M}$_{\mathrm{line,tot}}$}
\newcommand{\mtotb}{\textit{M}_{\mathrm{line,tot}}}
\newcommand{\mcore}{\textit{M}$_{\mathrm{line,core}}$}
\newcommand{\mwing}{\textit{M}$_{\mathrm{line,wing}}$}
\newcommand{\mcoreb}{\textit{M}_{\mathrm{line,core}}}
\newcommand{\mwingb}{\textit{M}_{\mathrm{line,wing}}}
\newcommand{\mcrit}{\textit{M}$_{\mathrm{crit}}$}

\newcommand{\mpc}{\textit{M}$_{\mathrm{\odot}}$\,pc$^{-1}$}
\newcommand{\av}{$A_{\mathrm{V}}$}
\newcommand{\nscale}{\textit{N}$_{\mathrm{scales}}$}
\newcommand{\fsample}{core-scale}
\newcommand{\core}{core}

\begin{document}


\title{Galactic Cold Cores\thanks{\MakeLowercase{\textit{\MakeUppercase{h}erschel} is an \MakeUppercase{ESA} space observatory with science instruments provided by \MakeUppercase{E}uropean-led \MakeUppercase{P}rincipal \MakeUppercase{I}nvestigator consortia and with important participation from \MakeUppercase{NASA}.}} VII: Filament Formation and Evolution}
\subtitle{Methods \& Observational Constraints}


\author{A.~Rivera-Ingraham\inst{{1},{2},{3}}\and 
I.~Ristorcelli\inst{1,2}\and
M.~Juvela\inst{4}\and
J.~Montillaud\inst{5}\and
A.~Men'shchikov\inst{6}\and
J.~Malinen\inst{4}\and 
V.-M.~Pelkonen\inst{4,7}\and
A.~Marston\inst{3}\and
P.~G.~Martin\inst{8}\and
L.~Pagani\inst{9,10}\and
R.~Paladini\inst{11}\and
D.~Paradis\inst{1,2}\and
N.~Ysard\inst{12}\and
D.~Ward-Thompson\inst{13}
J.-P.~Bernard\inst{1,2}\and
D.~J.~Marshall\inst{6}\and
L.~Montier\inst{1,2}\and
V.~T\'oth\inst{14}
}

\offprints{A. Rivera-Ingraham, \email{alana.rivera@esa.int}}

\institute{Universit\'{e} de Toulouse; UPS-OMP; IRAP;  Toulouse, France
\and
CNRS; IRAP; 9 Av. colonel Roche, BP 44346, F-31028 Toulouse cedex 4, France
\and
European Space Astronomy Centre ESA/ESAC, P.O.Box 78, 28691 Villanueva de la Canada, Madrid, Spain
\and
Department of Physics, P.O. Box 64, FI-00014, University of Helsinki, Finland
\and
Institut UTINAM, CNRS 6213, OSU THETA, Universit\'{e} de Franche-Comt\'{e},41 bis avenue de l'Observatoire, 25000 Besan\c{c}on, France
\and
Laboratoire AIM, CEA/DSM/Irfu –- CNRS/INSU –- Universit\'e Paris Diderot, CEA-Saclay, F-91191 Gif-sur-Yvette Cedex, France
\and
Finnish Centre for Astronomy with ESO, University of Turku, V\"ais\"al\"antie 20, 21500 Piikki\"o, Finland
\and
Canadian Institute for Theoretical Astrophysics, University of Toronto, 60 St. George Street, Toronto, ON M5S~3H8, Canada
\and
LERMA, Observatoire de Paris, PSL Research University, CNRS, UMR 8112, F-75014 Paris, France
\and
Sorbonne Universit\'es, UPMC Univ. Paris 6, UMR 8112, LERMA, F-75005, Paris, France
\and
Infrared Processing Analysis Center, California Institute of Technology, 770 S. Wilson Ave., Pasadena, CA 91125, USA
\and
IAS, CNRS (UMR8617), Universit\'{e} Paris Sud, Bat. 121, F-91400 Orsay, France
\and
Jeremiah Horrocks Institute, University of Central Lancashire, Preston, Lancashire, PR1 2HE, UK
\and
E\"otv\"os University, Department of Astronomy,P\'azm\'any P. s. 1/a,H-1117, Budapest, Hungary
}

\date{Received X March 2016 / Accepted X March 2016}


\abstract 
{The association of filaments with protostellar objects has made these structures a priority target in star formation studies.  However, little is known about the link between filament properties and their local environment.}
{The datasets of the \herschel\ Galactic Cold Cores Key Programme allow for a statistical study of filaments with a wide range of intrinsic and environmental characteristics. Characterisation of this sample can therefore be used to identify key physical parameters and quantify the role of environment in the formation of supercritical filaments. These results are highly needed in order to constrain theoretical models of filament formation and evolution.}
{Filaments were extracted from fields at distance $D<500$\,pc with the \getfilaments\, algorithm and characterised according to their column density profiles and intrinsic properties. Each profile was fitted with a beam-convolved Plummer-like function and quantified based on the relative contributions from the filament `\core', represented by a Gaussian, and `wing' component, dominated by the power-law behaviour of the Plummer-like function. These filament parameters were examined for populations associated with different background levels.}
{Filaments increase their \core\ (\mcore) and wing (\mwing) contributions while increasing their total linear mass density (\mtot). Both components appear to be linked to the local environment, with filaments in higher backgrounds having systematically more massive \mcore\ and \mwing. 
This dependence on the environment supports an accretion-based model for filament evolution in the local neighbourhood ($D\le500$\,pc). Structures located in the highest backgrounds develop the highest central \av, \mcore, and \mwing\ as \mtot\ increases with time, favoured by the local availability of material and the enhanced gravitational potential. Our results indicate that filaments acquiring a significantly massive central region with \mcore$\ge$\mcrit$/2$ may  become supercritical and form stars. This translates into a need for filaments to become at least moderately self-gravitating in order to undergo localised star formation or become star-forming filaments.}
{}

\maketitle


\section{Introduction} \label{sec:intro}
The unprecedented resolution and wavelength coverage of the 
\herschel\ Space Observatory (\herschel; \citealp{pilbratt2010}) has 
revealed a complex filamentary network present in a wide range of 
environments and spatial scales. 
The presence of filaments in the interstellar medium (ISM) and nearby clouds 
has indeed been investigated for many years with a variety of instruments 
and techniques (e.g., \citealp{bally1987}; \citealp{peretto2009}). 
\herschel\ Key Programmes such as the \herschel\ infrared Galactic Plane survey 
(Hi-Gal; \citealp{molinari2010}), the Gould Belt Survey (HGBS; \citealp{andre2010}), 
and the \herschel\ imaging survey of OB Young Stellar objects (HOBYS; \citealp{motte2010}) 
have shown their ubiquitous nature in both dense star forming complexes 
and diffuse, non-star forming fields (e.g., \citealp{getsources2010}; \citealp{miville2010}).

The above studies have also provided new clues regarding the role of filaments 
in the onset of star formation for low and high-mass stars 
(e.g. \citealp{arzoumanian2011}; \citealp{hill2012b}). 
Latest results point towards a scenario in which prestellar cores form by gravitational 
fragmentation of unstable filaments
\citep{andre2010} with quasi-constant width of $\sim$$0.1$\,pc in the solar 
neighbourhood \citep{arzoumanian2011}, although with possible variations at farther distances 
(\citealp{schisano2014}).
The processes that could give rise to such filaments are varied, ranging from shocks due 
to insterstellar magnetohydrodynamic (MHD) turbulence 
(e.g., \citealp{padoan2001})
to dynamical events, such as large scale compression (e.g., \citealp{peretto2012}).
Under the assumption of isothermality and no magnetic field, 
filament instability can be quantified by the mass per unit 
length (or linear mass \textit{density}; here denoted by \mline\ for consistency with previous studies) 
greater than a critical equilibirium value 
(\mcrit$=2c_{\mathrm{s}}^2/G$; e.g., \citealp{inutsuka1992}), 
a quantity exclusively dependent on 
temperature through the isothermal sound speed c$_{\mathrm{s}}$ \citep{ostriker1964}.
The tendency of \textit{bound} prestellar cores to be associated 
with those filaments in a supercritical state (\mline$ > $\mcrit, 
where \mcrit$\sim16.5$\,\mpc\ for a dust temperature of $T\approx10$\,K; 
e.g., \citealp{andre2010}) make the study of filament properties and evolution crucial 
for constraining the process of star formation.

The \herschel\ Galactic Cold Cores Key Programme 
(GCC; P.I: M. Juvela; \citealp{juvela2012a}) observed 116 fields that 
contained selected clumps from the Cold Clump Catalogue of Planck Objects 
(C3PO; \citealp{planck201122}; \citealp{planck201123}).
This unbiased sample covers a 
wide range of environments, Galactic positions, and physical conditions, which makes it ideal for 
the statistical investigation of properties associated with the compact source population 
(\citealp{montillaud2015}; M2015 hereafter), dust properties (e.g., \citealp{juvela2015b}; \citealp{juvela2015}) as well as 
star and structure formation in the most diffuse fields (Rivera-Ingraham et al; in prep.).

In this work we complement the analysis presented in \citet{juvela2012a} by carrying out 
an in-depth study of filamentary properties as a function of environment for the 
\herschel\ fields of the GCC Programme. 
The primary goal of this paper is to present the filament sample, the techniques, and 
the key observational results needed for constraining 
theoretical models of filament formation and evolution. 
Application of results to these models is the topic of a companion paper (Rivera-Ingraham et al. 2016; in prep). 
This combined study is important in order to quantify the physical processes associated 
with the origin of unstable filaments and the onset of star formation. 

In Sect. \ref{sec:data} we briefly describe the datasets used for this analysis.
Section \ref{sec:detection} introduces the \getfilaments\ algorithm used for 
filament detection and selection, while the techniques for filament profile analysis are 
included in Sect. \ref{sec:analysis}. 
Results are presented in Sect. \ref{sec:results}, and key observational 
constraints on the formation of star-forming supercritical filaments in 
gravitationally-dominated evolutionary scenarios are described in Sect. \ref{sec:discussion}.
We conclude with a summary of our main results in Sect. \ref{sec:conclusion}.

\section{Maps and Datasets} \label{sec:data}
The \herschel\ maps are those already introduced in previous GCC studies. 
A comprehensive description  of the method and techniques used for map creation and processing has 
been included in \citet{juvela2012a} and M2015).

The SPIRE maps ($250$\,\micron, $350$\,\micron, and $500$\,\micron; \citealp{griffin2010}) 
were reduced with the Herschel Interactive Processing 
Environment (HIPE\footnote{HIPE is a joint 
development by the Herschel Science Ground Segment Consortium, 
consisting of ESA, the NASA Herschel Science Center,and 
the HIFI, PACS and SPIRE consortia.}) v.10.0, 
using the official pipeline with the iterative destriper and the extended emission calibration.
The PACS $160$\,\micron\ maps \citep{pog2010} were created using \textsl{Scanamorphos} \citep{roussel2013}
version 20 with the galactic flag for the drift correction. 
All maps had colour and zero-point corrections applied as described in \citet{juvela2012a}.
 
Column density and temperature maps at a $40$\arcsec\ resolution were produced by 
fitting spectral energy distributions 
(SEDs) pixel-by-pixel to the three SPIRE datasets, assuming a dust 
opacity of $0.1$\,cm$^2$\,g$^{-1}$ at $1$\,THz \citep{hildebrand1983}, with a 
fixed dust emissivity index of $\beta=2$, and 
a mean atomic weight per molecule of $\mu=2.33$. 
While our assumed value for $\mu$ is consistent with previous filament papers (e.g., \citealp{arzoumanian2011}) 
we note that this number differs from 
the actual molecular weight per hydrogen molecule {($\mu=2.8$; e.g., \citealp{kauffmann2008}), and which would be more 
appropriate in the calculation of \nh\ maps. This 
choice does not affect, however, the main results of this work that depend on the 
relative properties of the filament population.

The catalog production techniques and final sample of compact sources associated with each field have been presented in M2015).
The source list was produced with the multi-scale, multi-wavelength source extraction 
algorithm \getsources\ (\citealp{getsources2010}; \citealp{getsources2012}), which was ran simultaneously 
on the colour and offset-corrected PACS and SPIRE brightness maps, 
and the column density map. 
The final catalog was corrected for galaxy contamination and all sources 
classified according to their stellar content and evolutionary state. 

\section{Filament Catalog: The \getfilaments\ Method for Filament Detection} \label{sec:detection}
Detection and characterisation of filamentary structures in our chosen fields was carried out 
with the \getfilaments\, algorithm v1.140127 \citep{getfilaments}, an 
integral part of the \getsources\, package. 

\subsection{The \getfilaments\ Approach}
Filament characterisation is highly complex because 
they constitute a hierarchical population in the ISM. 
The multi-scale  nature of filaments has already been observed in 
molecular studies, such as that presented in \citet{hacar2013}. 
The innovative detection technique of 
\getfilaments\ addresses this complexity by 
identifying different types of filaments according to the spatial scales at which they 
are detected. A detailed description of the source and filament extraction procedure has been 
published in \citet{getsources2012} and \citet{getfilaments}. 

In essence, the \getfilaments\ algorithm  decomposes the original map into 
a sequence of spatially filtered `single-scale' images, 
from the smallest to the largest scales. Each decomposed image contains signals from a narrow
range of spatial scales around one particular scale, all substantially larger or
smaller scales being filtered out. 
An important property of the decomposition is that the original image can 
be recovered by summation of all single-scale images.
In each single-scale image \getfilaments\ identifies, by means of an iterative thresholding
procedure and several morphological factors, all significantly elongated structures above
the image $1\sigma$ fluctuations level. 
In effect, the masking of all pixels below this level separates the filamentary structures from all 
other non-filamentary components (sources, background), determining the physical properties (e.g., 
length and width) of each filament in the single-scale image.
All filamentary structures above the threshold are preserved in the single-scale images, 
whereas all contributions of (compact) sources or background fluctuations are removed,
as they are (by definition) not significantly elongated.
This results in a set of single-scale images containing all the non-negligible filamentary emission at each particular 
scale, clean of noise/background contribution.  
The final reconstructed filament intensity map, containing filamentary information at all spatial scales, 
is produced by accumulating all the individual (clean) single-scale images of 
filamentary structures free of background and sources. 
This process of summation of all spatial scales therefore effectively recovers the complete structural 
properties of each filament (intensity, length, radial extent) in the map. 
 
As the filament extraction algorithm is part of the \getsources\ source extraction
method, the code also extracts all sources, separating them from filaments and
background. 
In essence, the \getfilaments\ method carefully separates 
the structural components (sources, filaments, isotropic background) into different images, 
which allows one to study the images of the filamentary component fully reconstructed 
over all spatial scales.
Each GCC \nh\ map was therefore decomposed into two images: a 
filament map (free of sources and background contributions), and a source map (free of filament and background). 
The image of background plus noise was obtained by subtracting both the source and filament images from 
the original map. 
For more information and clear illustrations of how the method works, 
we refer to the intensity profiles in Figs. 2, 15, 17 of \citet{getfilaments}, as well as to the images 
in Figs. 3--12, 14, 16 in that paper.

By quantifying the filamentary contribution at separate spatial scales, 
\getfilaments\ permits the analysis of filamentary substructures 
independently from their larger host filaments. 
This is critical for a better characterisation of the physics associated with 
filament formation and evolution, as the 
properties of filaments associated with a given scale might not 
necessarily resemble those of filaments reconstructed at substantially larger or smaller scales.
Key properties of a particular type of filaments could be missed when 
investigating just the average characteristics of the entire filament population.

In this work, the extraction and selection of filaments was based on our choice to 
focus only on those structures most relevant for prestellar core (star) formation 
(i.e., full-width at half maximum FWHM$<0.2$\,pc). 
Our extraction procedure was therefore tuned to identify structures that 
have non-negligible filamentary emission at these scales, effectively excluding 
others that can only be classified as filamentary when including information from larger scales. 
Examples of the latter could be a filamentary cloud,
itself possibly composed of smaller filaments, 
or other structures only appearing as filamentary-like 
when including diffuse emission at larger radii. 
In the multi-scale reality, it is therefore crucial to distinguish and separate these different 
types of filaments in order to be able to constrain the formation and evolution of a 
particular subtype. Here we are interested in investigating 
those filaments that represent the final link to prestellar core formation, or `core-scale' filaments. 
Filaments significant only when considering larger scales 
(filaments hosting smaller filaments), being a different 
type of structure, were excluded from our sample. 

In the absence of molecular data capable of confirming structural self-consistency 
based on velocity information, in the following analysis we define as `filaments' 
those detections that fulfil the following criteria: 

\begin{enumerate}
\item  significant elongation in the \nh\ maps: above 
the typical elongation in the compact source catalog of $\sim1.5$;

\item the potential for `core-bearing', 
i.e., comprising the last step in the hierarchical ladder of 
the filament population, within the resolution of the data, 
linked directly to compact source formation; 

\item well-behaved and self-consistent entities 
across the observed spatial scales relevant for star-formation 
(i.e. physical prestellar core sizes - FWHM$_{\mathrm{core}}<0.2$\,pc) 
at the resolution of the data. 

\end{enumerate}

These criteria reduced the number of available fields for the final 
analysis, due to the need to exclude distances at which our target filaments 
would not be resolved.
The distribution of distances of the GCC fields range from $D\sim100$\,pc 
to a few kpc (M2015), which at the resolution of the \nh\ 
maps implies that only those with distance less than $\sim500$\,pc are 
suitable for the detection of core-bearing filaments 
with widths comparable to those reported in previous studies 
($40$\arcsec$\sim0.1$\,pc at $D\sim500$\,pc). 
In this work we do not consider any field with an estimated distance above this value.

The process for selecting the target core-scale
filaments is summarised in the following paragraphs, 
but we refer to  \citet{getfilaments} for a more detailed 
description of the algorithm, its data products, and 
techniques employed.
\subsection{Preliminary Detection of Filamentary Structures}
The filament identification process is based on 
the decomposition of the filament emission according to 
its contribution to separate finely-spaced \textit{observed} spatial scales ($\theta$). 
Image decomposition, and ultimately, filament reconstruction, is  
carried in image units (arcseconds), therefore allowing for the possibility of a given field 
containing structures (clumps, filaments...) at different distances along the line of sight.

\getfilaments\ produces initially for each field a `skeleton' map for 
filaments reconstructed in pre-defined ranges of spatial scales $\theta/2$ to $\theta$, 
where $\theta$ is a multiple of the map resolution ($40$, $80$, $160$, $320$, $640$, and $1280$\arcsec). 
Each $\theta/2$\,--\,$\theta$ skeleton image contains 1-pixel lines tracing the length of  
filaments along their central part of highest column density (filament `crest' or `ridge'). 
However, each of these images comprises only the filamentary structures detected in the finely-spaced 
single-scale images in that $\theta/2$\,--\,$\theta$ range. 
For instance, skeletons in a $40-80$\arcsec\ scale image mark the position of the ridges 
of filamentary structures based exclusively on the emission detected in any of the various 
single spatial scales in that particular spatial scale range, 
independently of filament properties in spatial scales outside this range.  
Filaments could be detected in just a few or various single scales (\nscale) within a $\theta/2$\,--\,$\theta$ range. 
The set of maps ( $\theta=40-80$,  $\theta=80-160$\arcsec...) therefore progressively remove small spatial scales 
from the filaments, leaving wider, smoother, and more diffuse filaments. In consequence, 
skeletons derived from relatively large scales tend to lose precision 
in tracing the small-scale filament crests in favour of smoother underlying structures. 
This is advantageous in the large-scale analysis of filaments and ridges 
with considerable filamentary substructure 
(e.g., if interested only in the general, averaged properties of the filament group, 
or filament `bundle';  \citealp{hacar2013}).

Skeletons based on filaments reconstructed at small scales will, in general, trace the highest column density 
filaments better than those based on (smoother) filaments reconstructed just from larger scales.
The exclusive usage of small scales for filament detection, however, might also result in 
too conservative ridge tracing. A filament crest does not have to be continuous in nature 
but may contain regions of enhanced column density. Parts of the same filament that are less 
centrally condensed and smoother in nature will therefore be removed by the scale filtering, 
resulting in excessive sub-fragmentation of an otherwise single, long filament.
Filaments detected at the smallest scales can therefore be real sub-filaments without 
a larger scale component (i.e., present only at small scales, or `filaments within filaments'), or 
segments of a longer filament with inhomogeneities along its length in the column density maps.

In this study we restricted the range of spatial scales used for  
skeleton detection in order to create maps that best 
reproduced the most prominent filamentary structure as 
observed in the \nh\ maps. This would be in agreement with 
the filament detection techniques used in other
\herschel\, studies (e.g., \citealp{arzoumanian2011}). 
However, extra steps had to be taken in order to extract only 
those structures relevant for prestellar core formation, with non-negligible 
filamentary emission at the target spatial scales: 
\textit{physical} (intrinsic) spatial scales of 
$\theta_{\mathrm{int}}<0.2$\,pc at the distance of the field. 

\subsection{Filament Extraction}
\begin{figure}[ht!]
\centering
\subfigure[]{%
\label{fig:scale1}
\includegraphics[scale=0.5,angle=0,,trim=0cm 0cm 0cm 0cm, clip=true]{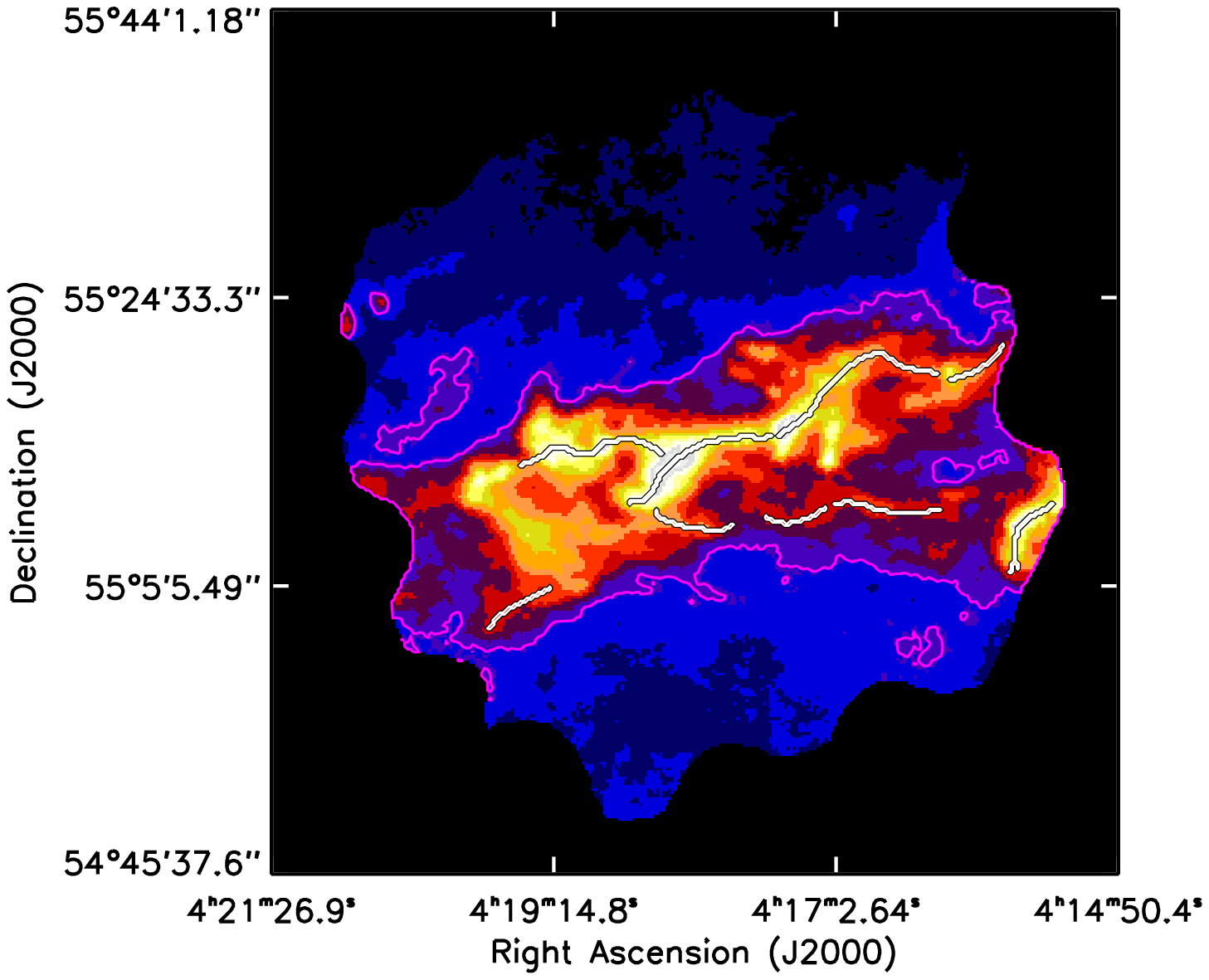}
}\\
\subfigure[]{%
\label{fig:scale2}
\includegraphics[scale=0.5,angle=0,trim=0cm 0cm 0cm 0cm, clip=true]{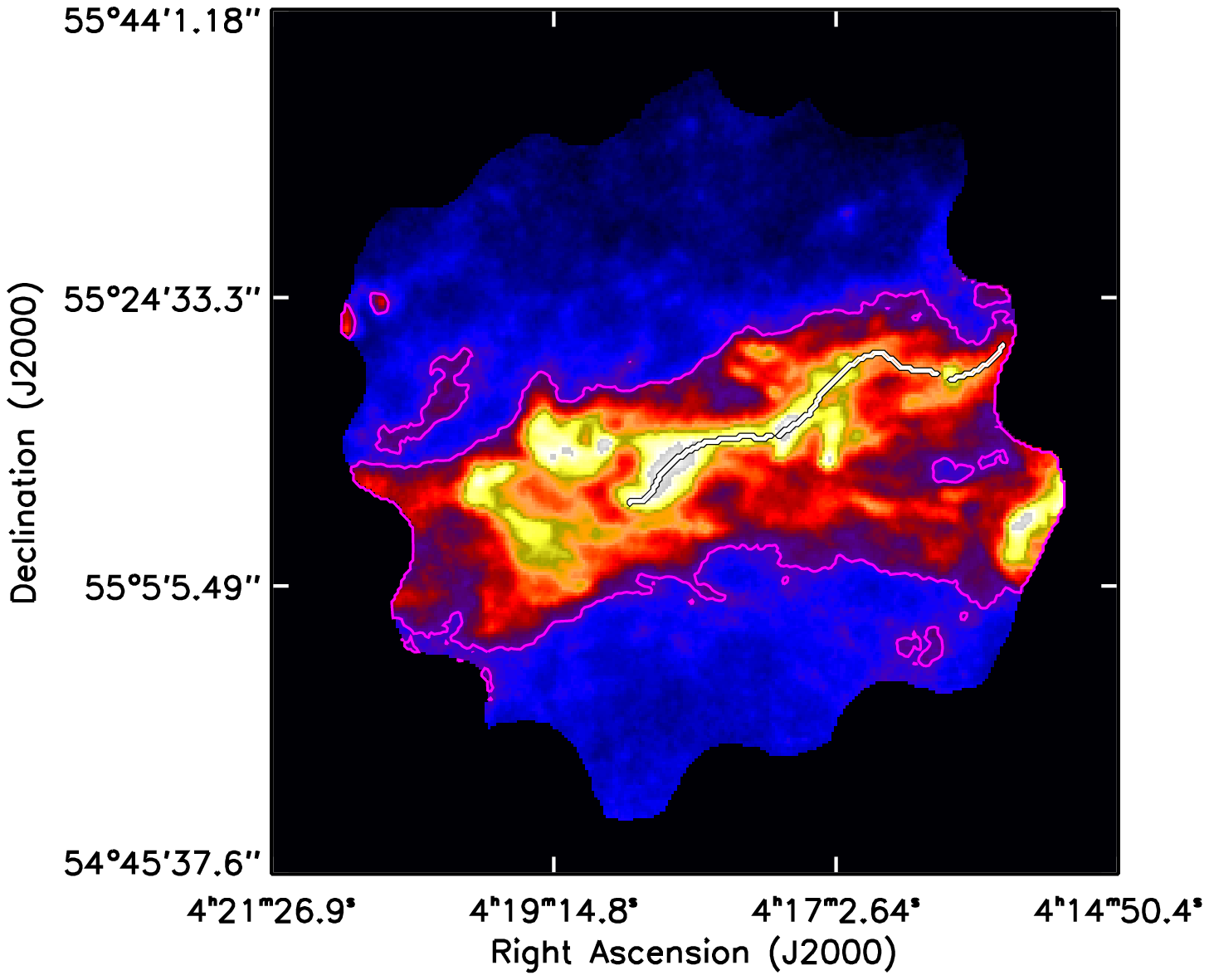}
}\\%
\subfigure[]{%
\label{fig:scale3}
\includegraphics[scale=0.5,angle=0,trim=0cm 0cm 0cm 0cm, clip=true]{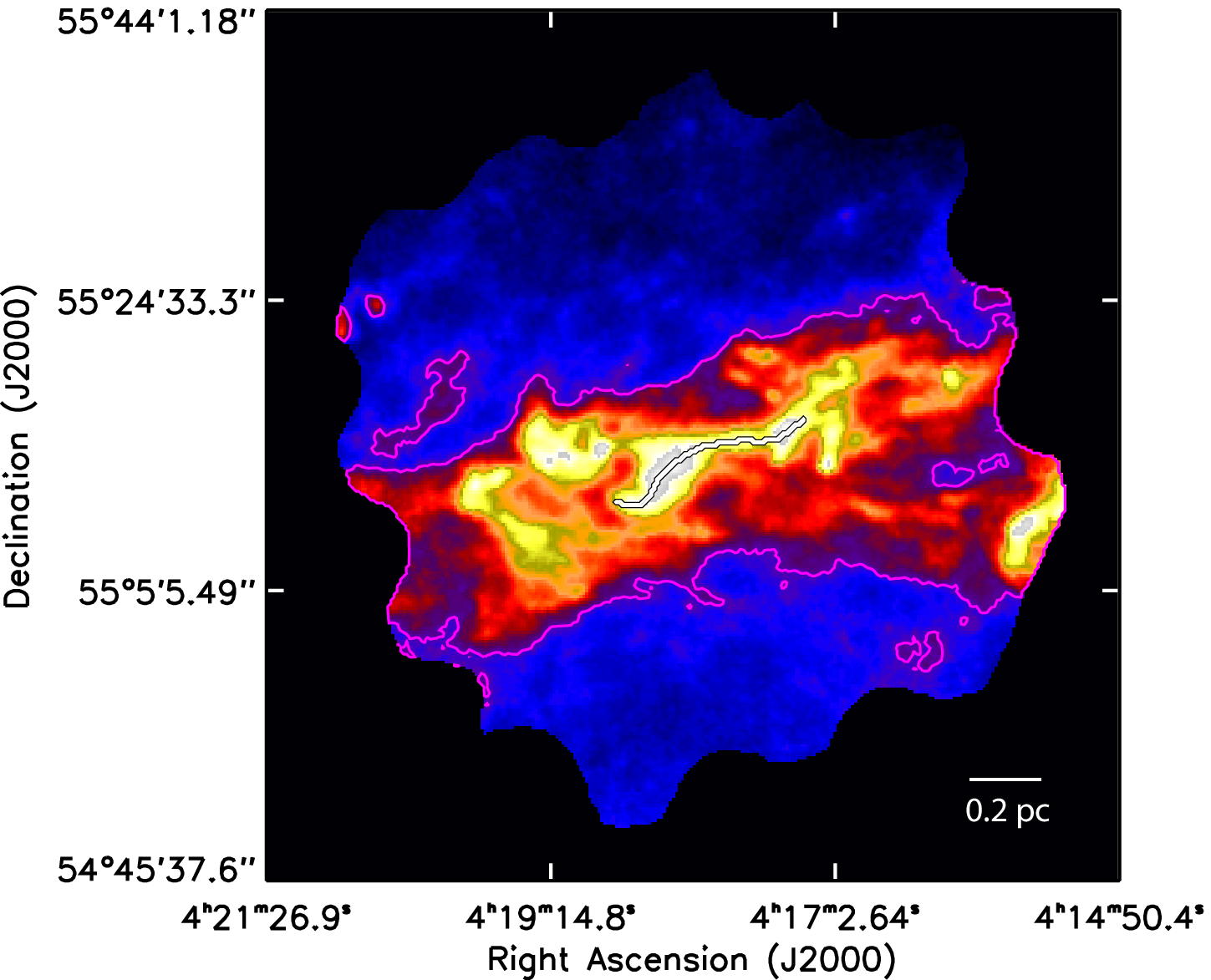}
}\\%
\caption{Column density images of the GCC field G149.67+3.56 with reliable filament skeletons superimposed. Detections are those satisfying the minimum selection and detection criteria (see text) when accumulated up to $320$\arcsec\ for a reliability level \nscale$=10$ (a), \nscale$=30$ (b), and \nscale$=50$ (c). The longest central filament in the field is an example of a well-behaved filament throughout the relevant spatial scales. The \av$=3$\,mag (magenta) contours are shown as reference.}
\label{fig:scale_image}
\end{figure}

Based on our filament definition and selection criteria, the field distance range ($D\sim100$\,--\,$500$\,pc), 
the compact source size ($0.2$\,pc), the spatial scale steps examined by default by 
\getfilaments, and the $40$\arcsec\, resolution of the \nh\ maps, the target core-scale 
filaments would be detected by examination 
of spatial scales up to $\theta=80$\arcsec\,--\,$320$\arcsec 
($\sim0.2$\,pc at the maximum and minimum distance limits, respectively). 
While the range of resolved filament physical sizes associated with a given spatial scale varies from field to field 
according to distance, extraction of all filaments with significant emission 
from any spatial scales at or below $\theta=320\arcsec$ 
guarantees that our final sample will 
contain the type of filaments that are the focus of this work.

To facilitate the identification and extraction of our target filaments, we first created a new single skeleton map 
that included all filamentary detections up to $\theta_{\mathrm{max}}=320$\arcsec. 
The skeletons of filaments detected in particular scale ranges  ($\theta/2$ to $\theta$) were initially provided 
by \getfilaments, but now these needed to be combined in order to produce one single accumulated skeleton map. 
As mentioned above, a skeleton image associated with the spatial scale range $\theta/2$\,--\,$\theta$ traces 
the crests of the filaments detected in the various single-scale images in that range. 
In this map, a skeleton pixel can therefore be quantified 
based on the number of single scales (\nscale\ between $\theta/2$ to $\theta$) the skeleton appears in. 
For practical purposes, the quantity \nscale\ becomes a measure of reliability: 
the higher the number of single scales in which the skeleton pixel appears, 
the higher the reliability of the skeleton. While this does not necessarily imply that 
skeletons with few \nscale\ are fake detections, if a filament is self-consistently 
identified in various scales it is more likely to be a robust filamentary structure, 
compared to another that looks like a filament only at one single spatial scale. In this case, its 
classification as filamentary in a particular scale could have been simply fortuitous if it 
no longer resembles a filament when its emission is examined at smaller or larger scales. 
Structures that appear as filamentary in just one or two scales are numerous, 
but this number decreases progressively as we increase the minimum \nscale. 
This behaviour can be observed 
in Fig. \ref{fig:scale_image}, which illustrates the differences in filament detection arising 
from application of different reliability (\nscale) limits. 
This is in fact the same effect observed when decreasing the signal-to-noise (S/N) detection 
limit when building a compact source catalog.  
Lowering the S/N (reliability threshold) increases the number of 
sources in the final list, but it also increases the 
chances of including spurious detections in the final sample.
Skeleton maps produced for each $\theta/2$ to $\theta$ scale range can therefore be 
associated with a reliability (significance) map, 
with pixel values representing the number of spatial scales, or \nscale, at 
which the pixel traces the ridge of a filament in this scale range.

To create an accumulated skeleton map at $\theta\sim320$\arcsec\, the significance maps 
at the spatial scale ranges of $\theta=20-40$\arcsec, $\theta=40-80$\arcsec, 
$\theta=80-160$\arcsec, and $\theta=160-320$\arcsec) were added to obtain 
a total significance image for the total $\theta=20-320$\arcsec\ range. 
In this map, the maximum number of scales associated with any given filament skeleton 
pixel then depends on its structural prominence along the entire $\theta=20-320$\arcsec\ range. 
In other words, each pixel in the significance maps represents the total number of spatial scales  
 in the $\theta=20-320$\arcsec\ range in which the pixel has been identified as a skeleton pixel of a filament.
In this work, the maximum \nscale\  (maximum reliability level) of any filament pixel in the accumulated $320$\arcsec\ 
significance map was found to be \nscale$\sim100$. 
New skeletons could then be derived from this accumulated significance map by 
thresholding at a chosen minimum \nscale\ reliability level.

\subsection{Filament Selection}
Application of a relatively low \nscale\ 
reliability level for thresholding the accumulated 
significance map allows for the detection of filamentary structures 
dominated by any scales up to $320$\arcsec. 
Depending on the distance, filaments only detected at small scales 
could be a particular type of filamentary substructure close to the resolution of the data 
without an external filamentary background. 
In the opposite extreme, detections would comprise diffuse filaments 
without prominent substructure, and that appear as `filaments' only when 
accounting for larger scales approaching $\theta\sim320\arcsec$.

According to the third criterion of our established filament definition, the 
goal of this work is to select those detections with 
`well-behaved' filamentary properties throughout the range of 
chosen (prestellar core-relevant) spatial scales. 
A high \nscale\ reliability level not only increases the robustness of the 
detection, but it also selects filaments that are more self-consistent 
throughout the entire range of scales. This results from the limited 
number of spatial scales detected in any given  $\theta/2$ to $\theta$ 
scale-range provided by \getfilaments\, and therefore the need of having 
contributions at more than one spatial scale range 
in order for the accumulated \nscale\ to reach the chosen higher minimum 
reliability level. 

We tested the robustness of our results by repeating our analysis 
using skeleton maps tracing filaments with 
minimum \nscale\ reliability levels at regular intervals of 
\nscale$=10$, $30$, $50$, and $70$. 
The default level used in our analysis was chosen as \nscale$=10$, which provided a 
filament sample that overlapped at the $\sim80$\% level with the structures traced by the filament 
detection algorithm from \citet{schisano2014}. We used the other reliability levels 
to test the robustness of the results derived with the \nscale$=10$ sample.

In addition to spatial self-consistency, 
we applied additional reliability criteria to the final skeletons 
to account for detection (sensitivity) limitations and our previously defined filamentary properties.
Pixels associated with filament skeletons required a S/N$>5$ on the original column density map 
with respect to their associated (column density) uncertainty.
These pixels were also required to be 
above local background level by 
at least a factor of 1.2 for a filament to be classified 
as a prominent enough structure with respect to its environment.
However, we imposed no limit on the minimum column density associated with 
background or filament in order to be able to account for the most diffuse fields 
in our sample (Table \ref{table:fields}).

Remaining reliable filamentary `segments' and pixels were 
removed by excluding all skeletons 
with length less than $4$ pixels ($\sim40$\arcsec). 
The final skeleton maps were visually inspected 
to ensure that they reproduced all major filamentary structures 
in any given field with the greatest fidelity.

The complexity of the filamentary nature of the ISM and 
the flexibility of the \getfilaments\ 
algorithm clearly allows for fine-tuning and variations of 
our chosen extraction technique. 
However, our approach that a filament should be classified as 
`significant' in the chosen accumulated 
spatial scale not only ensures the identification of reliable filaments relevant 
for the key star formation scales, but also minimizes the 
filament sub-fragmentation effect while keeping `real' filamentary substructure. 
Should contribution from scales larger than $320$\arcsec\ be necessary in order 
to make a filament `significant' (\nscale\ above minimum), 
such detection would not be considered reliable to enter our classification, even if 
contributions are nevertheless present at smaller scales.
While our filament sample might therefore not be necessarily complete, 
it contains the most robust and reliable detections of prestellar 
core-relevant filaments representative of the fields of the GCC Programme. 

Here we note that our filament selection technique would not result in 
biased results regarding filament properties such as width. 
The core-scale filament selection criterium ($\theta\la320$\arcsec) 
was used for \textit{detection} purposes only, i.e., keeping those filaments that are 
significant detections (i.e., satisfying the \nscale\ requirement) at core-scales. 
While this excludes filaments prominent only at larger scales, 
this does not preclude our final filament sample 
from having contributions at larger radii (larger widths). 
Filament characterisation and profile analysis was performed on the image of 
fully-reconstructed filaments at all scales (separated from the background), 
therefore taking into account all possible contributions 
from all available scales in any given field \citep{getfilaments}.

\begin{figure}[ht]
\centering
\includegraphics[scale=0.55,angle=0,trim=0cm 0cm 0cm 0cm,clip=true]{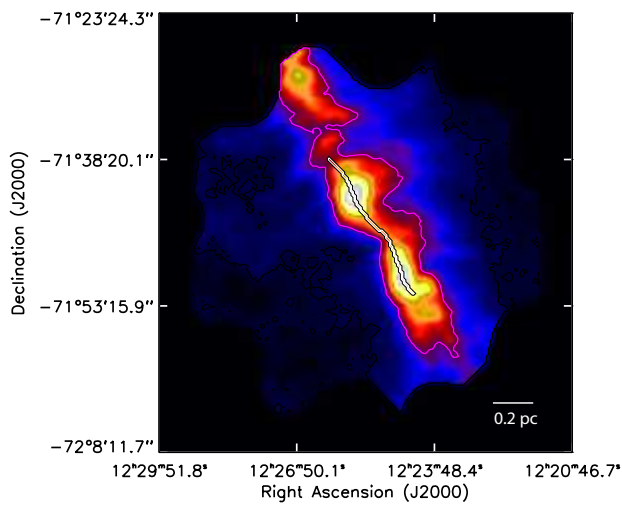}
\caption{Column density map of the GCC field G300.86-9.00 with the skeleton of the main filament superimposed. Skeleton was obtained with the \getfilaments\ algorithm and \nscale$=10$. The \av$=1$\,mag (black) and \av$=3$\,mag (magenta) contours are shown as reference.}
\label{fig:example}
\end{figure}

\section{Analysis: Filament Profiling and Characterisation}\label{sec:analysis}
Analysis of filament properties was carried out on the sample satisfying all 
the reliability requirements at the crucial star forming scales. 
Fig. \ref{fig:example} shows an example for the GCC field G$300.86-9.00$ 
with a minimum reliability level of \nscale$=10$. 
Filaments in other fields with this reliability level and 
satisfying all our reliability criteria are shown in Appendix \ref{sec:images}.

Radial column density profiles were obtained along directions 
perpendicular to each pixel of the filament skeleton (crest) using the \textit{fmeasure} utility 
(part of \getfilaments).
For profile derivation, we used two sets of background-free column density maps (see Section \ref{sec:detection}): 
the filament-only \nh\ map, and the filament$+$compact source \nh\ map, obtained by adding the filament-only and source-only 
maps provided by \getsources/\getfilaments. We produced two different filament catalogs from these sets of images, defined as the `source-subtracted' (SS) and `source-included' (SI) samples, respectively.

Profiles were averaged along the 
length of the filament to derive mean radial profiles, for the entire filament and separately 
for both of its sides. An example of such a profile is shown in Fig. \ref{fig:profile1}.
For consistency with previous studies, each averaged profile was fitted 
with an idealized model of a Plummer-like (\citealp{whitworth2001}; \citealp{nutter2008}) 
cylindrical filament (convolved with a $40$\arcsec\ beam) of the form 

\begin{equation}
\rho_p(r) = \frac{\rho_\mathrm{c}}{[1+(r/R_{\mathrm{flat}})^2]^{p/2}}.
\label{profile}
\end{equation}

Here, $\rho_{\mathrm{c}}$ is the central density, $R_{\mathrm{flat}}$ is the size of the inner flat portion of the filament profile, and \textit{p} is 
the exponent (\textit{p}$\sim2$) that characterises the power-law behaviour of the profile at larger radii. 
The inclination angle of the filament relative to the plane of the sky was assumed to be equal 
to zero.
The fitting process was carried out using a non-linear least squares minimization IDL routine 
based on {\sc mpfit} \citep{mpfit} tracing the entire profile as measured from the background-free 
\nh\ map or to the 
point of overlap with another filamentary structure.
Only those profiles with data extending past the half-maximum width 
of the filament were used in our analysis. 
This ensured that the overall shape of the profile and the parameter estimates obtained from the fit were reliably constrained.

\begin{figure}[ht]
\centering
\subfigure[]{%
\label{fig:profile1}
\includegraphics[scale=0.47,angle=0]{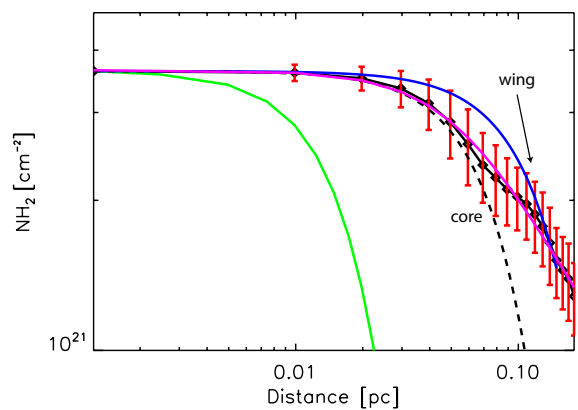}
}\\%
\subfigure[]{%
\label{fig:profile2}
\includegraphics[scale=0.47,angle=0]{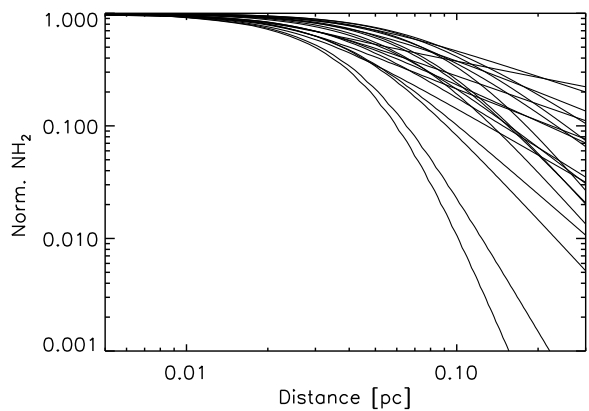}
}\\%
\caption{a) Skeleton-averaged filament column density profile (black curve) with best-fit (magenta) Plummer-like function. The best-fit (blue) Gaussian function to the data (no fitting radius restriction) is shown relative to the final best Gaussian function representing the innermost regions of the profile (black dashed-curve) and the $40$\arcsec\ telescope beam (green). Error bars show the dispersion of column density along the filament. The Gaussian-like inner regions of the profile (\core\ component) can be separated from the wing component of the filament, associated with the (outer) power-law regions of the Plummer profile not accounted for by the Gaussian function. b) Sample of 20 best-fit Plummer-like models (normalised: \nh$=1$ at radial distance from filament centre $R\approx0$) with varying proportions of \core\ and wing components.} 
\label{fig:profiles}
\end{figure}

\begin{figure}[ht]
\centering
\includegraphics[scale=0.50,angle=0, trim=1.5cm 0cm 0cm 0cm, clip=true]{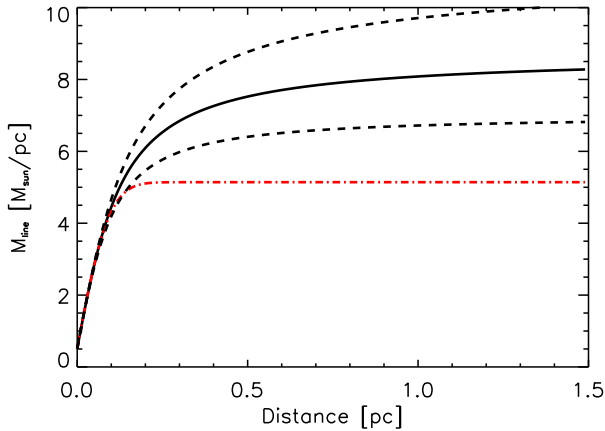}
\caption{Total linear mass density (\mtot) as a function of radial distance from the filament crest for a Plummer-like function (solid black curve) with $R_{\mathrm{flat}}=0.10\pm0.01$ and $p=2.9\pm0.2$, relative to a Gaussian with FWHM$=0.18\pm0.02$\,pc (dot-dashed red curve). Changes in the \mtot$-R$ relation depend on the uncertainties in the Plummer parameters $R_{\mathrm{flat}}$ and $p$ (black dashed-lines). The Gaussian FWHM was defined as the point where an increase in 
Gaussian width overestimates the linear mass density of the inner parts of the Plummer function.}
\label{fig:gaussian}
\end{figure}

The combination of a flat and a power-law component of a Plummer-like function can generally 
reproduce the observed profile accurately (Fig. \ref{fig:profile1}).
However, issues such as the correlation of $R_{\mathrm{flat}}$ and the \textit{p}-exponent, 
or the presence of profiles already accurately fitted by a simple Gaussian, can make the true physical 
meaning of the best-fit Plummer parameters, and their usability toward filament characterisation, 
questionable (see e.g., \citealp{juvela2012b}; \citealp{malinen2012}; \citealp{smith2014}). 
Rather than using the absolute values of $R_{\mathrm{flat}}$ and the \textit{p}-exponent, here we characterise the 
filament in terms of two alternative morphological descriptors: a \core\ component and a wing component.
Identification and separation of each of these two quantities relies in one main assumption, used already in 
previous studies such as that of \citet{arzoumanian2011}, which claims that the innermost central regions of the filament profile can be represented by a Gaussian function. 
This Gaussian-like inner component of the profile, which in this work we define as the filament `\core' component, 
can then be quantified separately from the `wing' component,
 associated with the power-law behaviour of the filament profile and which causes it to deviate
from a Gaussian-like shape at larger radii (Fig. \ref{fig:profile1}). 
The variety of filament \core-wing combinations can be observed in the sample of 20 filament Plummer-like profiles 
included in Fig. \ref{fig:profile2}.

The Plummer parameters ($\rho_{\mathrm{c}}$, $R_{\mathrm{flat}}$, and \textit{p}-exponent) 
are only used in deriving the model that fits the filament column density profile best. 
Such a model replaces the observational data 
when calculating the relative contributions of the \core\ and wing filament components to the profile. 
The total linear mass density, \mtot, can be calculated by integration of the model Plummer profile:

\begin{equation}
\mtotb = \int \Sigma_{\mathrm{model}}(r)dr=\mcoreb+ \mwingb,
\label{profile}
\end{equation}
with \mwing$=0$ for a purely Gaussian profile. Here, \mtot\ 
remains accurately determined regardless of the final value of the best-fit Plummer parameters, 
as long as the shape of the profile is well described by such parameters. 
Overall, integration of the background-free profiles beyond $\sim1.5$\,pc introduces 
variations in \mtot\ already within $1\sigma$ uncertainty of the linear mass density estimate.

\begin{table*}[t!]
\caption{GCC fields with filaments at $D\le500$\,pc$^*$ and filaments in SI-sample and \nscale$=10$}
\label{table:fields}
\centering
\begin{tabular}{l l l l l}
\hline \hline
Name&l&b&$<$BKG \nh$>^a$&$<$Filament \nh$>^b$\\ 
&[\degree]&[\degree]&[$10^{20}$\,cm$^{-2}$]&[$10^{20}$\,cm$^{-2}$]\\
\hline
G$0.02+18.02$&$0.03$&$18.03$&$9.5$&$1.1$\\
G$3.08+9.38$&$2.90$&$9.34$&$11.6$&$3.7$\\
G$25.86+6.22$&$25.86$&$6.22$&$29.7$&$3.5$\\
G$116.08-2.40$&$116.13$&$-2.45$&$20.3$&$2.1$\\
G$126.63+24.55$&$126.65$&$24.55$&$3.3$&$0.4$\\
G$150.47+3.93$&$150.36$&$3.95$&$29.6$&$9.8$\\
G$159.23-34.51$&$159.22$&$-34.24$&$8.8$&$3.7$\\
G$173.43-5.44$&$173.53$&$-5.27$&$10.9$&$0.8$\\
G$206.33-25.94$&$206.35$&$-26.08$&$2.7$&$2.2$\\
G$210.90-36.55$&$210.89$&$-36.55$&$7.0$&$2.0$\\
G$300.61-3.13$&$300.63$&$-3.02$&$17.0$&$1.3$\\
G$300.86-9.00$&$300.87$&$-9.00$&$11.87$&$5.8$\\
G$315.88-21.44$&$315.87$&$-21.45$&$3.4$&$1.4$\\
G$358.96+36.75$&$358.96$&$36.75$&$6.3$&$2.4$\\
\hline
\multicolumn{5}{l}{{$^*$ From M2015.}}\\  
\multicolumn{5}{l}{{$^a$ Average \nh\ of background.}}\\  
\multicolumn{5}{l}{{$^b$ Average \nh\ of filament$+$compact source (background-free).}}\\
\hline
\end{tabular}
\end{table*}

\begin{table*}[t!]
\caption{Parameter distributions for filament samples with reliability level \nscale$=10$}
\label{table:filas}
\centering
\begin{tabular}{l l l l l l l}
\hline \hline
Sample&Num. Detections&$<$\mcore$>$&$<$\mwing$>$&$<$\nh$>^a$&$<$BKG \nh$>$&FWHM\\ 
&&[\mpc]&[\mpc]&[$10^{20}$\,cm$^{-2}$]&[$10^{20}$\,cm$^{-2}$]&[pc]\\
\hline
SI&$42$&$5.52\pm0.63$&$7.36\pm1.32$&$21.48\pm2.90$&$21.11\pm1.79$&$0.13\pm0.01$\\
SI$_{\mathrm{sb}}^b$&$29$&$3.79\pm0.41$&$2.89\pm0.51$&$13.93\pm1.60$&$18.57\pm2.07$&$0.13\pm0.01$\\
SI$_{\mathrm{sp}}^c$&$13$&$9.37\pm1.31$&$17.34\pm2.44$&$38.31\pm6.72$&$26.79\pm3.01$&$0.13\pm0.02$\\
\hline
\hline
SS&$29$&$4.90\pm0.69$&$5.73\pm1.90$&$11.37\pm1.58$&$20.64\pm2.71$&$0.20\pm0.01$\\
SIS&$17$&$5.79\pm1.21$&$5.51\pm2.31$&$19.95\pm4.12$&$20.15\pm2.74$&$0.13\pm0.01$\\
\hline
\multicolumn{7}{l}{{$^a$ Average intrinsic (background-removed) \nh\ and standard error on the mean of crest.}}\\ 
\multicolumn{7}{l}{{$^b$ Subcritical filaments: \mtot$< $\mcrit$\sim16.5$\,\mpc.}}\\ 
\multicolumn{7}{l}{{$^c$ Supercritical filaments: \mtot$\ge$\mcrit$\sim16.5$\,\mpc.}}\\ 
\hline
\end{tabular}
\end{table*}

Quantification of the filament characteristic width (and therefore, the Gaussian \core-component) 
based on a best-fit $\chi^2$ Gaussian fit to the entire profile becomes, 
however, problematic for those cases dominated at larger radii by the power-law component of the 
Plummer-like function. Such a wing component will 
result in a poor (Gaussian) fit of the central (filament \core) region (e.g., Fig. \ref{fig:profile1}) 
that we wish to separate from the wing component.

To overcome this problem, other studies excluded the wing component
 by limiting the maximum radius used for the fit 
(e.g., fitting radius from centre of filament $R<0.5$\,pc; D. Arzoumanian; priv. comm.).
However, the variety of radii at which the background level is reached 
does not allow a common fitting radius to be defined that would work for all filaments in the GCC sample.
Furthermore, such an approach could easily introduce bias in the estimated widths 
depending on the range chosen for the fitting process (e.g., \citealp{smith2014}).
Instead of choosing arbitrary radii to constrain a Gaussian fit, we quantified the width of 
the filament Gaussian \core\ component by examining \mtot\ as a function of 
the distance from the centre of the fitted Plummer profile, comparing that to the value predicted for a Gaussian.
Fig. \ref{fig:gaussian} shows an example of the increase in \mtot\ with radius for a Plummer-like 
profile with $R_{\mathrm{flat}}=0.10\pm0.01$ and $p=2.9\pm0.2$, 
relative to that of a Gaussian, 
with FWHM$=0.18\pm0.02$\,pc. 
This value defines the width of the \core\ component. 
It is derived directly from the best-fit Plummer profile of the filament, and is 
defined as the maximum (deconvolved) FWHM that a Gaussian can have 
without overestimating the linear mass density of the derived Plummer profile. 
For larger FWHMs, it can be observed that the Gaussian \mtot-\textit{R} distribution would 
overestimate that of the Plummer profile. This occurs at the point where the power-law behaviour starts 
to dominate the shape of the Plummer distribution, at the outer parts of the filament profile.  

The linear mass density of the \core\ component, \mcore, was assumed to be equal to 
the integrated area of a Gaussian with the defined FWHM. 
The wing component thus stands for the material not accounted for by the Gaussian, 
i.e., \mwing$=$\mtot$-$\mcore.
Uncertainties on these quantities were derived by performing a similar analysis 
on Plummer profiles varying according to 
the uncertainties on the default best-fit parameters $R_{\mathrm{flat}}$ and \textit{p}-exponent, and 
which alter the dependence of \mtot\ with distance from the filament crest (Fig. \ref{fig:gaussian}). 

When our new approach for deriving the typical filament width (Gaussian FWHM) was applied to the Plummer parameters from 
\citet{arzoumanian2011}, the widths we obtained for their filaments were found to be in good agreement 
(overall well within their estimated $3\sigma$ errors) with those 
derived by these authors by performing a Gaussian fitting to their Plummer profiles for $R<0.5$\,pc.

In addition to profile fitting and linear mass density determination, each filament was 
characterised based on other intrinsic properties such as length, elongation, average crest 
column density and temperature, and local background. 
For the purpose of this analysis, the background level of a filament was assumed to be equal 
to the average value at the base of the filament crest. This quantity was obtained using the 
\nh\ background images provided by \getsources\ as secondary products for each GCC map 
(see Section \ref{sec:detection}). The background estimate was then calculated by averaging the 
values assigned in this map to the pixels coincident with the filament skeleton.
While the background component can include material in the line of sight, the location and 
proximity of the fields make this possible contribution a minor effect.

Additional filament parameters provided by \textsl{fmeasure} include an estimate of the mean curvature of the 
filament, as well as the width at half maximum of each averaged profile (not to be confused with the 
Gaussian FWHM). The averaged column density profiles were corrected when necessary for 
filament overlap or punctual substructure by averaging only those pixels unaffected by these effects.

\begin{figure*}
\centering
\subfigure[]{%
\label{fig:dis1}
\includegraphics[scale=0.40,angle=0]{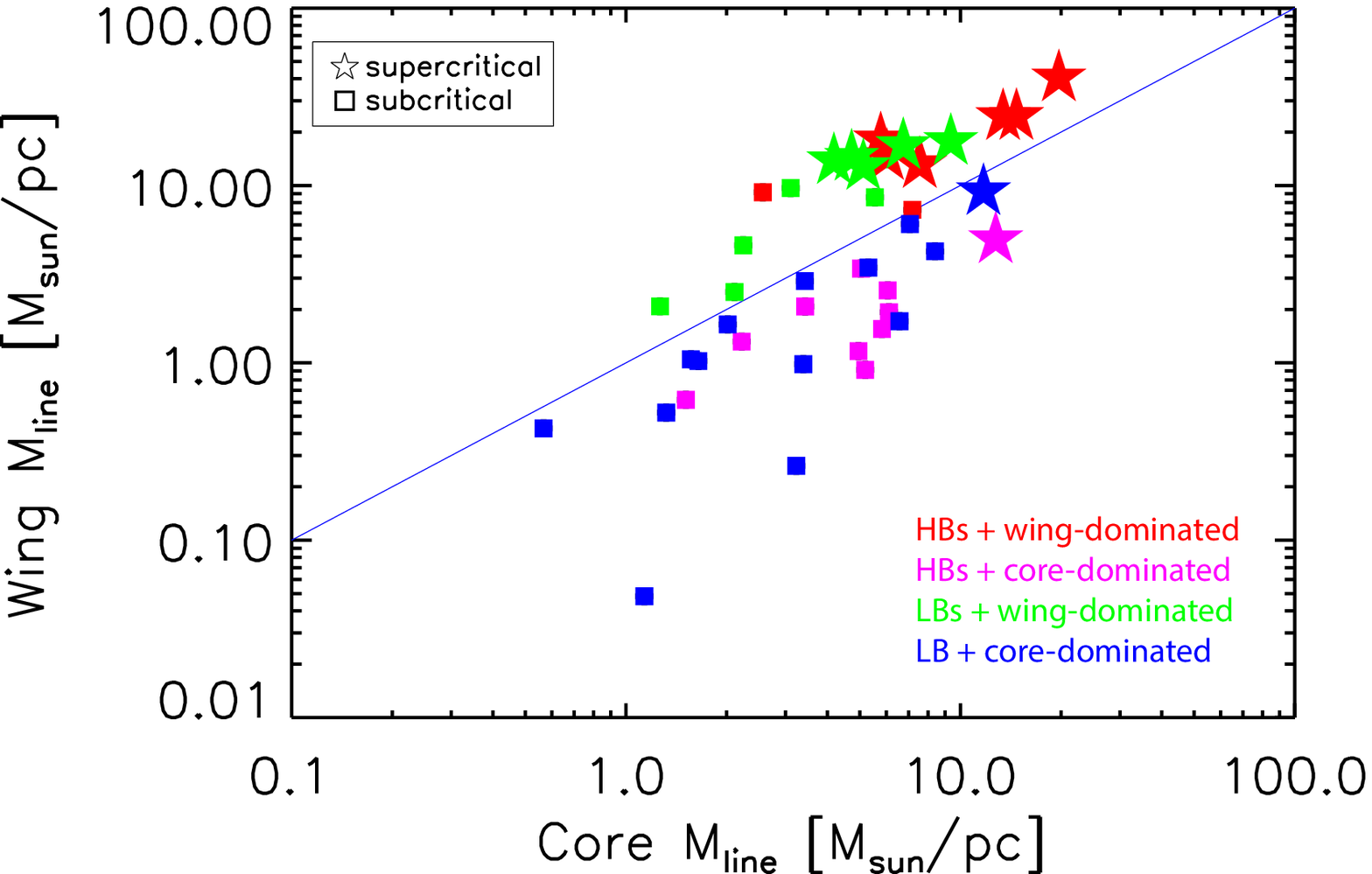}
}%
\subfigure[]{%
\label{fig:dis2}
\includegraphics[scale=0.40,angle=0]{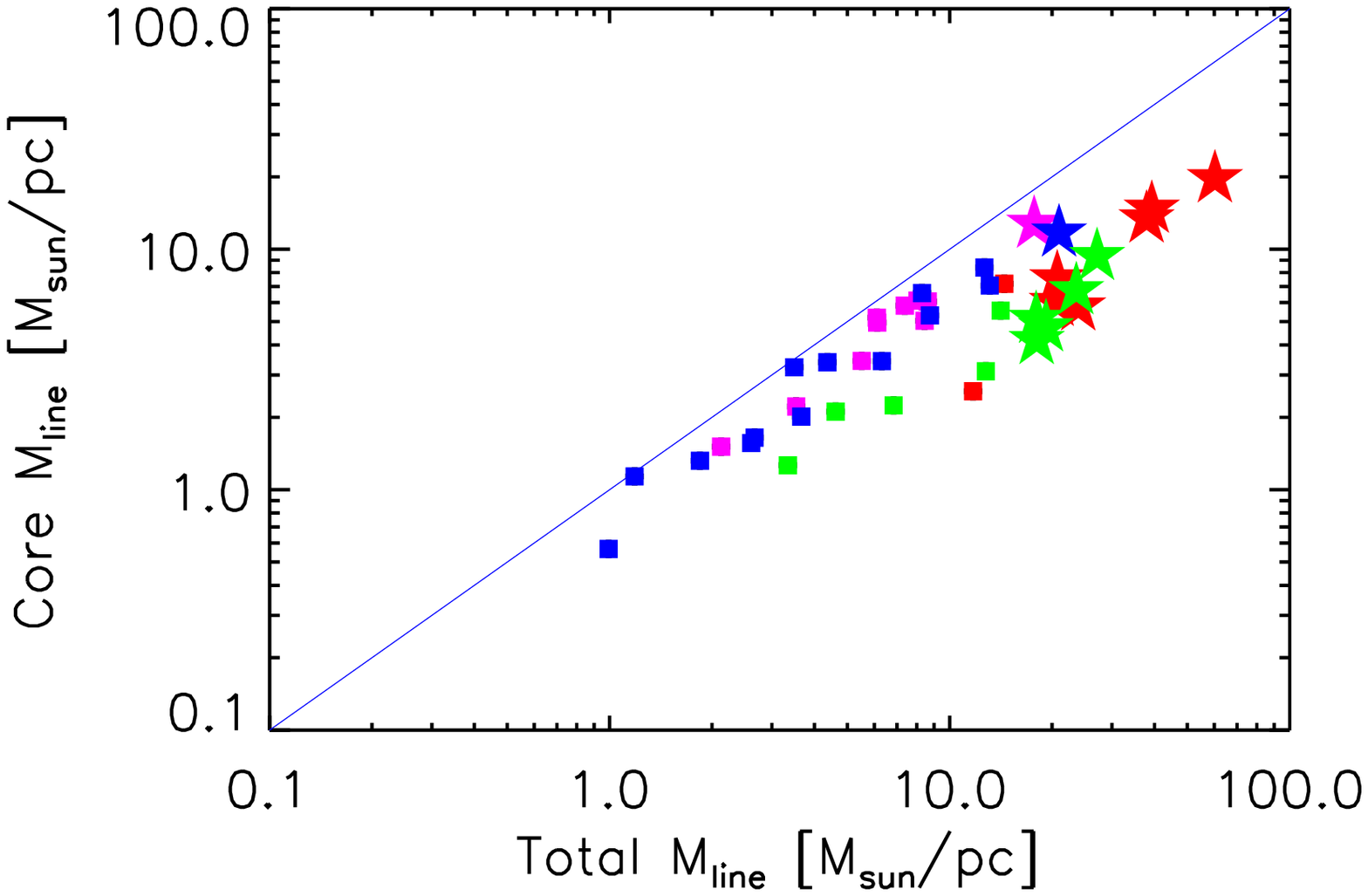}
}\\%
\subfigure[]{%
\label{fig:dis3}
\includegraphics[scale=0.40,angle=0]{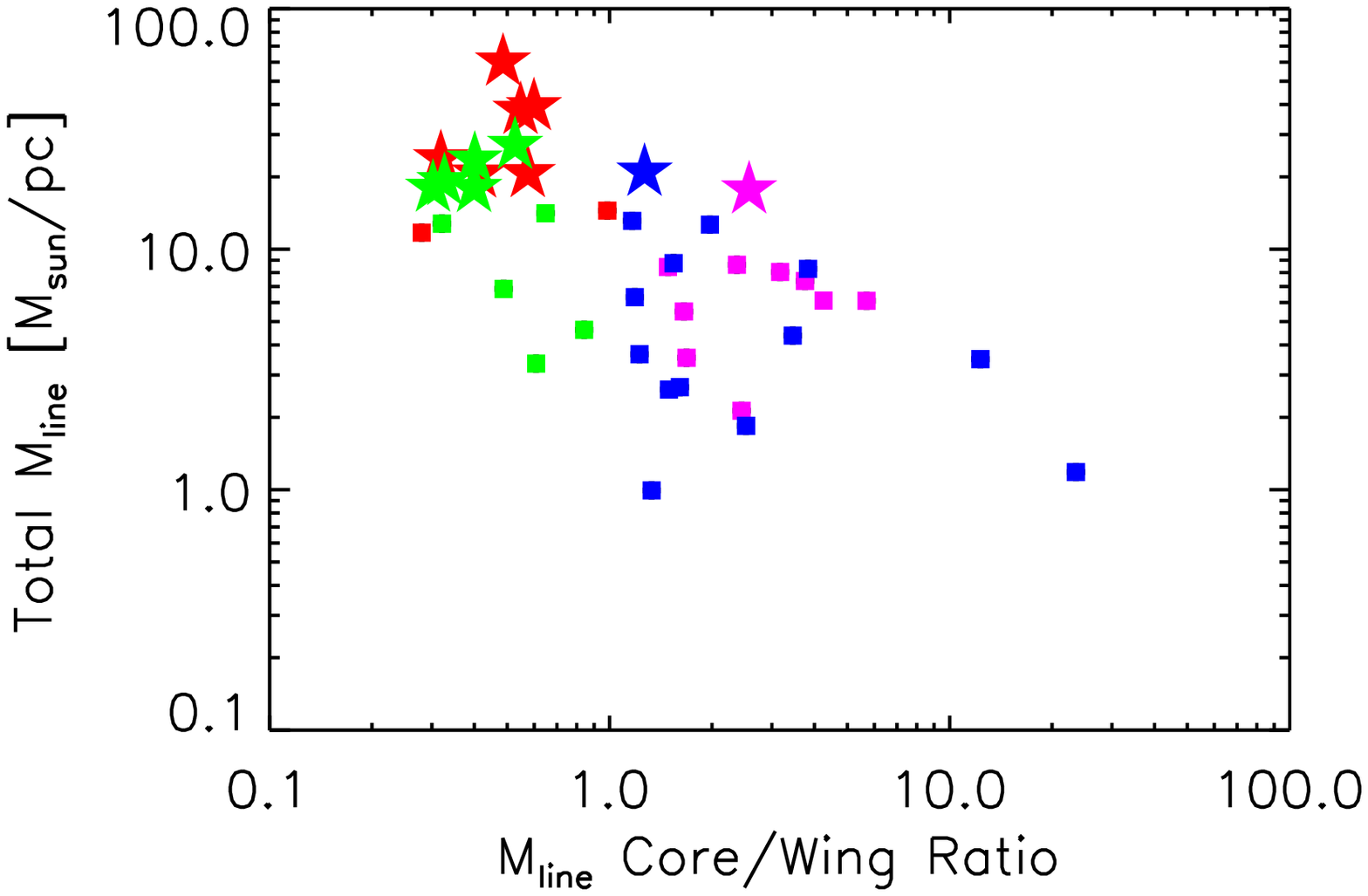}
}%
\subfigure[]{%
\label{fig:dis4}
\includegraphics[scale=0.40,angle=0]{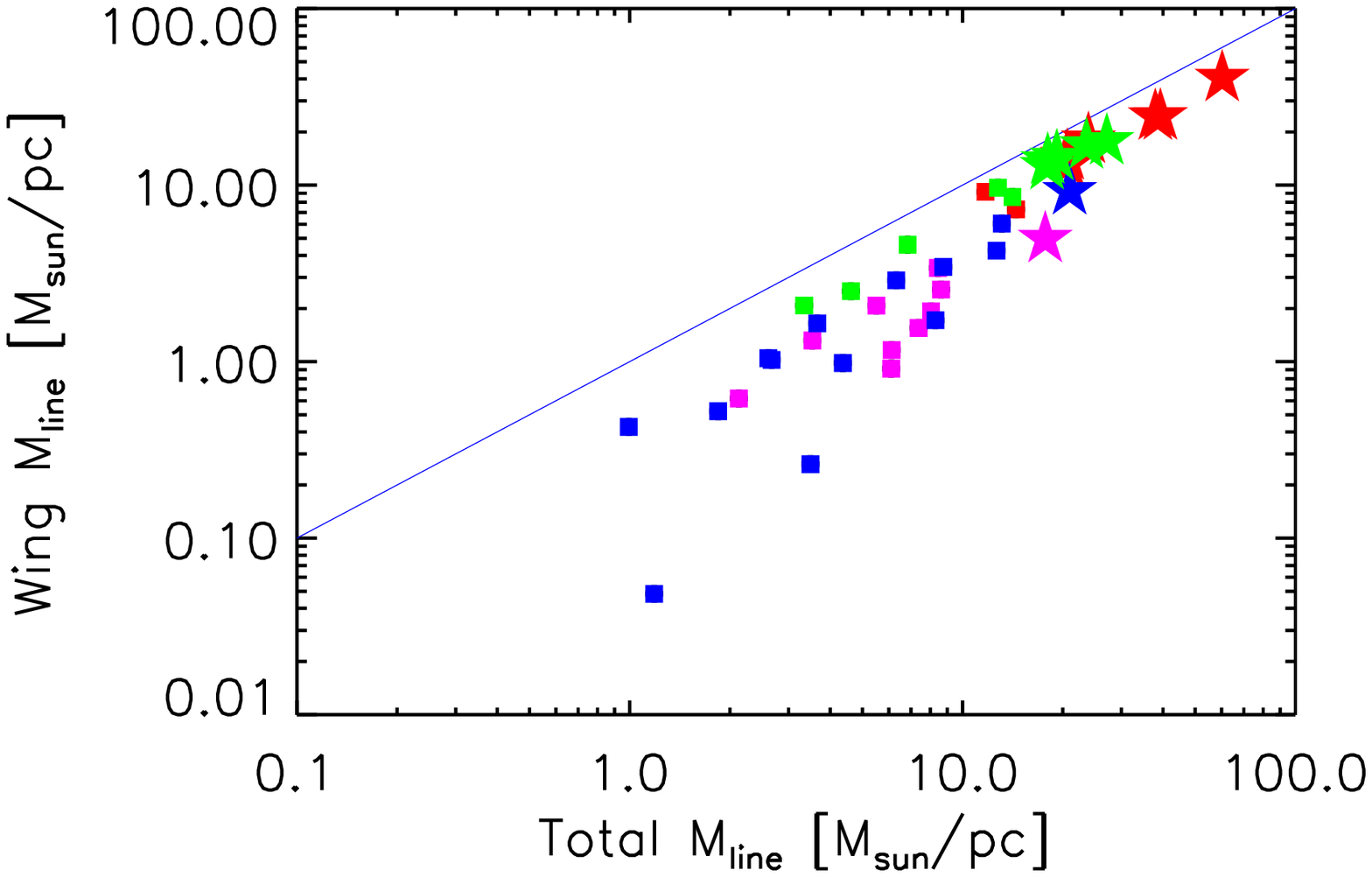}
}\\%
\caption{Distributions and correlations of key filament parameters for \fsample\ subcritical (squares) and supercritical (stars) filaments. Intrinsic total linear mass density \mtot, \core\ linear mass density \mcore, wing linear mass density \mwing, and ratio \mcore$/$\mwing. Low-background filaments (LBs: blue/green) are separated from those in denser (high-background) environments with \av$>2.2$\,mag (HBs: magenta/red) (see text). For each type of environment, filaments are further separated according to the relative contribution of \core/wing components to their \mtot: \core-dominated (blue/magenta for LBs and HBs, respectively), and wing-dominated (green/red).The 1:1 relation (blue solid line) is shown as reference.} 
\label{fig:corewing-si}
\end{figure*}

\section{Results}\label{sec:results} 
Of the 116 regions comprising the field sample of the GCC Programme 
only 38 have filamentary structure detected by \getfilaments.
The sample was further reduced with the application of our distance constraint, our filament definition, 
and the reliability criteria applied to the \getfilaments\ filament catalog. 
Only those fields at $D\le500$\,pc (including their $3\sigma$ uncertainty) 
and with a distance reliability flag$=1$ or $2$ (medium and high level of confidence; M2015) were considered for the analysis. 
We also excluded filaments that could not be accurately fitted 
by the Plummer function and those that were visually identified as not being filaments by our 
chosen definition. Examples of the latter case are 
those exclusively associated with the elongated head of cometary globules (although
 we kept those filaments trailing behind such structures). 

The final SS-sample comprises 13 fields at \textit{D}$\le500$\,pc, with 29 reliable filaments with \nscale$=10$. 
At the same reliability level, the SI-sample contains a larger number of reliable filaments ($42$ detections in $14$ fields). 
This is mainly because, without source subtraction, filaments 
are slightly narrower (satisfying our minimum elongation criterion) and have overall 
cleaner profiles that extend below half of the maximum value. 
This final sample remains highly conservative, comprising $\sim10$\% of the original filament population extracted 
from the GCC fields at  \nscale$=10$ and $D\le500$\,pc, satisfying the distance reliability criteria.
The final SI-sample filaments are shown in Appendix \ref{sec:images}, and the characteristics of their 
fields are presented in Table \ref{table:fields}.

A third filament group was created by selecting those filaments 
in the SS-sample also classified as reliable in the final SI-sample. 
This yielded 17 filaments, which we define as the `source-included' 
subsample, or SIS-sample. 
Both the SS and SIS samples were only used to 
quantify the effects of source removal on the parameters derived from the SI-sample, 
the one chosen as the default filament population for our analysis.
Table \ref{table:filas} presents and compares the average filament properties derived 
for the different samples. 
The final results, presented in more detail below, are used in the following sections to 
investigate possible environmental effects on filament formation and evolution.

Figure \ref{fig:corewing-si} provides an overview of the distribution of the \fsample\ filament 
population separated according to linear mass density and environmental criteria. 
The plots highlight the relationship between the filament structural components, 
as well as the relative dominance of each component (\core\ and 
wing) with respect to \mtot.
Our final sample comprises filaments with a wide range of intrinsic and environmental properties, 
as shown in Table \ref{table:filas} and the parameter histograms in Fig. \ref{fig:histograms}. 
Figure \ref{fig:corewing-si} also indicates that the filament sample can be 
classified into particular sub-populations depending on 
 specific structural properties. Wing-dominated filaments, for instance, 
 are more frequent at high \mtot\ than core-dominated ones, which dominate instead at 
 low \mtot.
The various filament properties are discussed in more detail in the following sections.

\begin{figure*}
\centering
\subfigure[]{%
\label{fig:histo2c}
\includegraphics[scale=0.40,angle=0]{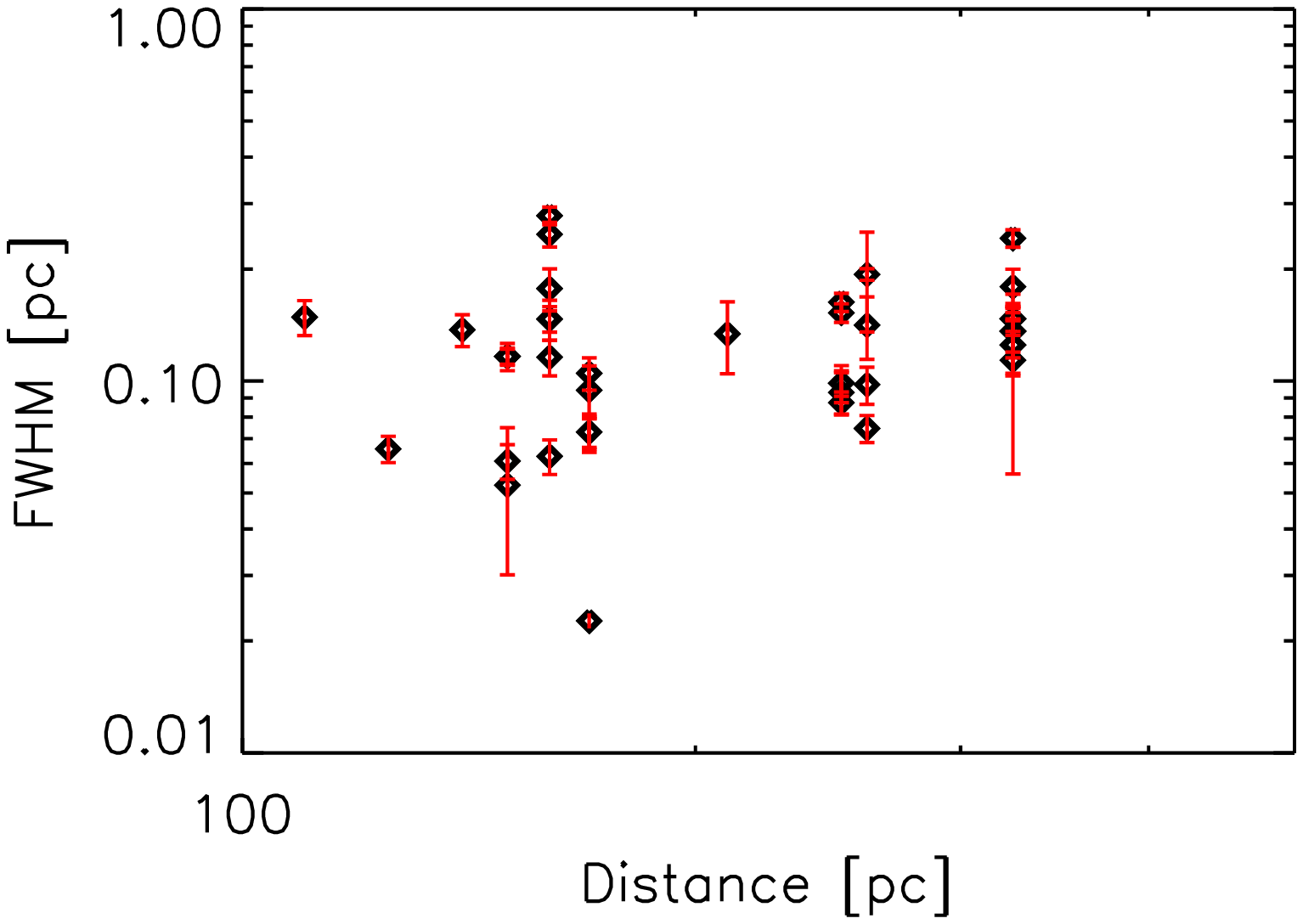}
}%
\subfigure[]{%
\label{fig:histo2cb}
\includegraphics[scale=0.40,angle=0]{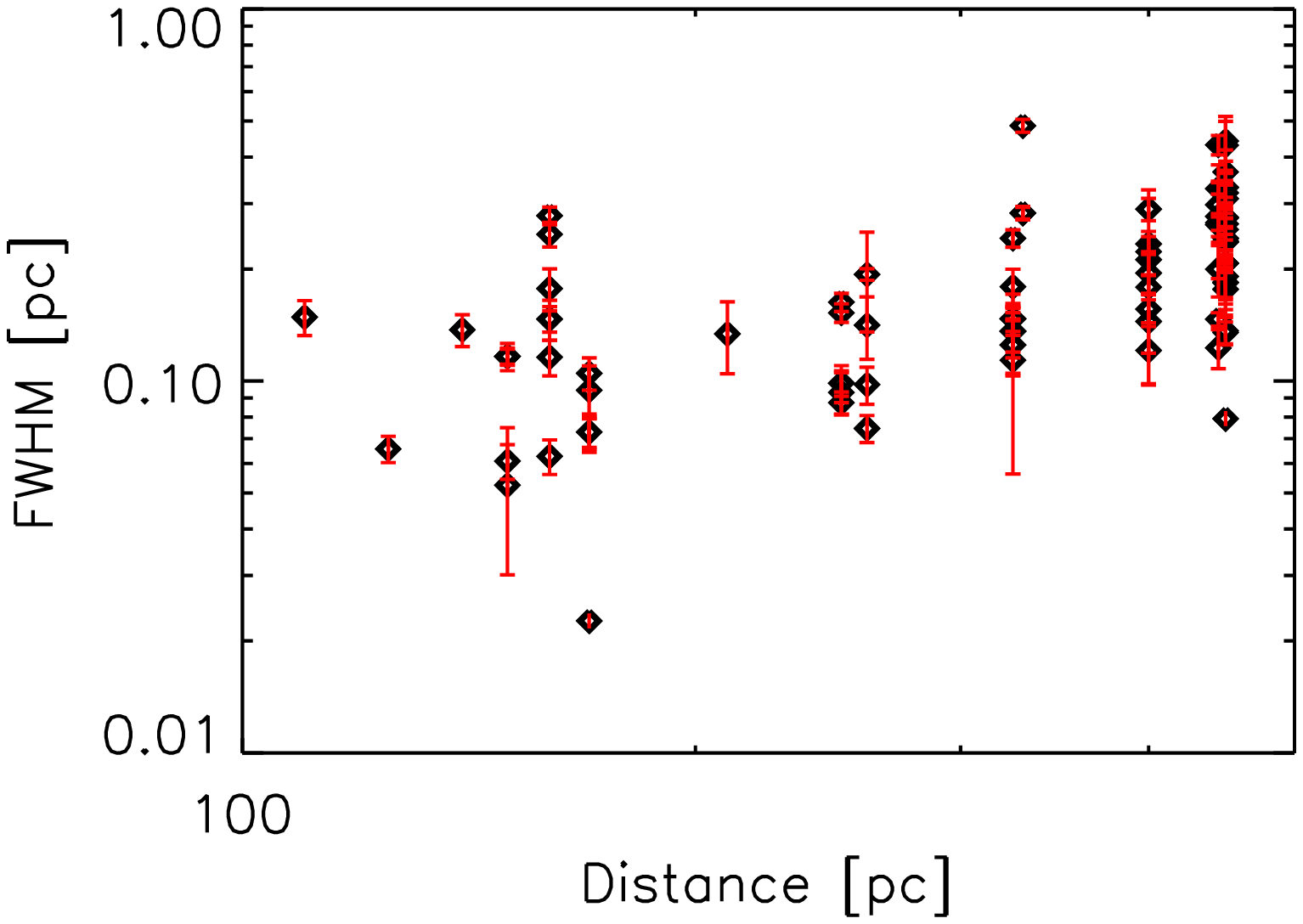}
}\\%
\subfigure[]{%
\label{fig:histo1a}
\includegraphics[scale=0.40,angle=0]{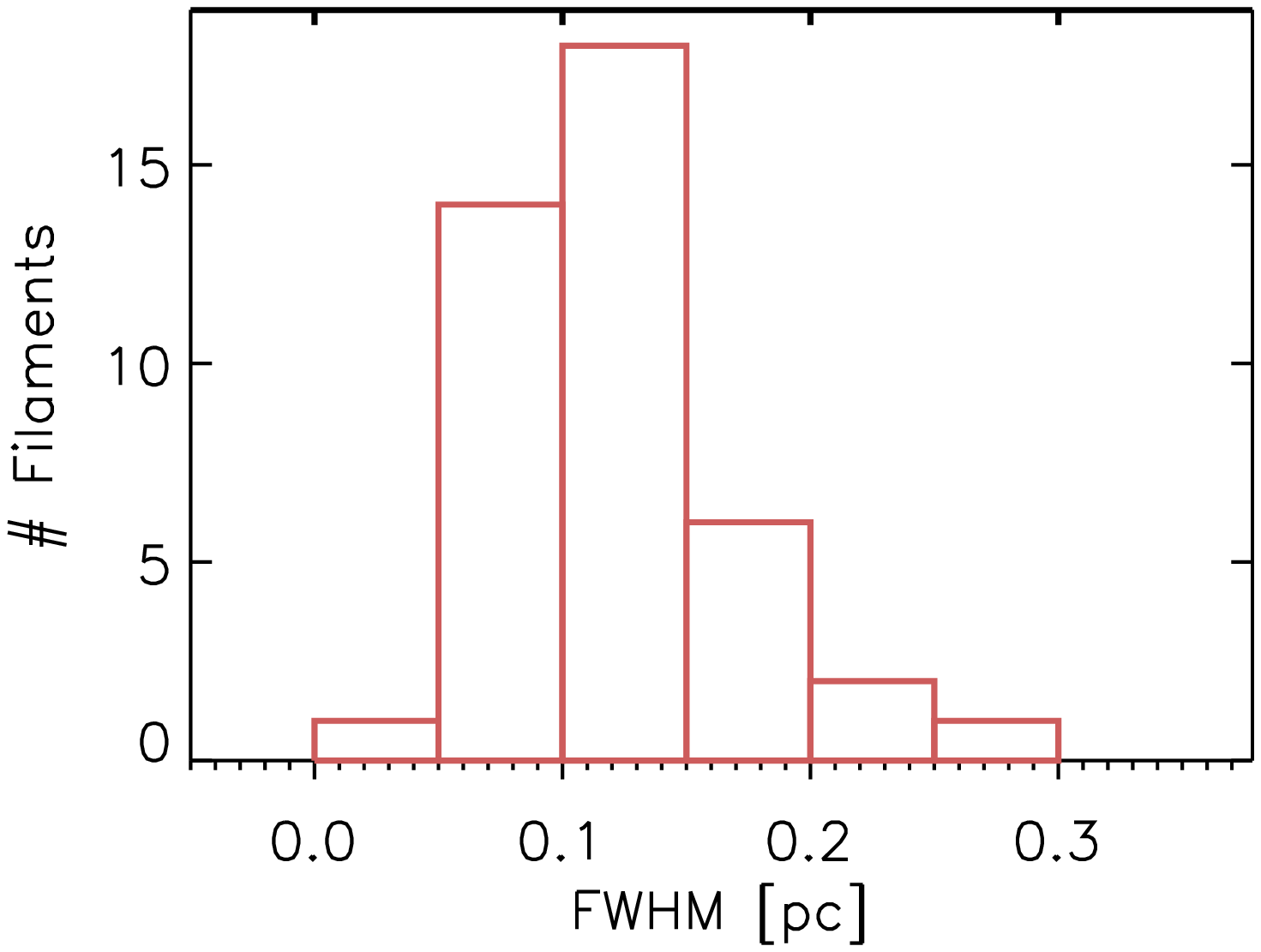}
}%
\subfigure[]{%
\label{fig:histo2d}
\includegraphics[scale=0.40,angle=0]{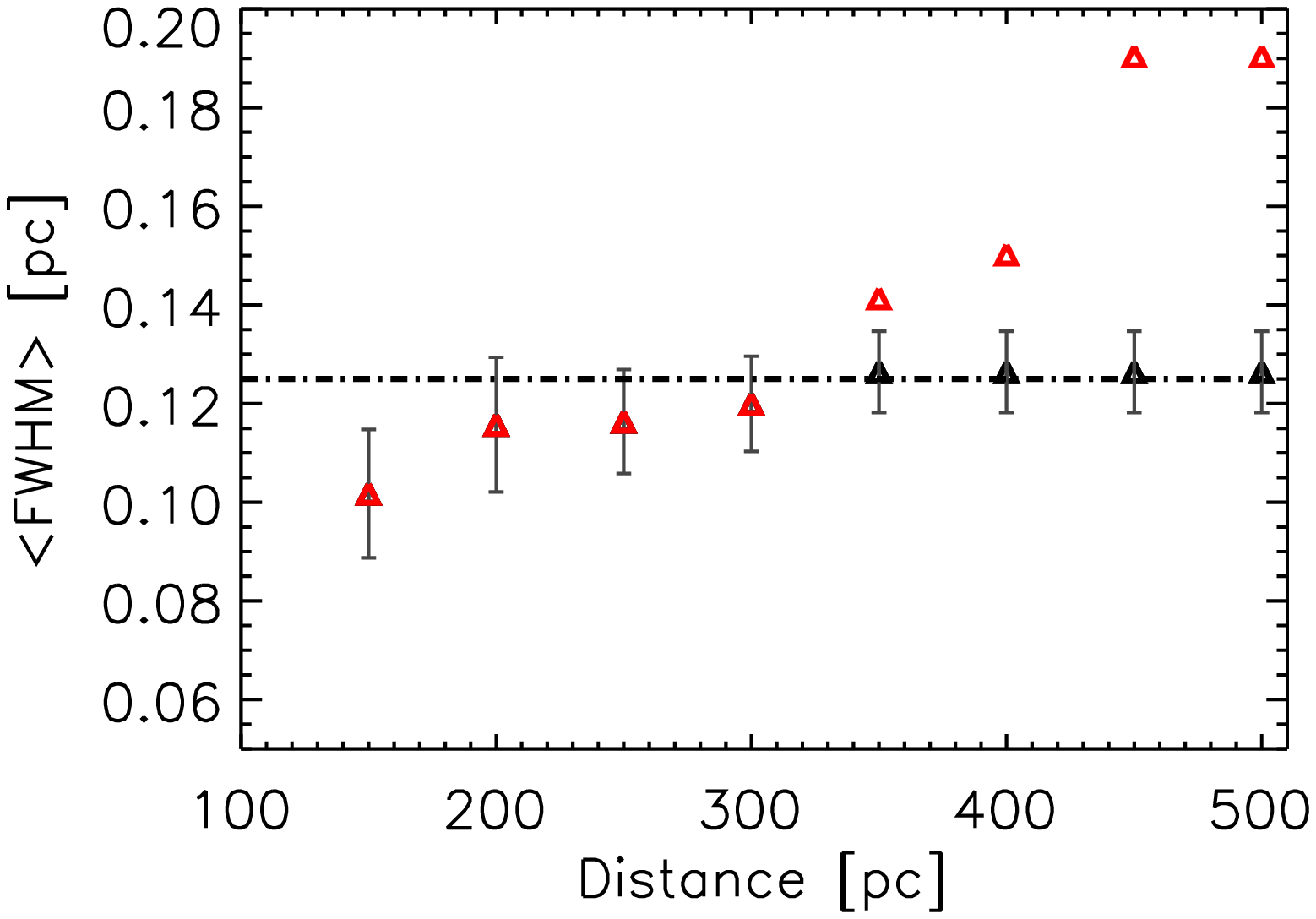}
}\\%
\caption{Filament width (Gaussian FWHM) properties for the final SI-sample: (a) filament width as a function of field distance; (b) same as Fig. \ref{fig:histo2c}, but without the $3\sigma$ distance uncertainty constraint (see text); (c) width distribution histogram; (d) variation of the (accumulated) average FWHM of filaments with distance for the final sample (triangles with error bars) relative to that of the filament sample in Fig. \ref{fig:histo2cb} (red triangles). Dash-dotted line in Fig. \ref{fig:histo2d} marks the median FWHM of the SI-sample.} 
\label{fig:histo_widths}
\end{figure*}

\subsection{Filament Widths}
The width distribution and its dependence on distance for the SI-sample are shown in Fig. \ref{fig:histo_widths}.
As seen in Fig. \ref{fig:histo1a} (bin size$=0.05$\,pc) the filament population is highly peaked, with a median value of $\approx0.13$\,pc and a 
standard deviation of $0.05$\,pc. These characteristic width and dispersion are somewhat larger than those quoted for 
filaments in nearby fields of the Gould Belt Survey (e.g., $0.09\pm0.04$\,pc; \citealp{andre2013}), possibly more in tune 
with predictions from other observational and theoretical studies (e.g., \citealp{juvela2012a}; \citealp{kirk2015}).

\subsubsection{Effects on the Width Distribution}
Many factors can affect the observed filament width distribution, from intrinsic differences between populations 
to more systematic effects, such as distance, filament selection, and source removal.

Distance appears to have a clear influence on the measured width, with the mean FWHM of the population increasing when approaching 
the telescope resolution limit. This is evident in Figs. \ref{fig:histo2cb} and  \ref{fig:histo2d} (red curve), which include all the fields with $D\le500$\,pc 
in M2015 with filaments detected by \getfilaments\ (without excluding fields not satisfying the criteria of $D\le500$\,pc when including their $3\sigma$ distance 
uncertainty. The average width of the population remains close to constant (FWHM$\sim0.12$\,pc) up to $D\sim300$\,pc, after which 
the mean FWHM increases with distance. The same trend of increasing filament width with distance, albeit less pronounced, is still observed in Fig. \ref{fig:histo2c}. 
This result is likely due to a combination of two main effects.

First, resolution and confusion can decrease the number of detections of $\sim0.1$\,pc-wide filaments at large distances. 
However, and considering the common hierarchical nature of filaments, unresolved (or barely resolved) filaments could have 
been detected but only as part of their larger scale (filamentary) host (see e.g., \citealp{juvela2012a}; \citealp{hacar2013}). 
In this case, these could appear as resolved filaments, 
albeit with larger widths, therefore producing the increase in 
average FWHM with distance.  
Similarly, small asymmetries in shape and orientation  
along the filament length (wiggles) might also be undistinguishable
at large distances. This effect could result in such filaments appearing 
more  `straight' and with larger average widths, due to the inclusion of the 
unresolved asymmetries in the overall profile.

Second, our filament detection method was fine-tuned to ensure the extraction of 
all filaments that are significant detections at the key \textit{physical} (linear) core-scales. 
However, the large range of distances considered ($\sim100-500$\,pc) might still have led 
to the inclusion of some large-scale filamentary structures in our final sample if present in the GCC maps.
This is due to the use of a common \textit{observed} (angular) spatial scale threshold for all fields, 
which means that (physically) large scale filaments that would be excluded in the nearest fields could 
have been included in those at larger distances (same angular scale in both cases). 
Such filaments could then make it to the final sample without being relevant at core-scales if 
they fulfil the \nscale\ requirement by using their contributions from larger physical scales. `Contamination' by 
these structures would affect primarily the fields at the farthest distances (e.g., Fig.\ref{fig:histo2cb}), but could also affect other fields at 
intermediate distances to a lesser degree.

The magnitude of the effect on the filament width caused by source removal is most likely dependent on the proportion (and location) of 
the source contribution relative to that of the host filament. Its relevance would also depend on the choice to 
include or exclude compact sources as part of the filamentary structure and evolution. 
Based on our findings, however, an influence of the source component 
on filament modelling could then explain (or significantly contribute to) the presence of 
a wider width distribution when treating a source-subtracted sample 
composed of filaments with different degrees of source contribution 
(e.g., SS vs SIS sample; Table \ref{table:filas}). 

\subsubsection{Identification of Core-scale Filaments}
In order to best constrain an evolutionary process leading to star-forming filaments it is crucial to minimize 
all possible systematic effects. With the effects of resolution and our filament selection method affecting primarily 
those fields at larger distances, a possible solution would be to reduce the distance range to a maximum of $\sim300$\,pc 
(e.g., the point of increase in $<$FWHM$>$ in Fig. \ref{fig:histo2d}). 
However, based on telescope resolution, our target filaments could in principle 
still be detected up to $D\sim500$\,pc. Furthermore, 
as mentioned above, these effects can also still impact the filament population of nearby fields.
Based on this analysis we therefore chose not to reduce the distance upper limit. 
Instead, we excluded those fields neighbouring the resolution limit of $D\le500$\,pc 
in which this effect is expected to be most prominent, 
and whose distance uncertainty ($3\sigma$) could actually place them beyond this limit.
As seen in Fig. \ref{fig:histo2d}, this process significantly minimizes 
the prominent increase in width at large distances observed for the filament sample, resulting in a 
relatively flat  $<$FWHM$>$-$D$ distribution that argues in favour of a characteristic (average) 
filament width for regions of low-mass star formation in the solar neighbourhood. A similar conclusion was already 
reached by \citet{arzoumanian2011} and similar \textit{Herschel}-based studies (e.g., \citealp{andre2013}). 

Application of the $3\sigma$ distance uncertainty requirement shifts the 
average filament width from 
$0.19\pm0.01$\,pc to $0.13\pm0.01$\,pc for the final SI sample. 
However, we note that these values cannot be compared with those 
found when treating the entire (mixed) filament population. 
If considering all types of filaments at all scales, and 
taking into account their well known hierarchical nature, the mean of the entire filament population would 
most likely shift to a higher value above, therefore more in line with the findings of other studies 
(e.g., \citealp{juvela2012a}; \citealp{schisano2014}; \citealp{smith2014}). 

\begin{figure*}
\centering
\subfigure[]{%
\label{fig:histo_all1}
\includegraphics[scale=0.4,angle=0]{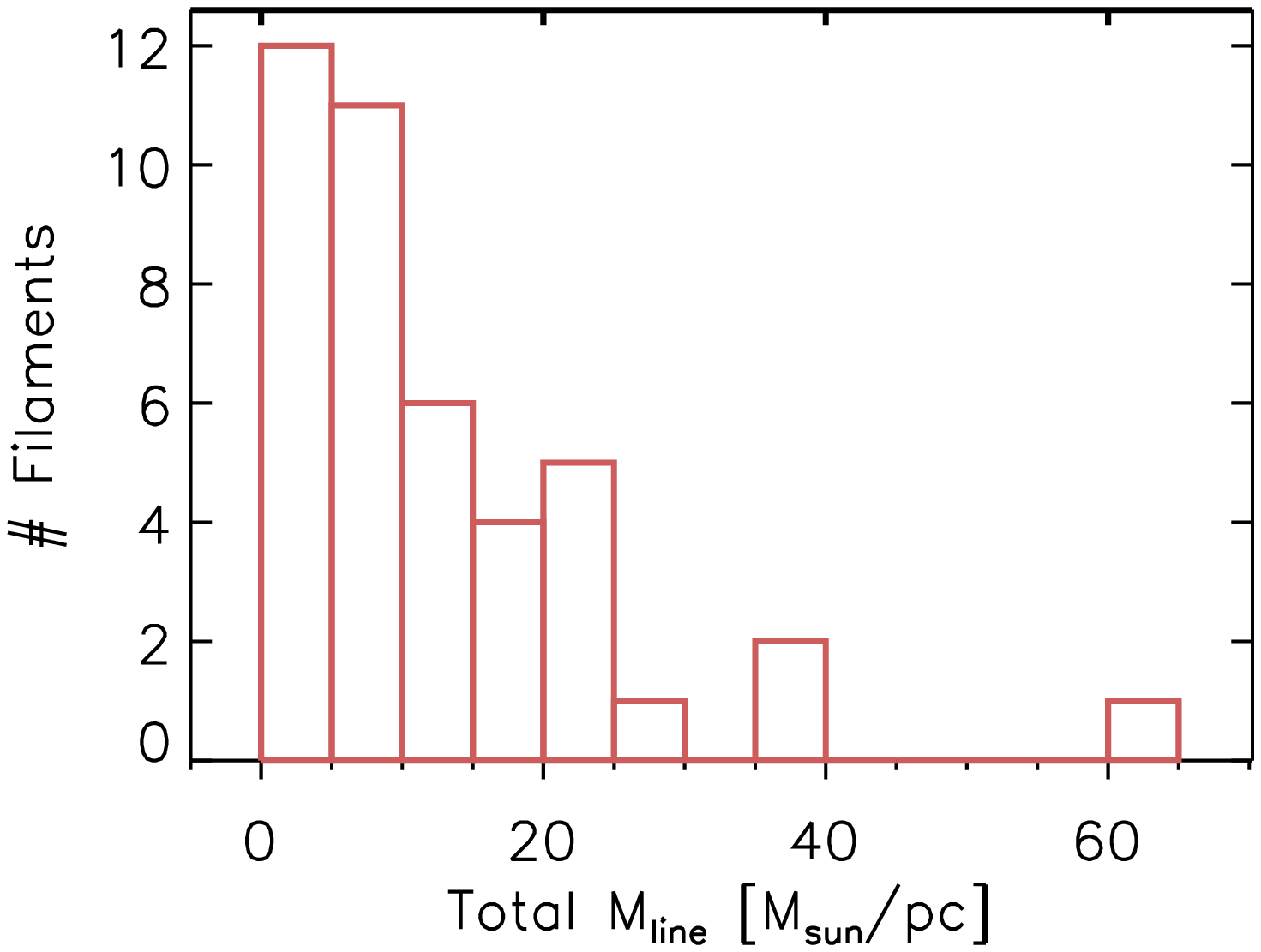}
}%
\subfigure[]{%
\label{fig:histo_all2}
\includegraphics[scale=0.4,angle=0]{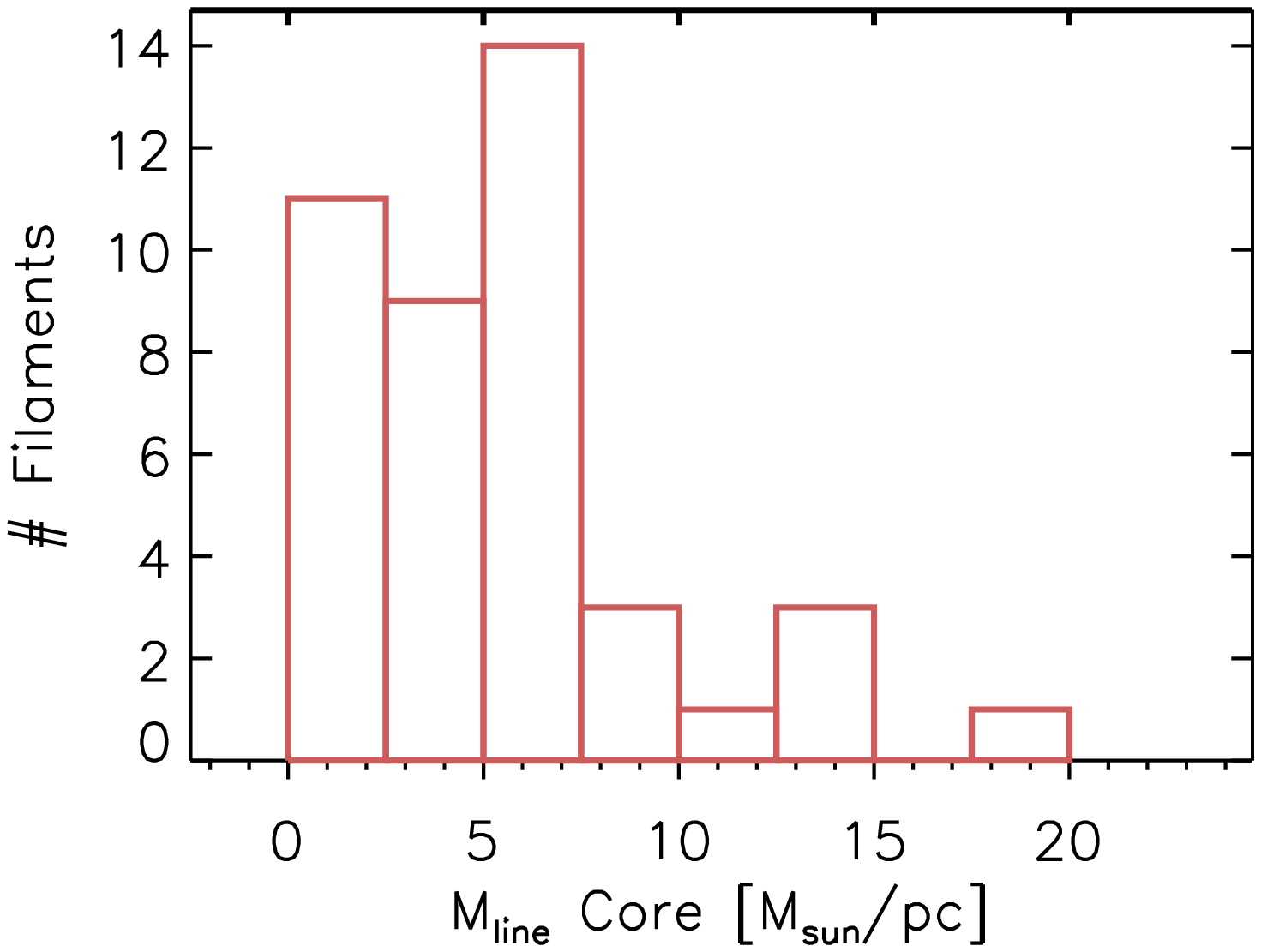}
}\\%
\subfigure[]{%
\label{fig:histo_all3}
\includegraphics[scale=0.4,angle=0]{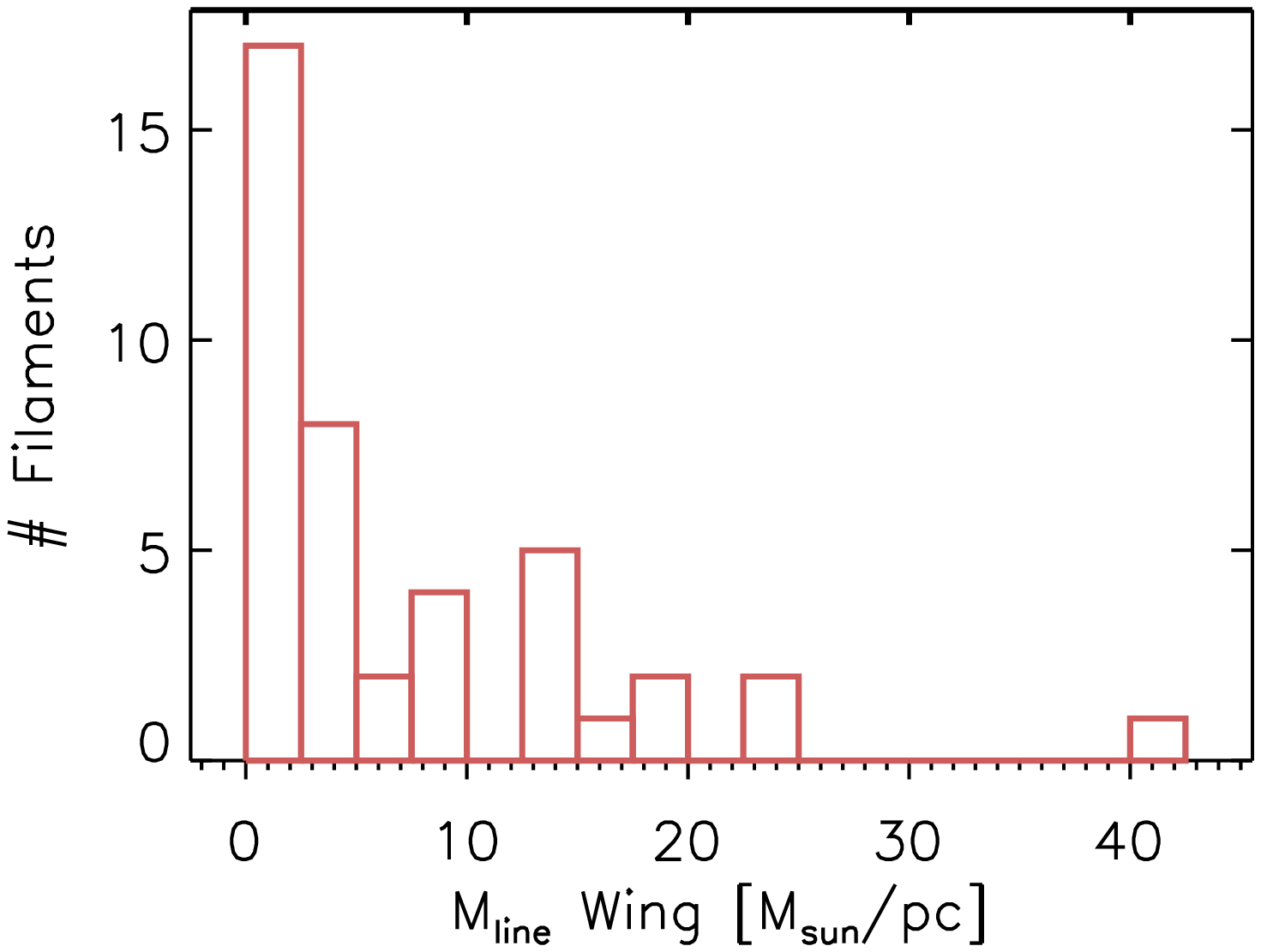}
}%
\subfigure[]{%
\label{fig:histo_all4}
\includegraphics[scale=0.4,angle=0]{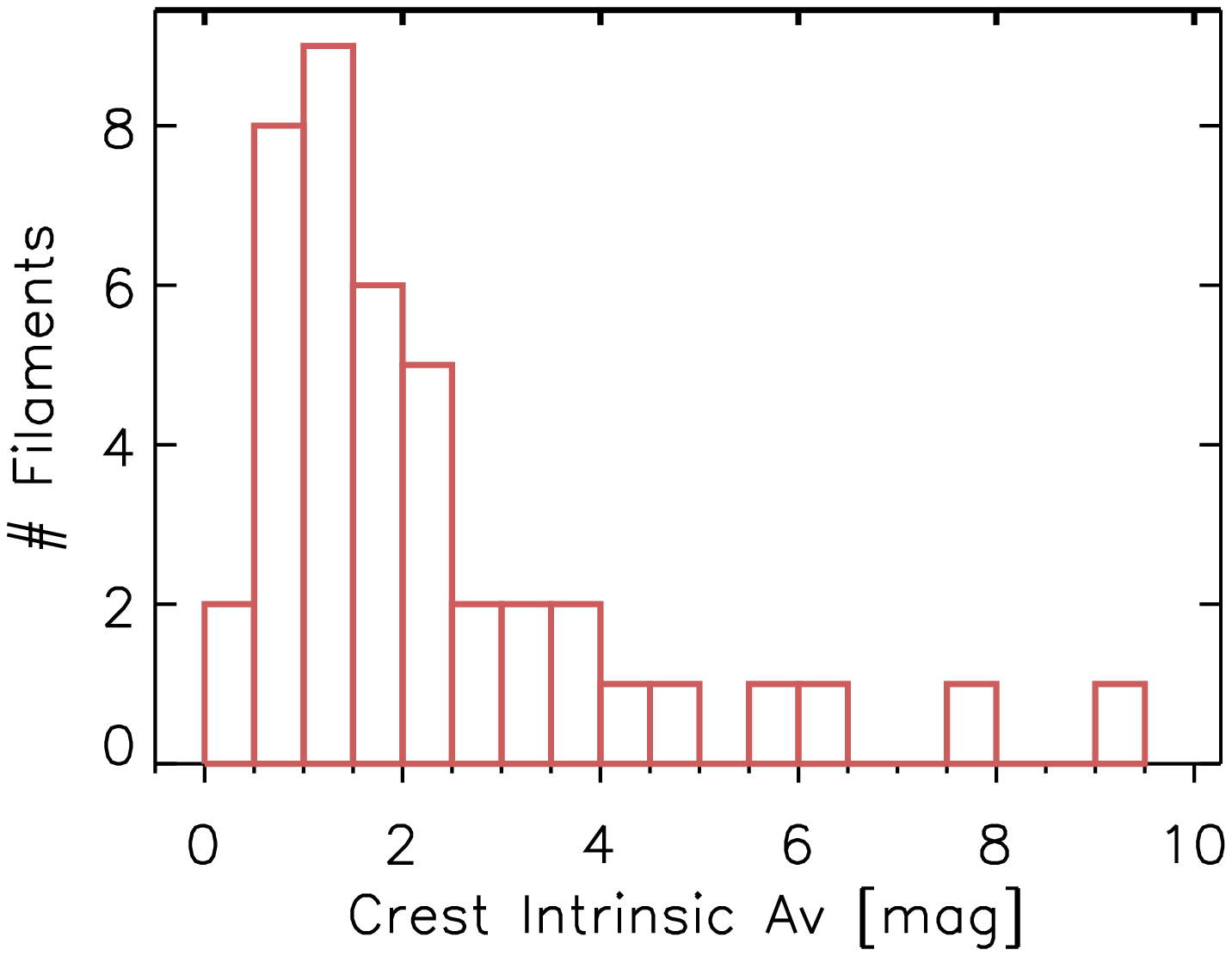}
}\\%
\subfigure[]{%
\label{fig:histo_all5}
\includegraphics[scale=0.4,angle=0]{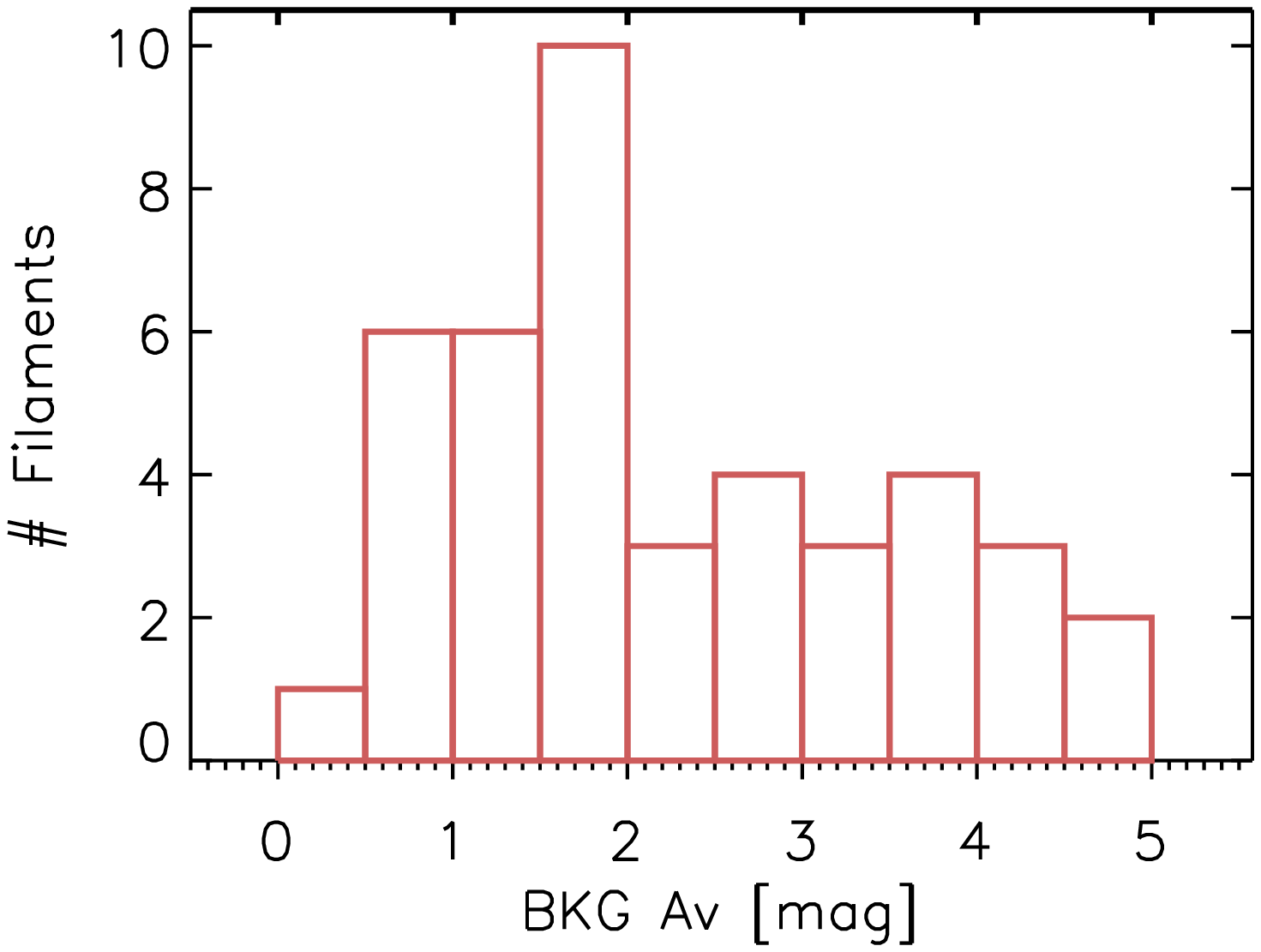}
}%
\subfigure[]{%
\label{fig:histo_all6}
\includegraphics[scale=0.4,angle=0]{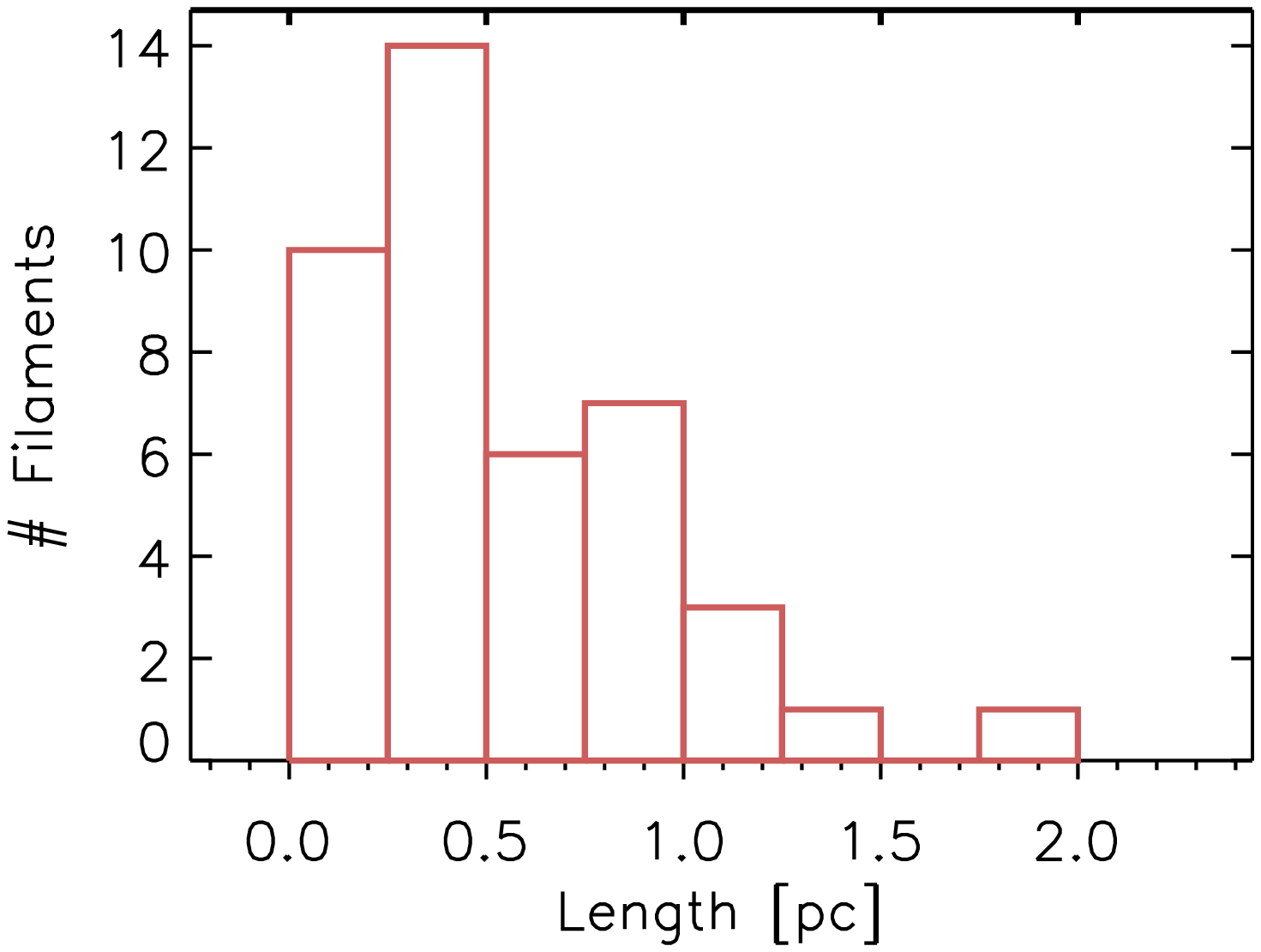}
}\\%
\caption{Histograms of key filament parameters for reliable filaments including source contribution (SI-sample): (a) Total linear mass density (\mtot); (b) \core\ linear mass density) (\mcore); (c) wing linear mass density (\mwing); (d) crest central \av; (e) background \av; (e) filament length. Blue shading represents the distribution for the \fsample\ filament sample.} 
\label{fig:histograms}
\end{figure*}

\subsection{Filament Length}
With a mean length of $\sim0.5$\,pc (Fig. \ref{fig:histograms}: \ref{fig:histo_all6}), 
our population clearly differs from the typical pc-scale filaments investigated in 
other studies (e.g., \citealp{hennemann2012}; \citealp{palmeirim2013}; \citealp{schisano2014}). 
Such short average length could be associated with real filamentary substructure, 
but it may not be representative of the overall true mean of the population. 
The size of the GCC fields (e.g., M2015) 
already imposed an upper limit on the length of the detections. 
In addition, the reliability criteria applied in our 
filament extraction method also frequently resulted in the extraction of just the most reliable `segments' 
of otherwise longer filamentary structures. The result is a sample more consistent 
with the sub-pc `fibers' \citep{andre2013} analysed in \citet{hacar2013}, 
or the `branches' of the main filaments presented in \citet{schisano2014}. 
As mentioned in the latter study, these shorter structures could be, 
however, more revealing than the larger ones. The study of more localised regions should be more 
sensitive to small variations in physical filament properties than results averaged 
over scales many times above that of a typical prestellar core or clump.

\subsection{Stability}
Crest \nh\ values and total linear mass densities are overall of 
the same order as those estimated in previous studies 
(e.g., \citealp{arzoumanian2011}; \citealp{schisano2014}). 
Based on total linear mass density criteria, most GCC filaments ($\sim70$\%) are subcritical in nature (\mtot$<16.5$\,\mpc), 
while only one filament would be classified as supercritical based on its \mcore. 
Mass estimates from SED fitting are known to underestimate the true mass of high extinction regions in the ISM, 
including the densest (core) part of the filament profile (see e.g., \citealp{pagani2015} and reliability discussion below). 
However, the presence of a few young stellar objects (YSOs) could also be explained by 
localised and sporadic star formation. This could happen in segments of those same filaments but 
with temperature below and column densities above the mean values, 
and therefore more favourable for collapse. 
A more in-depth study of the filament properties of the GCC sample relative to the YSO and compact source population will be 
presented in a follow-up study.

\subsection{Filament Components \mcore\ and \mwing: Intrinsic Properties}
\mcore\ and \mwing\ can both 
vary by several orders of magnitude, with the \core\ and wing components within the ranges 
of $\approx0.5-19.5$\,\mpc and $\approx0-40.5$\,$M_{\odot}$\,pc$^{-1}$, respectively. 
Overall, \mtot\ varies between $1.0$ and $60$\,$M_{\odot}$\,pc$^{-1}$. 
Filaments were broadly classified according to the 
relative contribution of the \core\ and wing components to their \mtot.
Detections dominated by the \core\ component, \mcore$>$\mwing, 
were classified as `\core-dominated' filaments. Those whose \mtot\ could be accounted for mainly by 
the contribution from their wing component (\mcore$<$\mwing) were classified as `wing-dominated'. This allowed 
for a direct quantification of the relevance and influence of the \core\ 
component (the region concentrating the highest column densities and the most relevant for star formation). 
Here we chose to remain consistent with the standard approach used in \citet{arzoumanian2011} and later 
studies, and classified as `subcritical' those filaments with \mtot$<$\mcrit\ ($\sim16.5$\,\mpc\ at $T=10$\,K), 
and `supercritical' when \mtot$\ge$\mcrit. 
While convenient for the purpose of this work, such terminology is, however, likely incomplete and inappropriate to fully describe the stability state 
of a filamentary structure (e.g., \citealp{fischera2012a}). 
Even when using this standard stability description, it is also crucial 
to distinguish massive filaments with a high \mcore, with the highest potential for 
local collapse, from other (low \mcore) structures that appear 
supercritical only because of their very extended wing component. 
A very extended wing (flat profile at large radii) could contribute significantly to the mass but might 
not be (or lead to) a star-forming filament if associated with a very low \mcore. This difference is 
particularly important in the investigation of a possible filament evolutionary scenario.

\subsection{Reliability and Robustness of the Results}
Filament studies are subject to well-known
caveats that should be taken into consideration.

The stability criteria based on filament linear mass density is dependent on observables 
that are generally difficult to constrain, such as the filament (wing) radius of integration 
and the background level. 
Simplifying assumptions, such as one (isothermal) filament temperature 
or a Gaussian-like morphology for the innermost parts of the filament, 
might not necessarily be accurate approximations.
Similarly, due to the lack of molecular data we cannot confirm that all of our filament 
detections are also self-consistent (single) structures in velocity as 
they seem to appear in the \herschel\ dust maps (and not due to a convenient superposition of structures in the line of 
sight with the general appearance of a filament). Inclination effects \citep{arzoumanian2011}, which have not 
been included in the present analysis, might affect the observed range of linear mass densities, 
but should not affect conclusions based on the relative behaviour (formation/evolution) 
of the different filament populations in different environments.

Prominent and dense filaments in star-forming complexes can be, overall, 
much better constrained, identified, and characterised due to, e.g., a higher \nh\ 
contrast with the environment, or the presence of YSOs tracing the structure itself. 
For the same reasons, 
we expect a higher degree of uncertainty in diffuse and 
non-star forming environments. Here we have aimed to minimize the impact of 
such detections on our (statistical) 
conclusions by testing our results on filament populations with 
increasing reliability (\nscale) level. An increase in \nscale\ 
reduces the number of filaments in a sample, but it also increases the 
robustness of the derived filament properties. 
In this study, all our results and conclusions hold for samples with different reliability levels and are 
consistent within $3\sigma$.
Source removal has negligible impact on the main results of this work, 
the major effects being, however, an increase of 
the average filament width (Table \ref{table:filas}) and a systematic decrease of 
the filament crest column density by a factor of $\sim2$. 

All filaments and measurements are ultimately affected by the assumptions made in the 
creation of the original column density maps, and from which the filaments are extracted. 

First, our neglect of radiative transfer in the derivation of our \nh\ and $T$ maps in favour 
of SED fitting has been known to underestimate the true column density. This is caused by 
the presence of a dust population consisting of dust grains with a mixture of 
properties (e.g., \citealp{ysard2012}). 
Indeed, a considerable fraction of cold dust (T$<10$\,K) can be missed simply due to warmer dust 
dominating the modelling of the SED \citep{pagani2015}.
These effects would be predominantly associated with the filament \core\ component, 
due to it being associated with material with higher extinction. 
Similarly, the properties of dust grains can also evolve depending on local 
environmental conditions (e.g., \citealp{roy2013}; \citealp{ysard2013}). 
This implies that the use of a constant opacity cannot be appropriate for describing the 
evolution of a dust-based structure formed by regions (e.g., \core, wing components) 
with different (and varying) physical properties 
(e.g., an accreting filament growing in mass with time). 
All these effects are difficult to quantify, and their 
inclusion in the filament analysis is beyond the scope of this work 
(albeit consistent with similar previous studies mentioned in this work and that 
also ignore these effects). 
These uncertainties are expected to be minimized for our main results, 
based on the observed relative trends and properties 
of the population as a whole, but such 
effects remain crucial for any in-depth model associated 
with structures in the ISM.

Based on these caveats, the current approach remains simple, especially 
considering the complexity of filament detection and characterisation 
(e.g., selection effects, star formation history, and the currently undistinguishable 
changes in the \core\ and wing contributions for filaments in 
formation and those already in dispersal stage). 
Filaments are also subject to many processes (general dynamics, turbulence, magnetic field contributions, etc.), all capable of 
affecting the observed properties of the filament population. 
This complexity should always be kept in mind when interpreting the results.

\begin{figure*}[ht]
\centering
\subfigure[]{%
\label{fig:evol1}
\includegraphics[scale=0.40,angle=0]{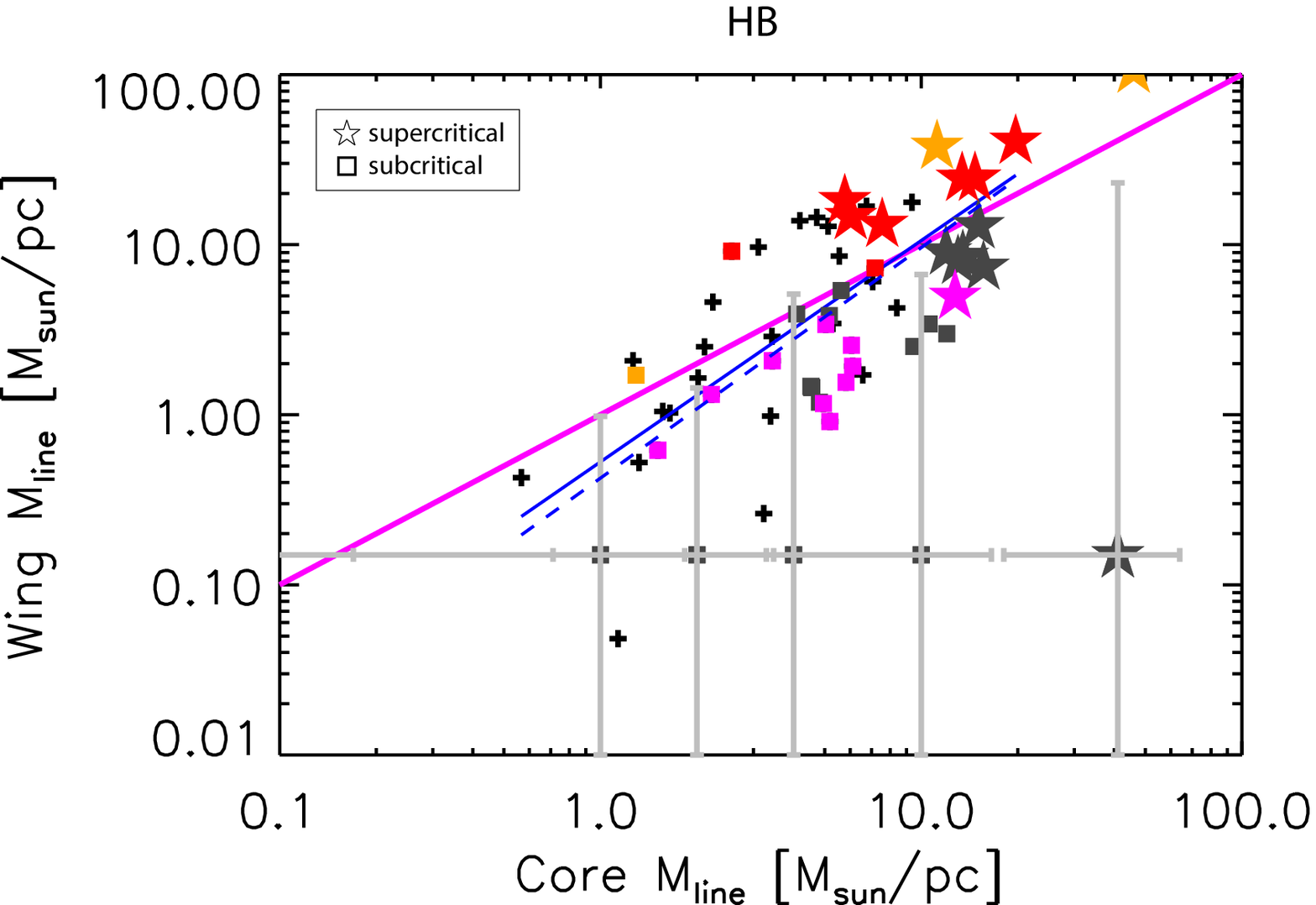}
}%
\subfigure[]{%
\label{fig:evol2}
\includegraphics[scale=0.40,angle=0]{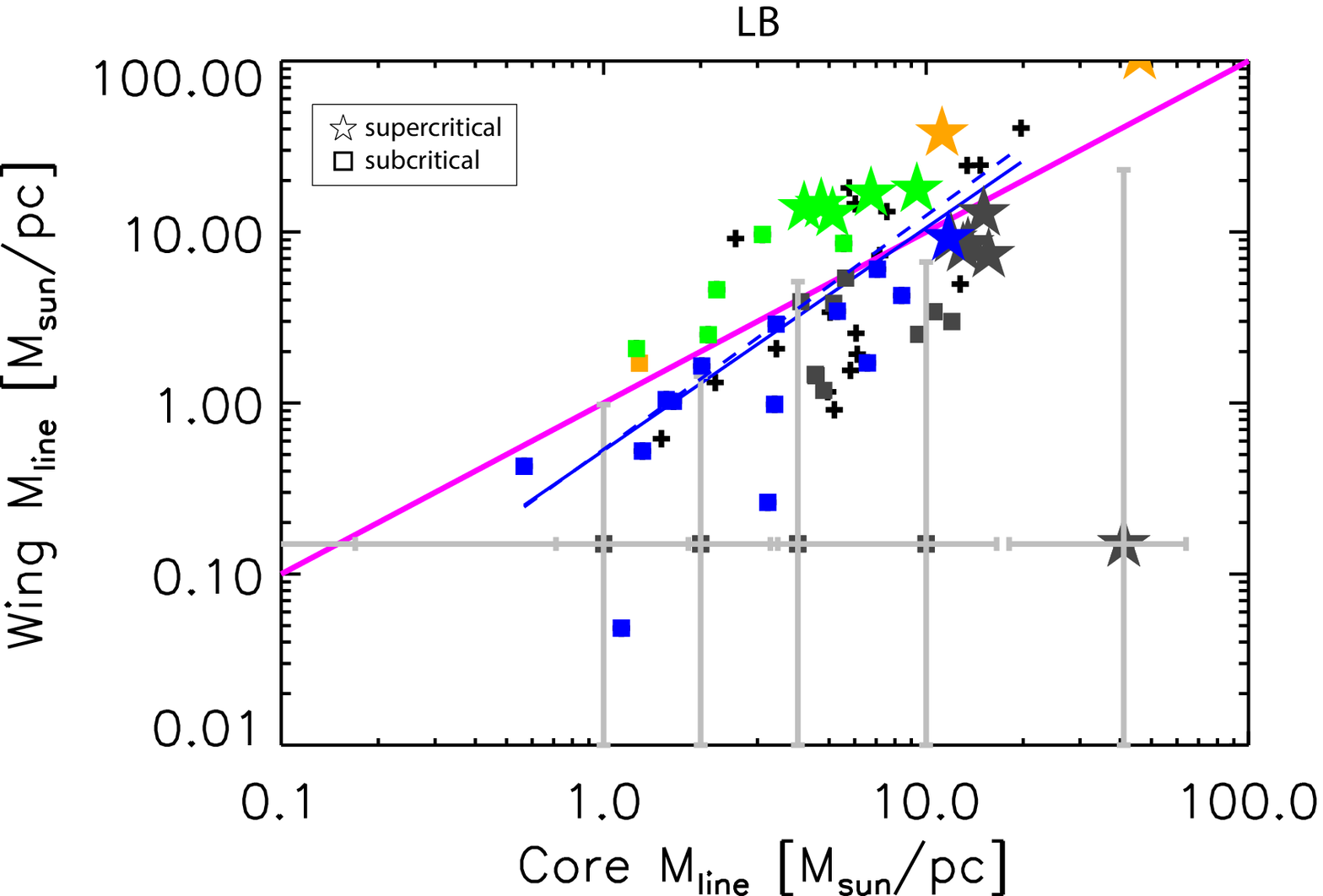}
}\\%
\subfigure[]{%
\label{fig:evol3}
\includegraphics[scale=0.40,angle=0]{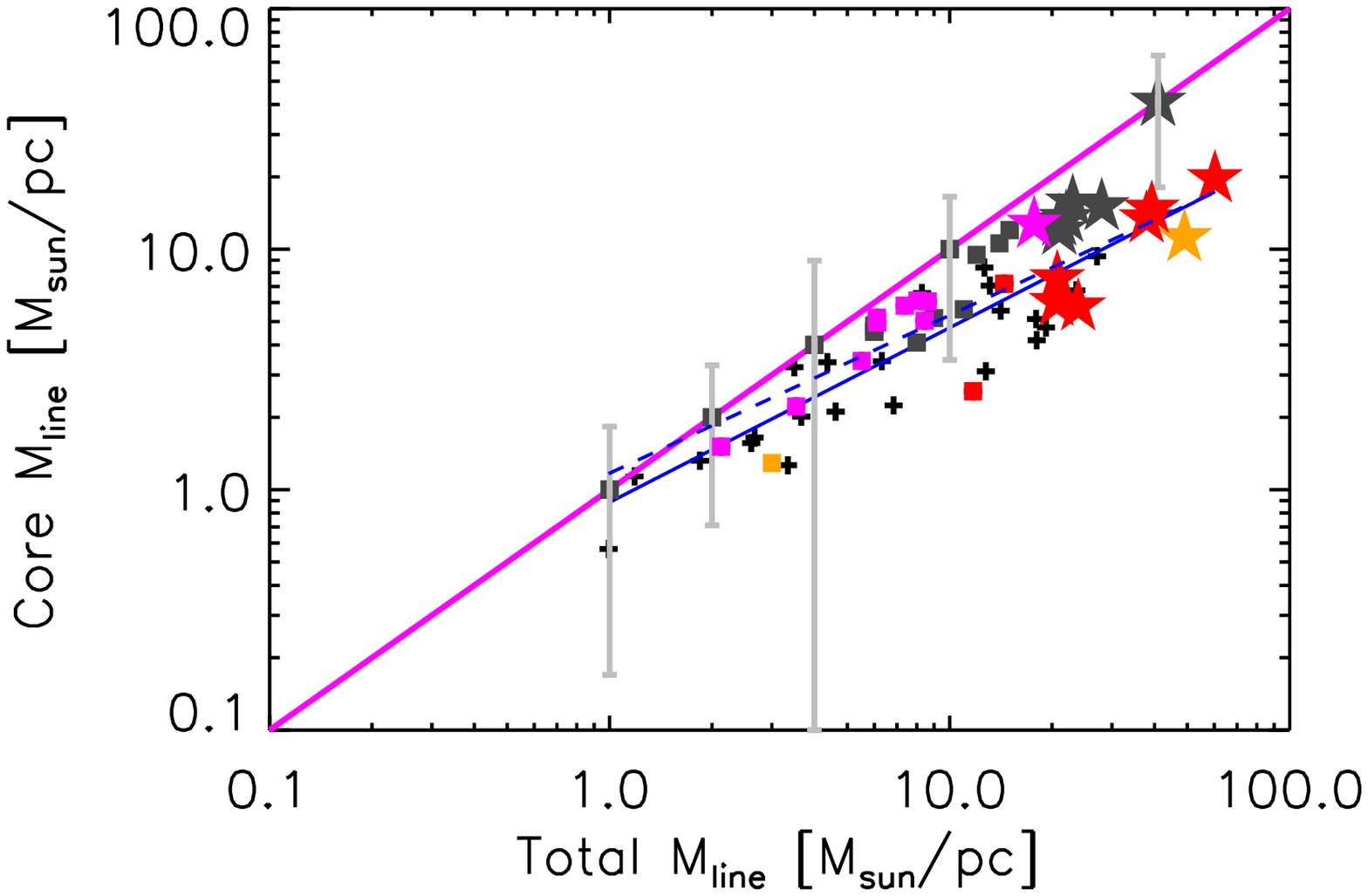}
}%
\subfigure[]{%
\label{fig:evol4}
\includegraphics[scale=0.40,angle=0]{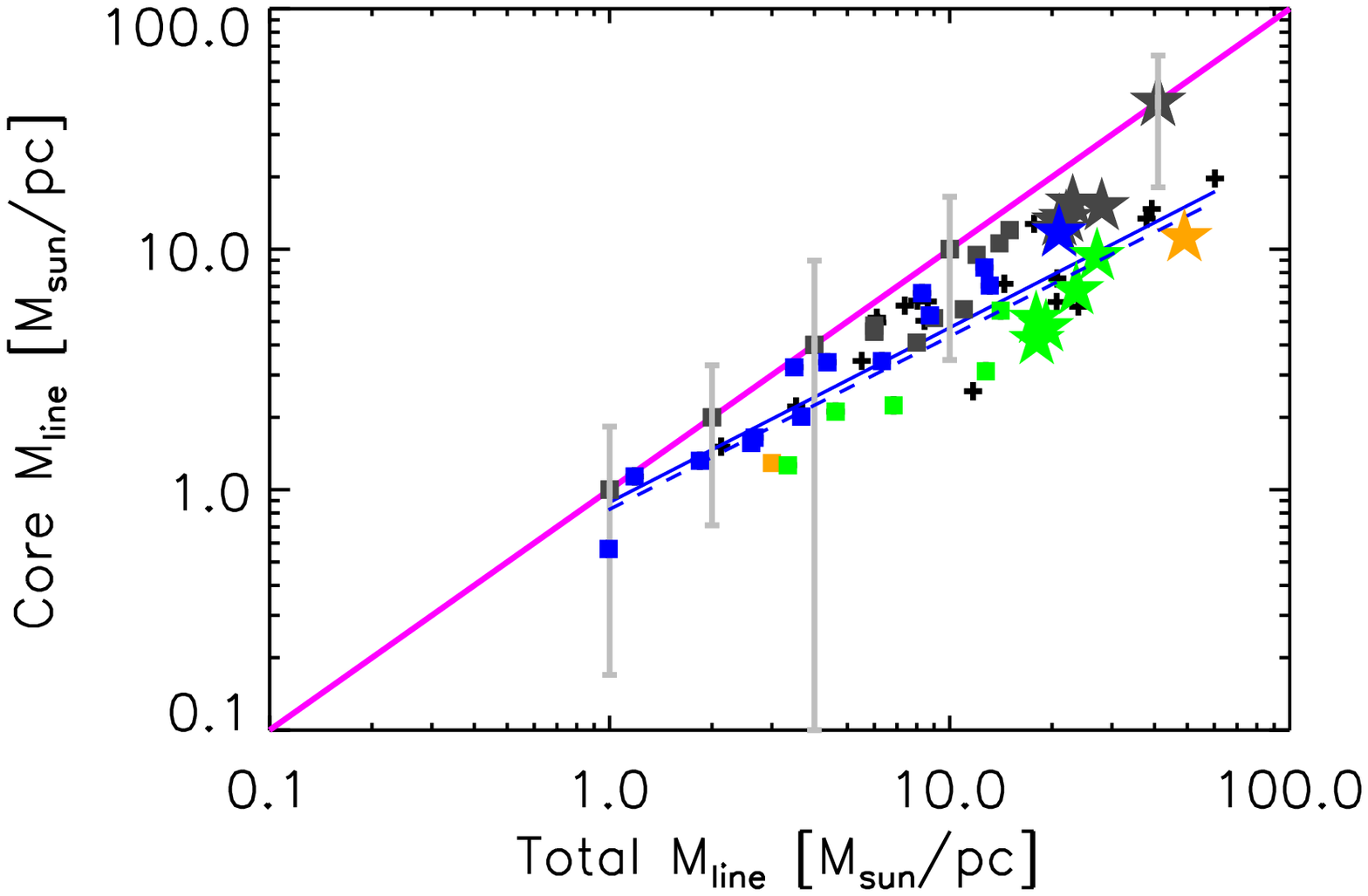}
}\\%
\subfigure[]{%
\label{fig:evol5}
\includegraphics[scale=0.40,angle=0]{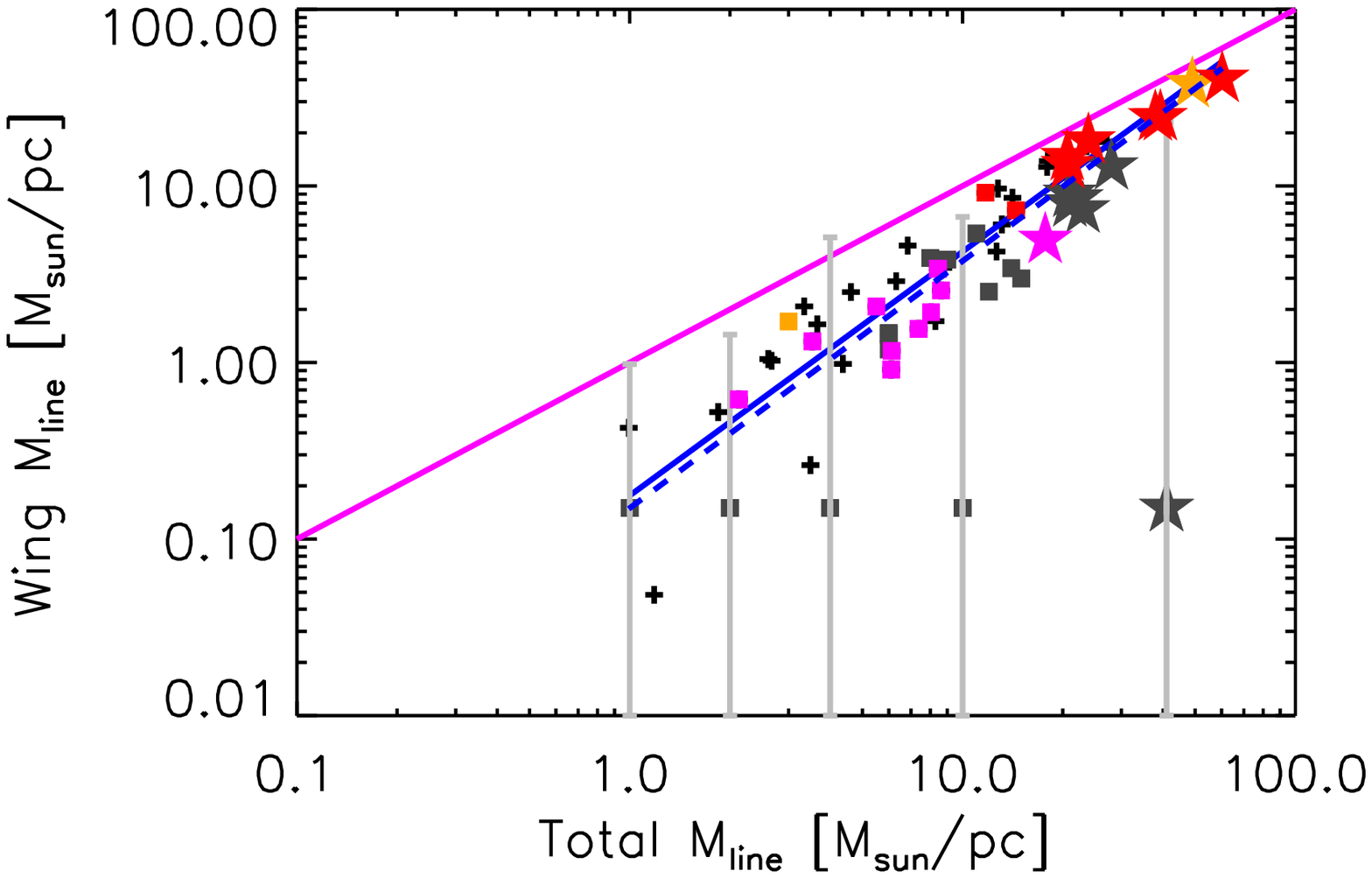}
}%
\subfigure[]{%
\label{fig:evol6}
\includegraphics[scale=0.40,angle=0]{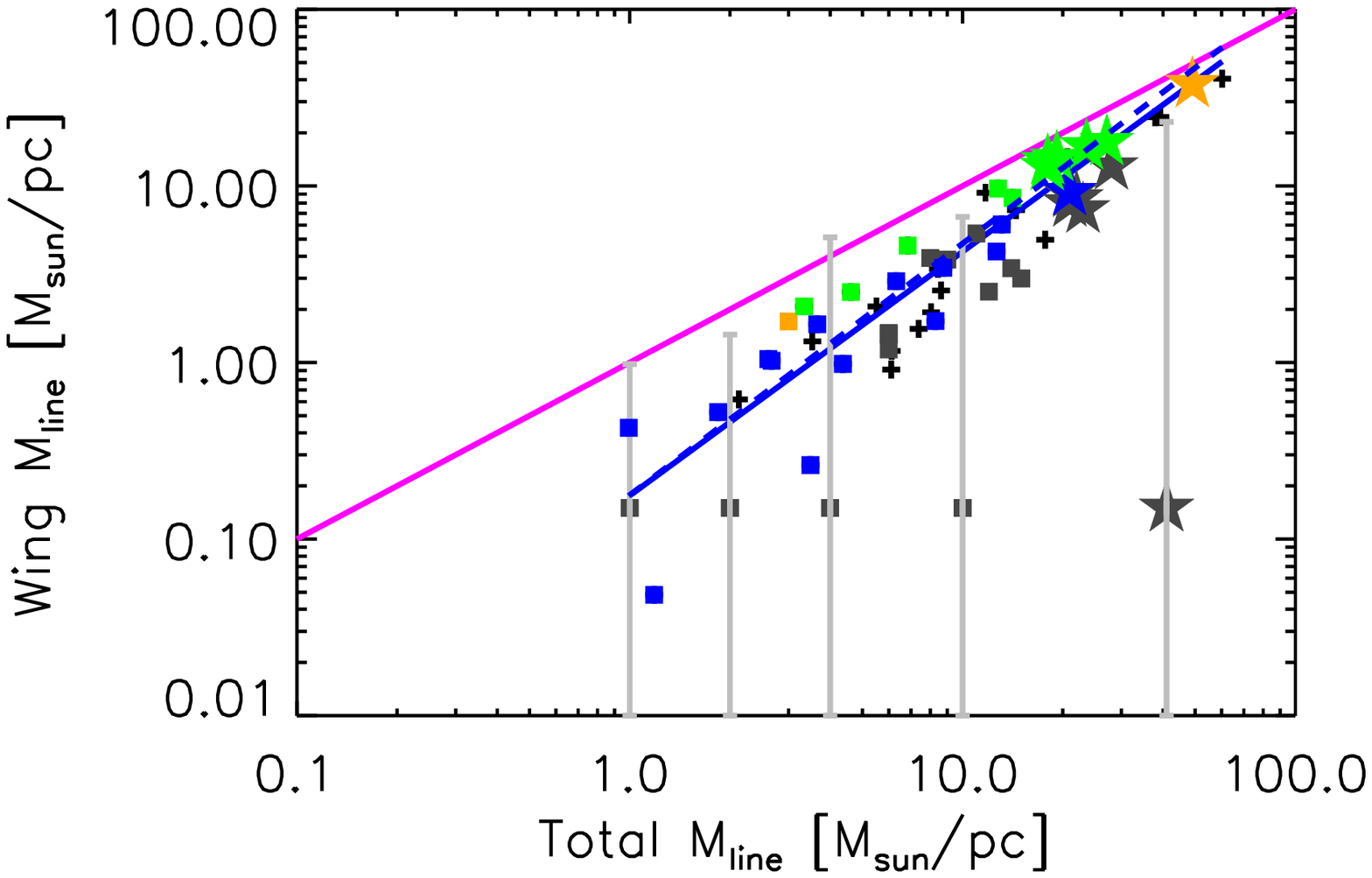}
}\\%
\caption{Correlation of \mcore, \mwing, and \mtot\ for \fsample\ filaments in high backgrounds (filled red and magenta symbols in Figs. \ref{fig:evol1}, \ref{fig:evol3}, \ref{fig:evol5}) and low backgrounds (filled blue and green symbols in Figs. \ref{fig:evol2}, \ref{fig:evol4}, \ref{fig:evol6}). Best linear fits according to environment and for the entire population (background-independent) are shown  as (blue) dashed and solid lines, respectively. Plots include the parameters derived for filaments from \citet{arzoumanian2011}: \core-dominated (dark grey) and wing-dominated (orange) symbols. Errors bars are shown only for filaments consistent with having no wing contribution (assigned a default value of \mwing$=0.1$\,$M_{\odot}$\,pc$^{-1}$ for plotting purposes). Here, uncertainties for the \core\ and wing contributions are those estimated for the \core\ component based on their quoted uncertainties on width and column density. As reference, black crosses in Figs. \ref{fig:evol1}, \ref{fig:evol3}, \ref{fig:evol5} mark the position of filaments in low-backgrounds. Crosses in Figs. \ref{fig:evol2}, \ref{fig:evol4}, \ref{fig:evol6} mark those in high-backgrounds. Other symbols, lines, and colours as in Fig. \ref{fig:corewing-si}.} 
\label{fig:fits-si}
\end{figure*}

\section{Discussion: Observational Constraints for Filament Models}\label{sec:discussion} 
The GCC filament sample covers structures with a wide range of physical and environmental conditions. 
We use this sample to identify and constrain particular properties of the 
filament population that can serve as an observational basis for theoretical models addressing the origin of star-forming filaments.

\subsection{Diversity of Filaments: Global Characteristics}
Under the standard assumption of star formation in filaments with linear mass density close to \mcrit, 
we can distinguish clear structural properties that filaments must satisfy (e.g., by evolution with increasing \mline) 
in order to achieve criticality. 
  
Figure \ref{fig:corewing-si} reveals distinct types of filaments based on their \mcore\ and \mwing\ values. 
A link between the filament \core\ and wing parameters is evident in Fig. \ref{fig:dis1}, 
with \mwing\ increasing with \mcore\ following a best linear 
fit ($Y=bX+c$) of $b=1.3\pm0.2$ and $c=-0.3\pm0.1$\,$M_{\odot}$\,pc$^{-1}$, 
as derived from the logarithmic distribution. 
The trend of increasing  \mwing\ with \mcore\ therefore conveys a shift from a regime dominated 
by subcritical filaments to a supercritical one, i.e., an increase of \mtot (Figs. \ref{fig:dis2} and \ref{fig:dis4}). 

Filaments in the low end of the \mcore$-$\mwing\ distribution (Fig. \ref{fig:dis1}), or equivalently, in the 
\mtot$-$\mcore\ or \mtot$-$\mwing\ diagrams (Figs. \ref{fig:dis2} and \ref{fig:dis4})
are also characterised by a dominant \core\ component (i.e., `\core-dominated' filaments). 
However, the overall faster increase in \mwing\ than \mcore\ with \mtot, 
as shown by the steeper trend for the former ($b_{\mathrm{wing}}=1.38\pm0.09$, c.f., $b_{\mathrm{core}}=0.73\pm0.06$), 
results in a tendency for filaments to \textit{lower their \mcore$/$\mwing\ ratio with increasing \mtot}, even with 
the simultaneous increase of both parameters (Fig. \ref{fig:dis3}). 
The final result is a clear tendency of subcritical filaments to be \core-dominated ($\sim75$\% 
of the subcritical population), 
while supercritical filaments are predominantly ($\sim85$\%) wing-dominated.

The highlighted behaviour of the derived filamentary properties allows us to define 
three main `Regimes' in the population. 
The bulk characteristics of each Regime are 
summarised in Tables \ref{table:tsummary} and \ref{table:tsummary2}. 

\begin{itemize}
\item \textit{Regime 1}: 
Massive supercritical filaments are predominantly ($\sim85$\%) wing-dominated, 
with minimum \mcore\ and \mwing\ of 
$\approx4.2$\,$M_{\odot}$\,pc$^{-1}$ and $\approx5.0$\,$M_{\odot}$\,pc$^{-1}$, respectively. 
Filaments with \mcore$<4.2$\,$M_{\odot}$\,pc$^{-1}$ are therefore exclusively subcritical and predominantly ($70$\%) \core-dominated.

\item \textit{Regime 2:}
The filament population with \mcore$\ge4.2$\,$M_{\odot}$\,pc$^{-1}$ contains 
a mixture of supercritical and subcritical filaments 
with mixed proportions of \core\ and wing components, environments, and crest column densities.  
The maximum \mcore\ derived for any subcritical 
filament is found to be \mcore$\approx8.4$\,$M_{\odot}$\,pc$^{-1}$, above which only supercritical structures are found.
The region in the range $4.2\le$\mcore$\le8.4$\,$M_{\odot}$\,pc$^{-1}$ therefore defines a
 `transition' regime, comprised of wing-dominated supercritical 
($\sim34$\% of the filaments in this regime) and \core\ and wing-dominated subcritical filaments. 
A total of $\sim53$\% of all filaments in this Regime are \core-dominated filaments.

\item \textit{Regime 3}: 
A third region, comprised exclusively of supercritical filaments, 
is also characterised by having the most massive \core\ components and all core-dominated supercritical filaments. 
This regime contains 
all filaments with \mcore$>8.4$\,$M_{\odot}$\,pc$^{-1}$ and with properties 
consistent with those associated with the actively star-forming supercritical filaments 
investigated in \citet{arzoumanian2011}. Such centrally massive filaments, associated with the 
highest column densities, hold the greatest potential for star formation relative to other apparently massive 
structures with lower \mcore\ in Regime 2. 
As with Regime 2, the sample is comprised of filaments with mixed proportions of \mcore\ and 
\mwing\ components ($\sim33$\% \core-dominated).

\end{itemize}

\begin{table*}[ht!]
\caption{Overview of Average Filament Intrinsic \& Environmental Properties in Each Regime.}
\label{table:tsummary}
\centering
\begin{tabular}{c c c c c c}
\hline \hline
Regime&BKG Class$^a$&$<$BKG$>^b$&$<$BKG$>^c$&$<$Ridge \nh$>^d$&$<$FWHM$>$\\ 
& &[$10^{21}$\,cm$^{-2}$]&[mag]&[$10^{21}$\,cm$^{-2}$]&[pc]\\
\hline
1&LB$+$HB&$1.4\pm0.2$&$1.5\pm0.2$&$0.9\pm0.1$&$0.11\pm0.01$\\
2&LB$+$HB&$2.6\pm0.3$&$2.7\pm0.3$&$2.3\pm0.3$&$0.14\pm0.01$\\
3&LB$+$HB&$2.8\pm0.4$&$2.9\pm0.5$&$5.2\pm0.8$&$0.12\pm0.01$\\
\hline
\multicolumn{6}{l}{{$^a$ High-background [HB] or low-background [LB].}}\\ 
\multicolumn{6}{l}{{$^b$ Average \nh\ of the environment with standard error on the mean.}}\\ 
\multicolumn{6}{l}{{$^c$ \av$=$\nh$/(9.4\times10^{20})$}}\\ 
\multicolumn{6}{l}{{$^d$ Background-removed.}}\\ 
\hline
\end{tabular}
\end{table*}

\begin{table*}[ht!]
\caption{Overview of Average Filament [\mline] Properties in Each Regime.}
\label{table:tsummary2}
\centering
\begin{tabular}{c c c c c c}
\hline \hline
Regime&Wing/\core$^a$&Criticality$^b$&\mcore&$<$\mcore$>$&$<$\mwing$>$\\ 
& & &$[M_{\odot}$\,pc$^{-1}$]&[\% \mtot]&[\% \mtot]\\
\hline
1&\core&SB&$<4.2$&$58$&$42$\\
2&\core$+$wing&SB$+$SP&$4.2\le M \le 8.4$&$52$&$48$\\
3&wing&SP&$>8.4$&$45$&$55$\\
\hline
\multicolumn{6}{l}{{$^a$ Filament type characterising population ($\ga60$\%): wing-dominated [wing]}}\\ 
\multicolumn{6}{l}{{or \core-dominated [\core].}}\\ 
\multicolumn{6}{l}{{$^b$ Subcritical [SB] or supercritical [SP] filaments.}}\\ 
\hline
\end{tabular}
\end{table*} 

These results establish the direction (\mcore\ and \mwing\ behaviour with \mtot) 
and final conditions (supercritical filaments in Regime 2 and Regime 3) that must be 
accounted for by a potential evolutionary process leading to the formation of star-forming filaments. 
The identification of (structurally) distinct filament regimes could also indicate variations and/or limitations in the 
formation and evolution process of filaments in gravitationally-dominated scenarios.

\begin{figure}[ht!]
\centering
\subfigure[]{%
\label{fig:bkg1}
\includegraphics[scale=0.40,angle=0]{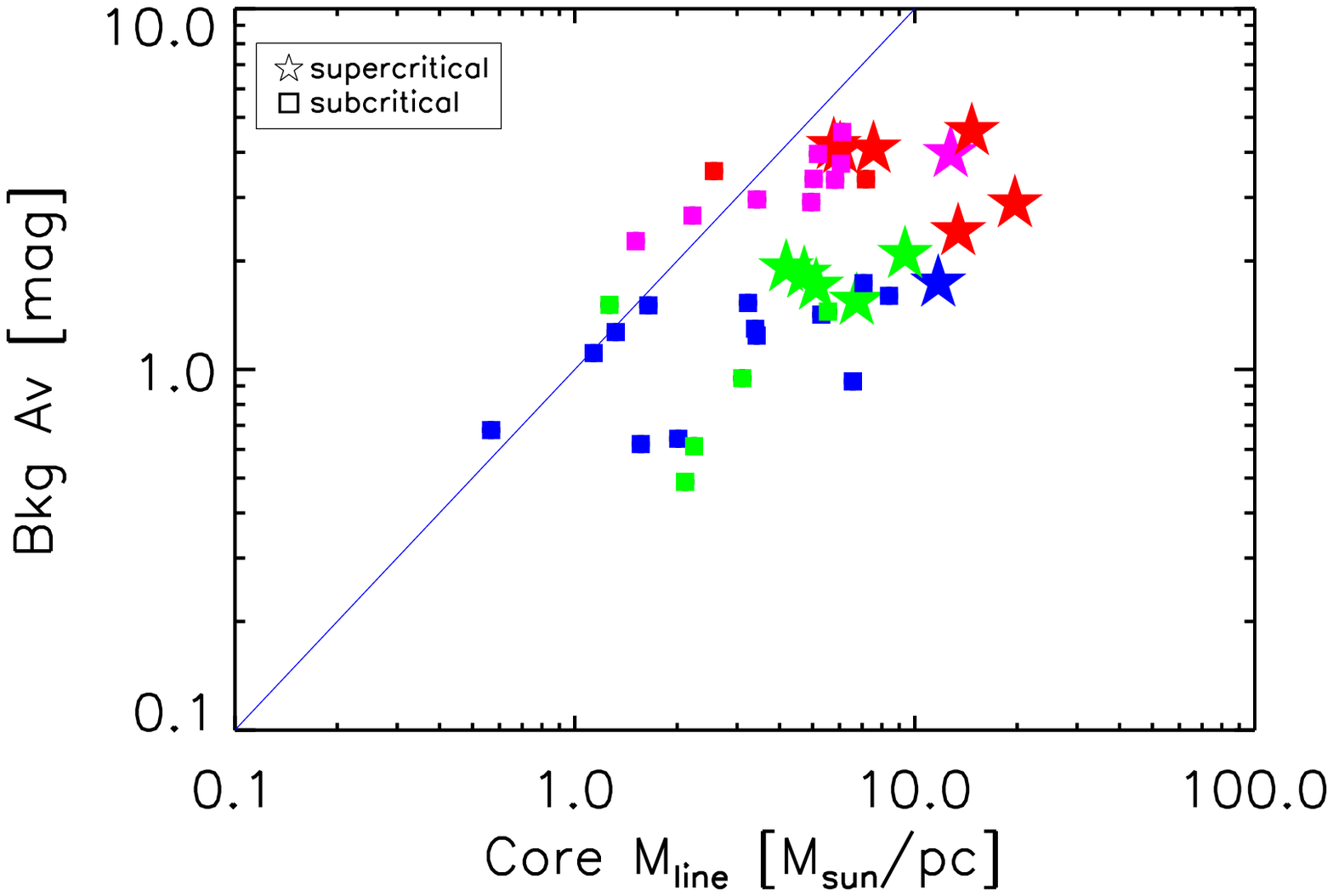}
}\\%
\subfigure[]{%
\label{fig:bkg2}
\includegraphics[scale=0.40,angle=0]{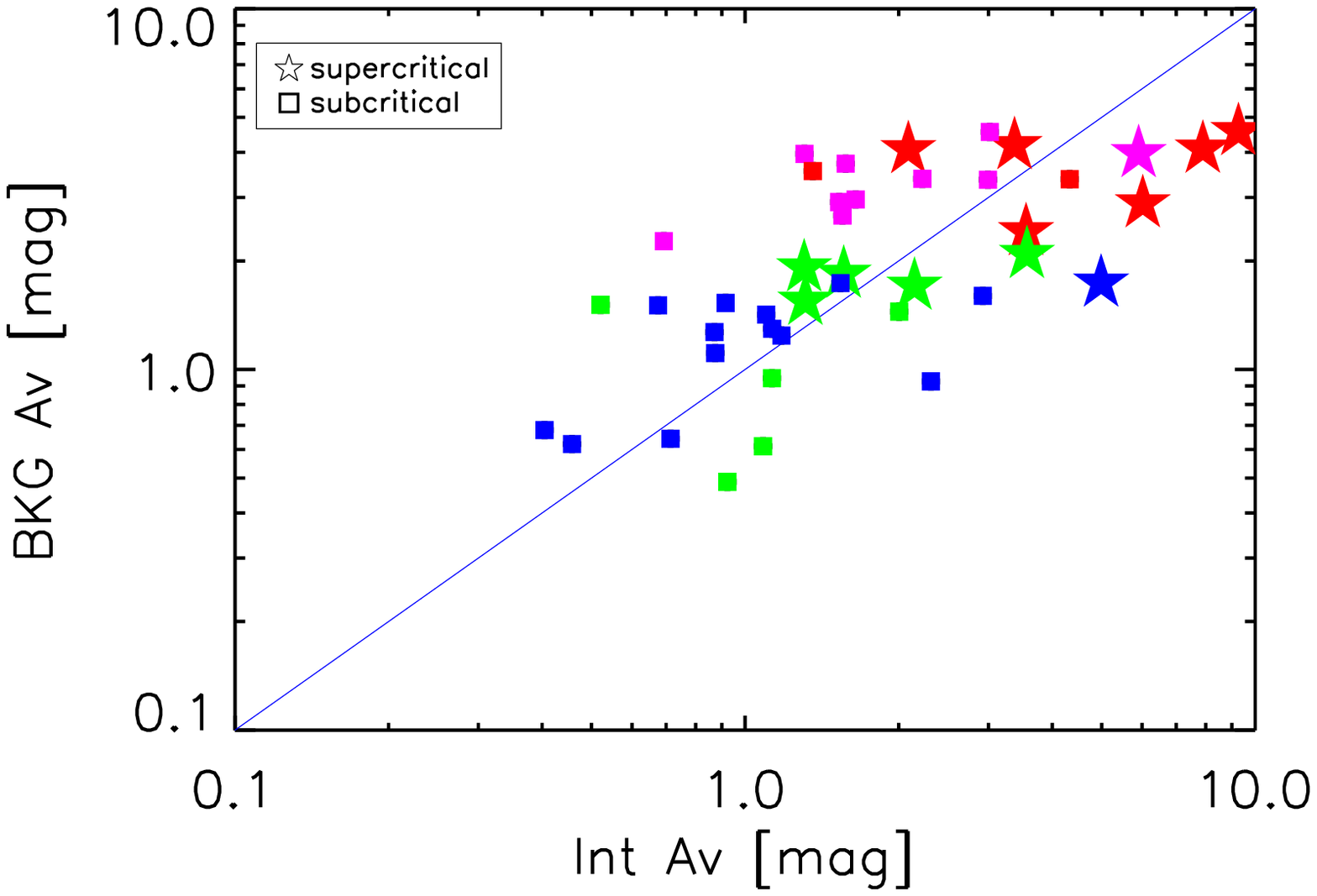}
}\\%
\caption{Distribution of filament background level as a function of \mcore\ (a) and intrinsic (crest) \av\ (b) for the \fsample\ filament sample. Symbols, lines, and colours as in Fig. \ref{fig:corewing-si}.}
\label{fig:bkg}
\end{figure}

\subsection{The Role of Environment In Filamentary Properties and Evolution}
In order to constrain the influence of the environment on 
the properties and/or evolution of the filament \core\ and wing components, 
filaments were divided into low-background (LB) 
and high-background (HB) populations. LB filaments were classified as those with average 
background level below the mean of the population of \nh$\approx2\times10^{21}$\,cm$^{-2}$
(\av$\approx2.2$\,mag)\footnote{\nh$=9.4\times10^{20}$\,cm$^{-2}$ \av$/$mag; \citet{bohlin1978}}, 
while HB filaments are associated with environmental column density above this value.

The main results of this analysis are summarised in Fig. \ref{fig:fits-si}, which investigates the 
behaviour of the HB and LB populations individually by performing a $\chi^2$ fitting to 
the different parameter distributions of the \fsample\ filament sample.
The figure also shows the \core\ and wing component distributions estimated for the filament sample from 
\citet{arzoumanian2011}. Here, the \core\ component was derived assuming a Gaussian with the FWHM and 
central \nh\ quoted by these authors\footnote{The filament central column density values from \citet{arzoumanian2011} are not background-subtracted. However, their estimates were derived from column density maps without an offset correction. Their estimates are therefore only weakly affected by background contribution, which is estimated to be of a few $10^{20}$\,cm$^{-2}$; D. Arzoumanian, priv. comm.} in their Table 1. The wing component was calculated by subtracting this \core\ contribution from their derived \mtot.
 
Overall, independent fits to the HB and LB subsamples reveal negligible environmentally-based differences 
(considering the $1\sigma$ error of the best-fit parameters) 
for the correlations between \mcore, \mwing, and \mtot.  
With the GCC filaments covering a range of background levels 
that can differ by a factor of $\sim10$ (\nh$\sim0.5-4.5\times10^{21}$\,cm$^{-2}$; e.g., Table \ref{table:filas}) 
this is suggestive of a common dominant process (taking place in a wide range of environments, 
albeit not necessarily at the same magnitude) driving 
the formation and/or evolution of the majority of our filament population (e.g., turbulence, shocks, or gravity).

Evidence for an environmental dependence of \mcore\ might be observed, however, 
based on its correlation with background level, as shown in Fig. \ref{fig:bkg1}. 
At similar width, \mcore\ depends exclusively on the Gaussian 
peak of the filament profile. This would naturally lead to the correlation in Fig. \ref{fig:bkg2} 
between background and crest column density.
Indeed, the limits of the parameter range for both \mcore\ and crest \nh\ increase with the \nh\ of the environment. 
The maximum \mcore\ (or crest \nh) for HB filaments 
can double those of the LB sample, with supercritical filaments clustering at the highest 
backgrounds of $<$\nh$>$$\approx(2.7\pm0.3)\times10^{21}$\,cm$^{-2}$ ($<$\av$>\sim3$\,mag; Fig. \ref{fig:bkg}).

While the shift of the minimum intrinsic (crest) \nh\ and \mcore\ to 
higher values with environment could 
artificially arise from our initial filament detection criteria, 
these should not have an effect on the upper limits. 
Indeed, the real origin of our result is further supported by the findings of \citet{schisano2014}, 
who observed that denser filaments appear to be associated with denser environments. 
The filament Regimes, separated according to increasing \mcore\ and therefore, 
according to increasing star-forming potential, are also 
associated with increasingly higher environmental average column density values (Table \ref{table:tsummary}).

Compared to \mcore, \mwing\ is observed to 
have a stronger dependence with \mtot, regardless of the environment and 
stability state of the filament (Fig. \ref{fig:corewing-si}). 
This leads to the already mentioned tendency  of subcritical filaments 
to be \core-dominated.
The preference for the most massive wing components to be 
associated with the most massive \mcore\ and \mtot\ (e.g., Figs. \ref{fig:dis1}; \ref{fig:dis4}), 
can be explained if the wing (power-law) component dominates at a later stage of evolution 
(under the assumption of \mtot\ increasing with time).
Furthermore, the clear tendency of \mwing\ to reach systematically 
higher values in high backgrounds (Figs. \ref{fig:evol1}, \ref{fig:evol2}) seems 
to suggest that the formation of the wing component, 
together with that of supercritical filaments, is intimately linked to 
processes and conditions primarily associated with such dense environments. 
A filament evolution and wing origin driven by gravity (accretion, collapse) 
would be consistent with these results. 
The availability of mass and the increase in gravitational potential of the filament with time 
would lead to filament growth and late stages of evolution associated with 
massive core and wing components. 

Overall, we conclude that filament behaviour appears to be dominated by 
a significant correlation between the various structural components, 
with the most massive \mwing\ being preferentially associated with 
the most massive \mcore\ and \mtot. 
For similar filament widths, the Gaussian function explains the correlation between \mcore\ and 
crest \av. 
However, the observed dependence of \mcore\ (and crest \av) on the column density 
of the environment, together with the identification of 
filament regimes with structural characteristics depending on the background level, strongly suggests that 
filament formation and evolution is intimately linked to the conditions set by their environment.


\section{Conclusions}\label{sec:conclusion}
In this work we have presented an extensive characterisation of the filament population 
present in the \herschel\ fields of the Galactic Cold Cores Programme at $D<500$\,pc. 
The sample was used to identify and quantify key observational constraints, in regards to the structure and environment of filaments, 
needed for the development of theoretical models addressing filament formation and evolution.

Filaments were identified and extracted with the \getfilaments\ algorithm, 
and classified according to the spatial scales at which they dominate.
Filament morphology was characterised by fitting a Plummer-like function to the 
column density profiles. However, in order to avoid the inherent uncertainties 
associated with the physical meaning of the Plummer parameters, the structure 
of the filament was analysed instead with an approach that quantifies 
the Plummer-like shape according to the relative contribution to the profile (linear mass density) 
from two components: a central Gaussian-like region, or \core\ component, 
and a wing component, represented by the power-law tail at larger 
radii. The filament morphology and intrinsic properties (column density distribution, width, stability) 
were then examined as a function of local column density background. 

(i) We find that the filament characteristic width is highly dependent on distance and compact source association. 
This value can also be affected by the intrinsically complex hierarchical nature of filaments, which can 
lead to radically different values depending on the type of filaments being examined. Selection 
of filaments associated with prestellar core formation, or `core-scale' filaments, 
reveals a characteristic mean width of $\sim0.13$\,pc for low-mass star forming regions in the local 
neighbourhood ($D\le300$\,pc). 
The combined analysis of all types of filaments and without distance correction would 
lead to a larger mean for our sample of FWHM$\ga0.2$\,pc.

(ii) The \core\ and wing filament components appear to be environment-dependent, 
with filaments at higher backgrounds systematically reaching higher \core, wing, and total linear mass densities. 
The association of the most massive wing components with the most massive \core\ components, densest environments, and highest total linear mass densities, 
support a (late) wing formation driven by accretion and enhanced by the combined effects of large 
gravitational potential and availability of material. 
The relative contribution of the core and wing components to \mtot\ varies significantly, but 
all filaments with a central component \mcore$\ga8.5$\,$M_{\odot}$\,pc$^{-1}$ ($\sim$\mcrit/2) are supercritical.

(iii) The distribution of linear mass densities of the \core\ (\mcore) and wing (\mwing) components of the core-scale 
filament sample was used to identify three main filament regimes: 
a \core-dominated subcritical region (Regime 1; \mcore$<4.2$\,$M_{\odot}$\,pc$^{-1}$), 
a transition region (Regime 2; $4.2\le$\mcore$\le8.4$\,$M_{\odot}$\,pc$^{-1}$), and 
a supercritical-only region (Regime 3; \mcore$>8.4$\,$M_{\odot}$\,pc$^{-1}$). 
Each regime is characterised by a progressively higher background column density level, 
clearly indicating that the environment is key for the development of the filament structure and, 
ultimately, the formation of supercritical filaments.

\begin{acknowledgements}
A.R-I. acknowledges the French national program PCMI and CNES for the funding of her postdoc fellowship at IRAP. 
A.R-I. is currently a Research Fellow at ESA/ESAC and acknowledges support from the ESA Internal Research Fellowship Programme. 
The authors also thank PCMI for its general support to the `Galactic Cold Cores' project activities.
J.M. and V.-M.P. acknowledge the support of Academy of Finland grant 250741. 
M.J. acknowledges the support of Academy of
Finland grants 250741 and 1285769, as well as the Observatoire Midi-Pyrenees (OMP) in Toulouse for 
its support for a 2 months stay at IRAP in the frame of the `OMP visitor programme 2014'. 
We thank the anonymous referee for detailed comments, suggestions, and corrections that have significantly improved the content and results presented in the paper. 
We also thank J. Fischera, D. Arzoumanian, E. Falgarone, and P. Andr\'{e} for useful discussions.
SPIRE has been developed by a consortium of institutes led by Cardiff Univ. (UK) and including: Univ. Lethbridge (Canada); NAOC (China); CEA, LAM (France); IFSI, Univ. Padua (Italy); IAC (Spain); Stockholm Observatory (Sweden); Imperial College London, RAL, UCL-MSSL, UKATC, Univ. Sussex (UK); and Caltech, JPL, NHSC, Univ. Colorado (USA). This development has been supported by national funding agencies: CSA (Canada); NAOC (China); CEA, CNES, CNRS (France); ASI (Italy); MCINN (Spain); SNSB (Sweden); STFC, UKSA (UK); and NASA (USA). 
PACS has been developed by a consortium of institutes led by MPE (Germany) and including UVIE (Austria); KU Leuven, CSL, IMEC (Belgium); CEA, LAM (France); MPIA (Germany); INAF-IFSI/OAA/OAP/OAT, LENS, SISSA (Italy); IAC (Spain). This development has been supported by the funding agencies BMVIT (Austria), ESA-PRODEX (Belgium), CEA/CNES (France), DLR (Germany), ASI/INAF (Italy), and CICYT/MCYT (Spain).
\end{acknowledgements}



\bibliographystyle{aa}

\clearpage

\appendix

\section{Filament Analysis: Figures}\label{sec:figures}
\begin{figure*}[ht!]
\centering
\subfigure[]{%
\label{fig:evol1lm}
\includegraphics[scale=0.40,angle=0]{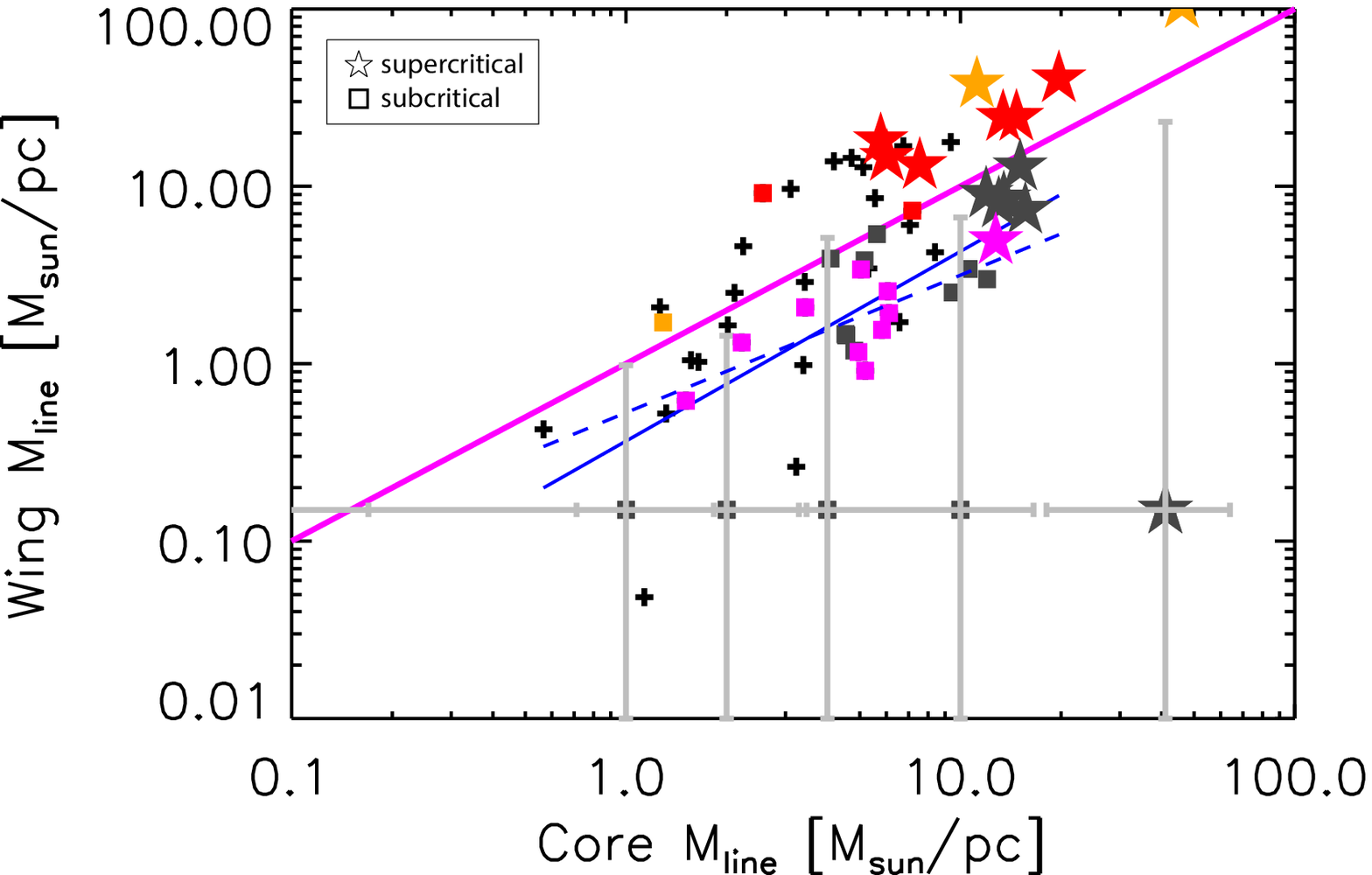}
}%
\subfigure[]{%
\label{fig:evol2lm}
\includegraphics[scale=0.40,angle=0]{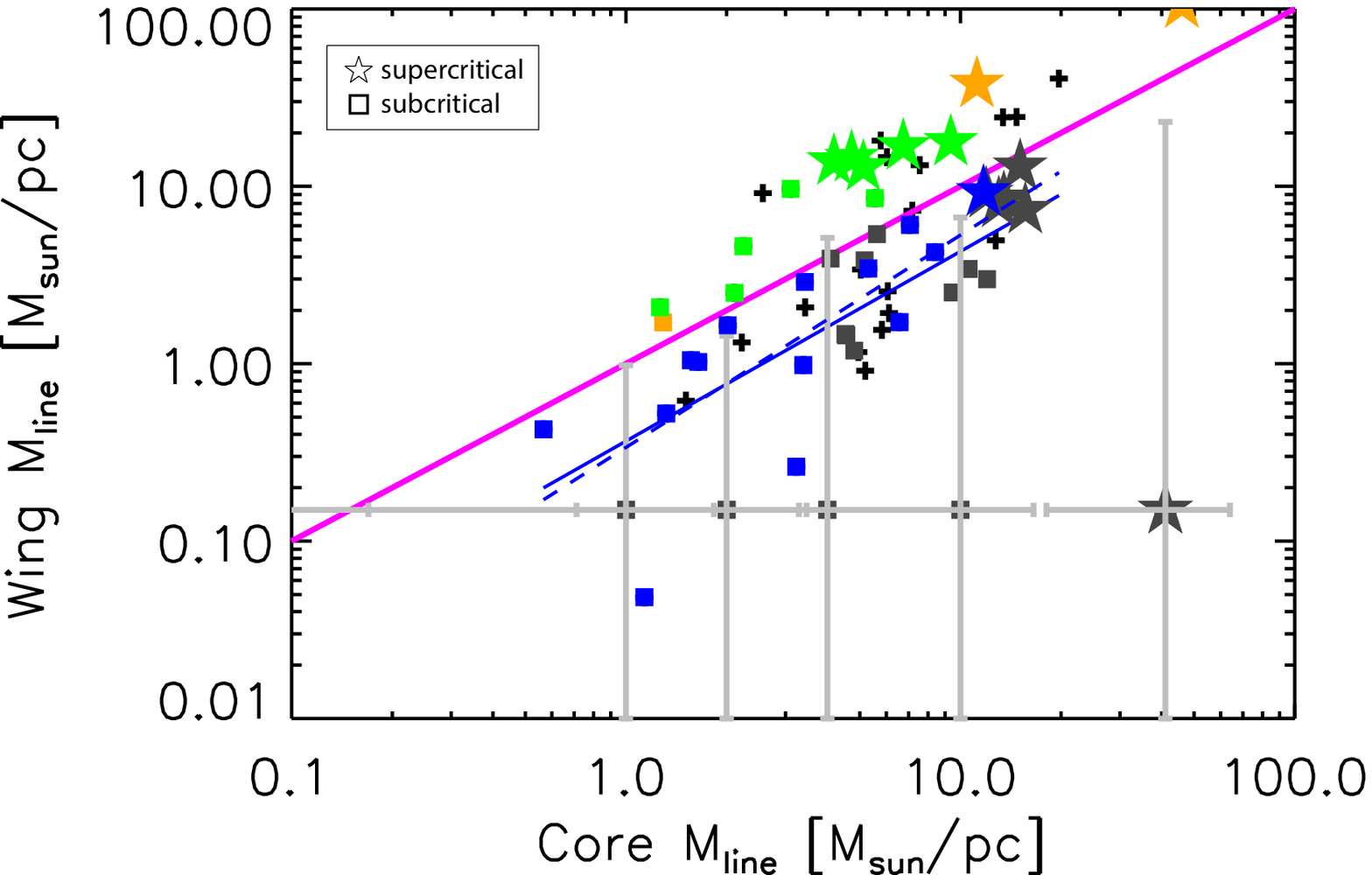}
}\\%
\subfigure[]{%
\label{fig:evol3lm}
\includegraphics[scale=0.40,angle=0]{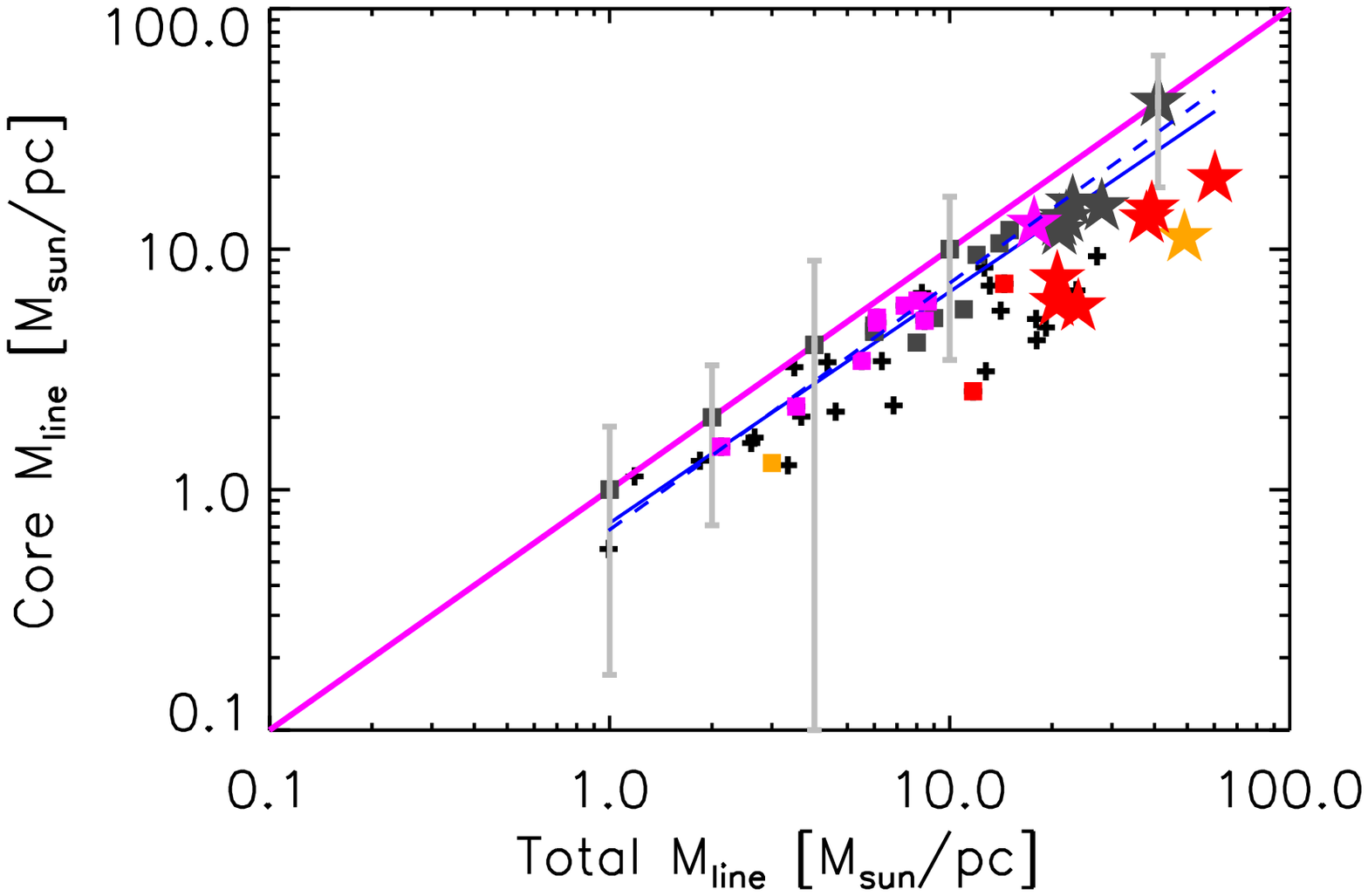}
}%
\subfigure[]{%
\label{fig:evol4lm}
\includegraphics[scale=0.40,angle=0]{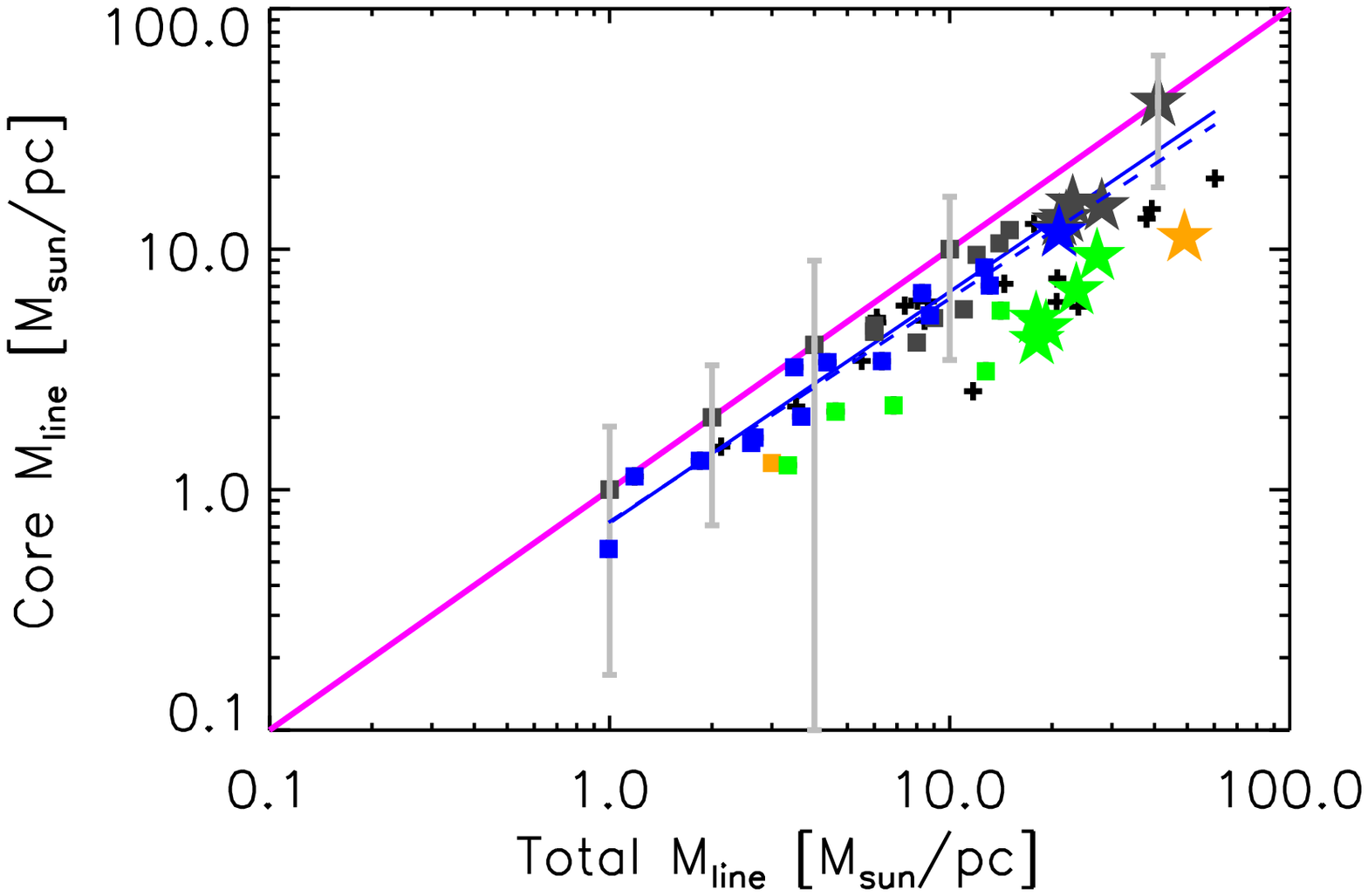}
}\\%
\subfigure[]{%
\label{fig:evol5lm}
\includegraphics[scale=0.40,angle=0]{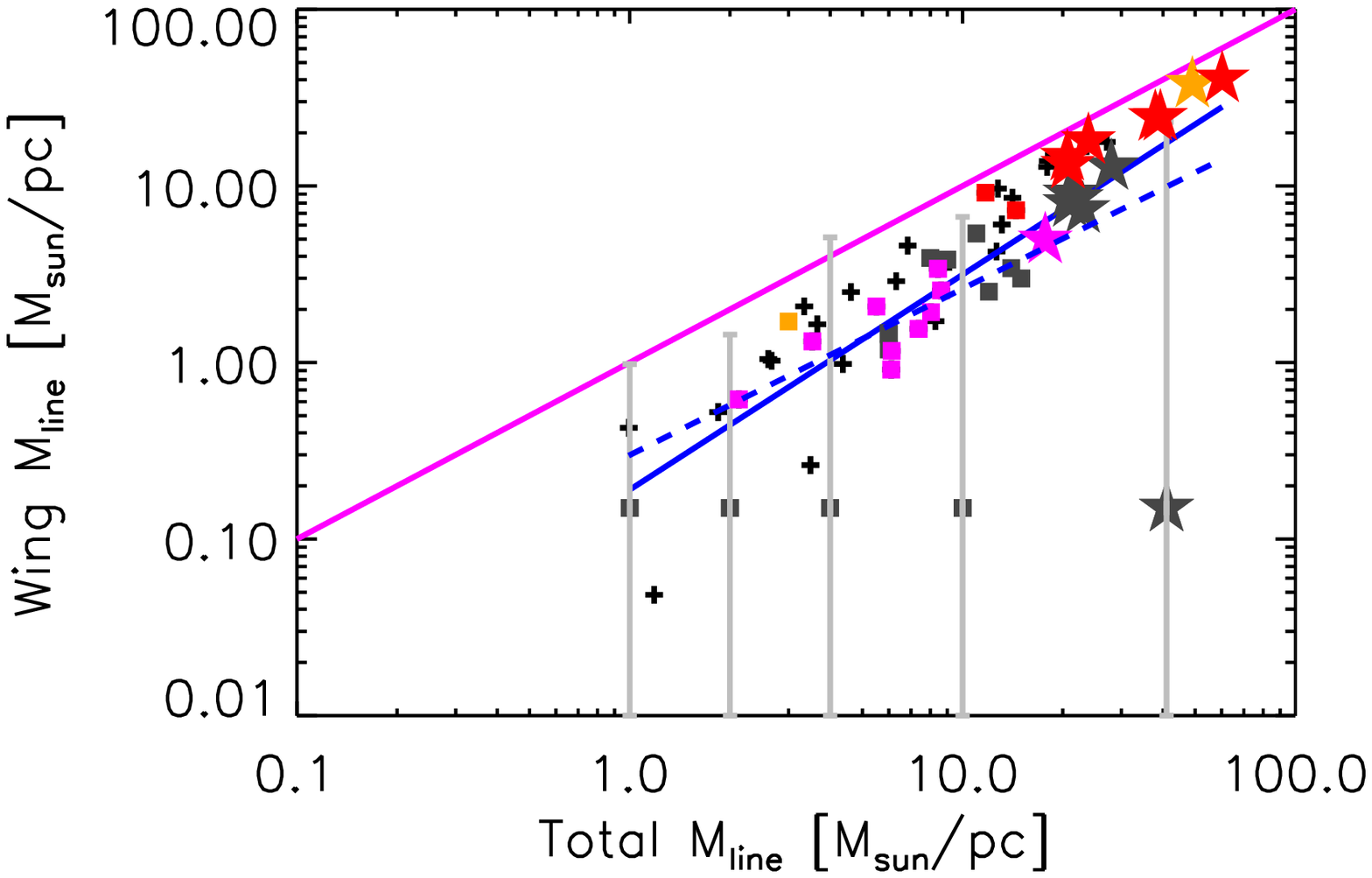}
}%
\subfigure[]{%
\label{fig:evol6lm}
\includegraphics[scale=0.40,angle=0]{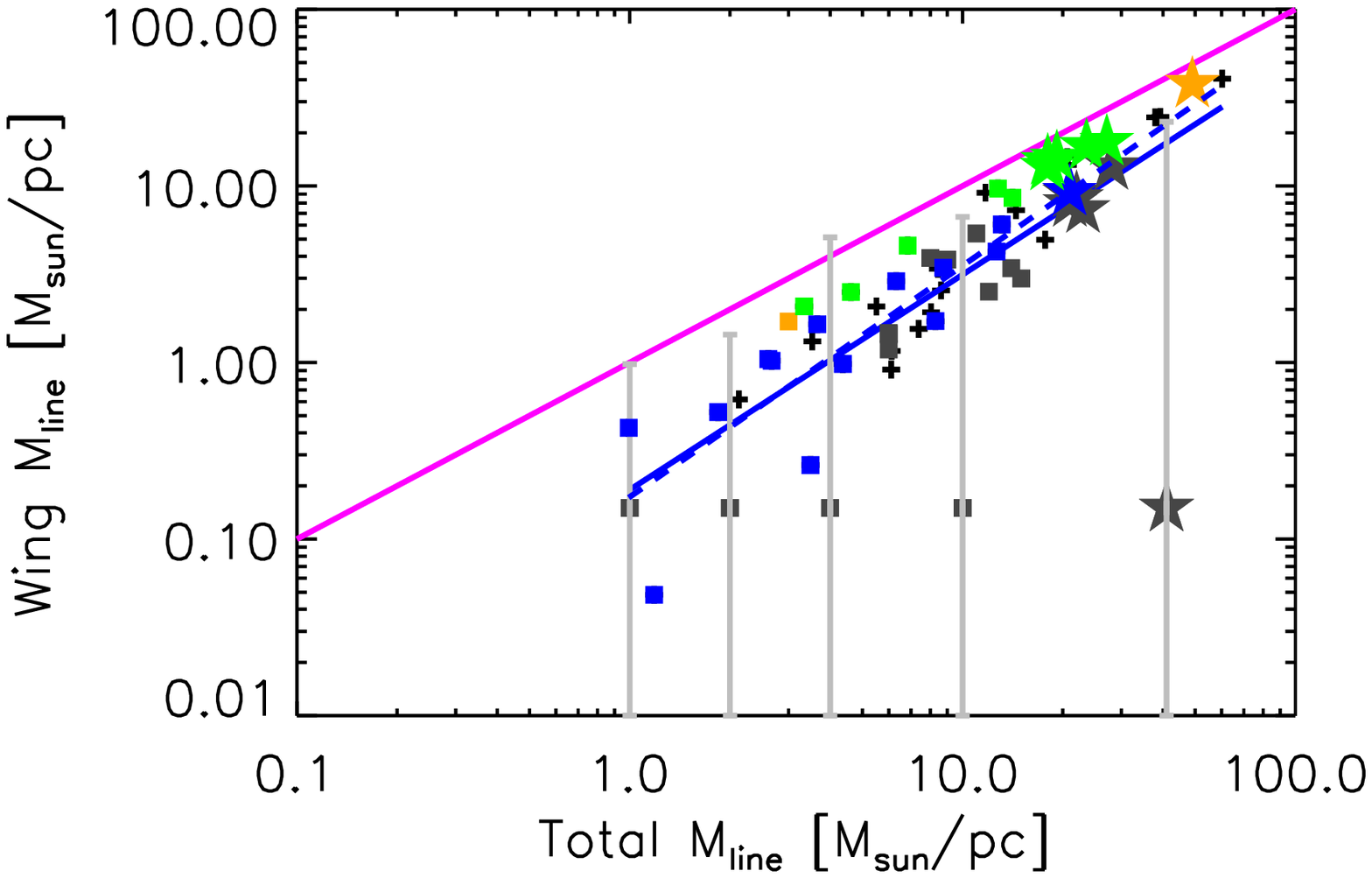}
}\\%
\caption{Same as Fig. \ref{fig:fits-si}, but with fits performed on the \core-dominated sample exclusively.} 
\label{fig:fits-si2}
\end{figure*}

\clearpage

\section{GCC Fields with SI-sample filaments and \nscale$=10$}\label{sec:images}
\begin{figure*}[h!]
\centering
\subfigure[G0.02+18.02]{%
\label{fig:m1}
\includegraphics[scale=0.40,angle=0,trim=0cm 1.5cm 0cm 0cm,clip=true]{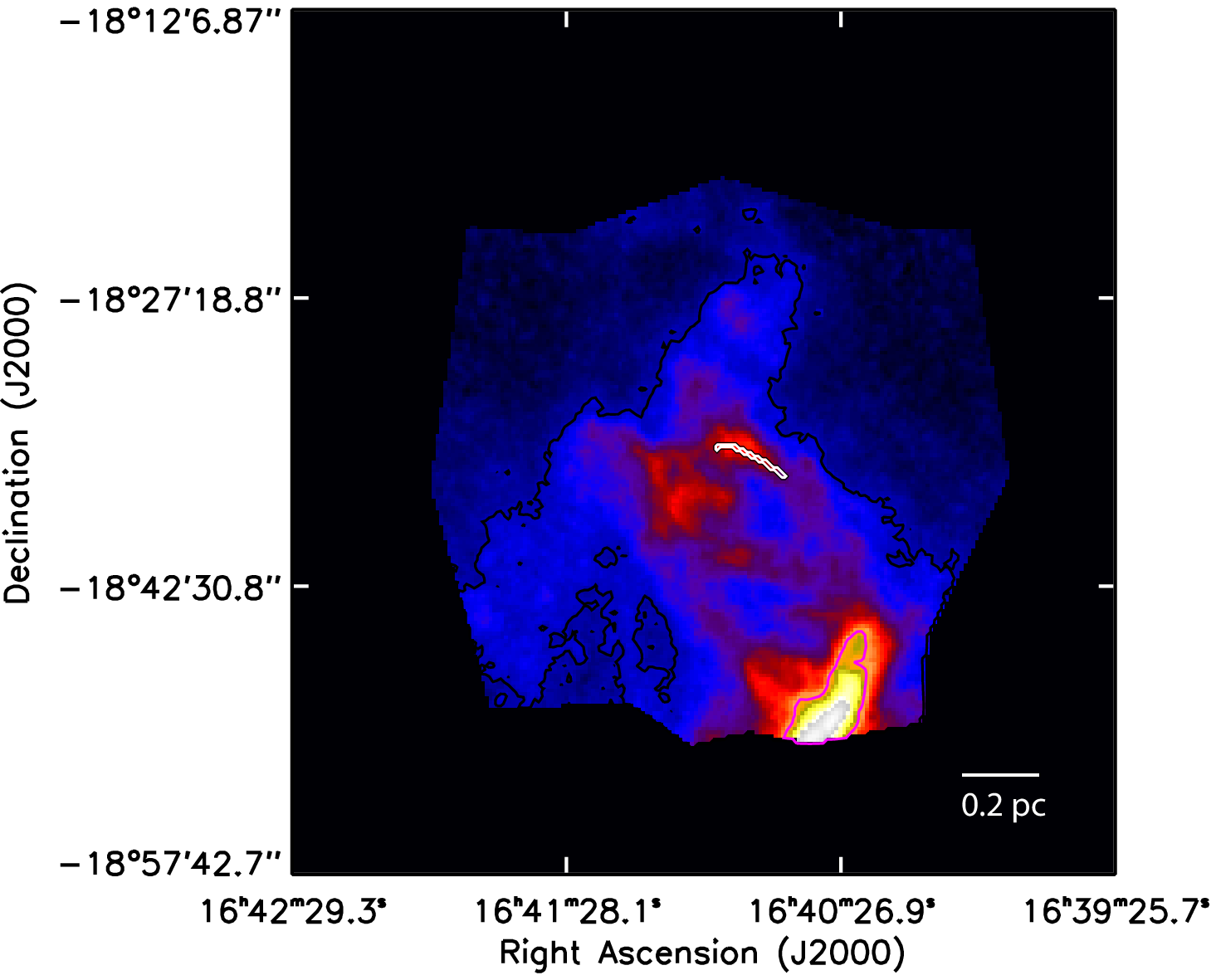}
}%
\subfigure[G3.08+9.38]{%
\label{fig:m2}
\includegraphics[scale=0.40,angle=0,trim=0cm 0cm 0cm 0cm,clip=true]{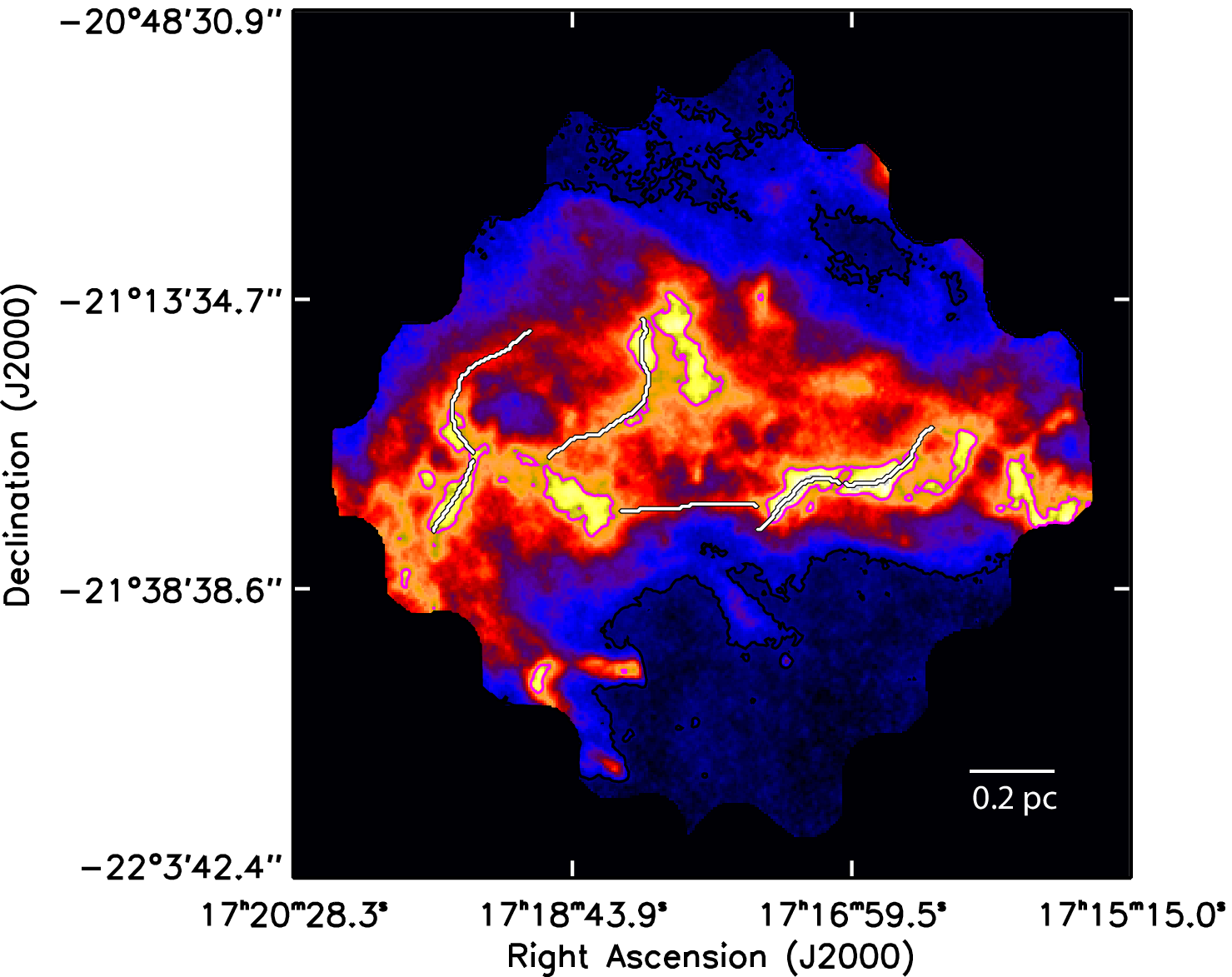}
}%
\subfigure[G25.86+6.22]{%
\label{fig:m3}
\includegraphics[scale=0.40,angle=0,trim=2cm 0cm 0cm 0cm,clip=true]{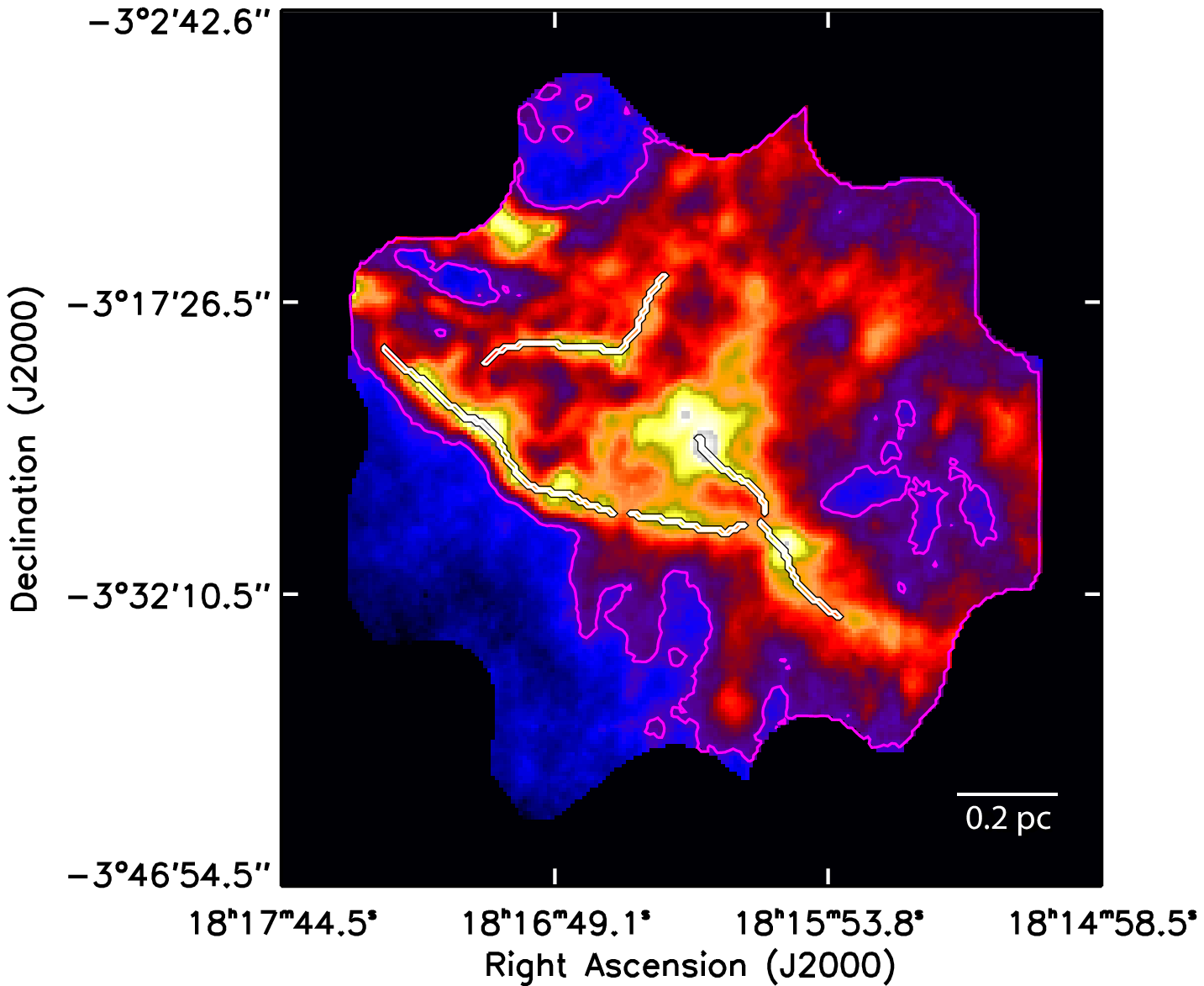}
}\\%
\label{fig:maps1}
\end{figure*}

\begin{figure*}[h!]
\centering
\subfigure[G116.08-2.40]{%
\label{fig:m4}
\includegraphics[scale=0.40,angle=0,trim=0cm 0cm 0cm 0cm,clip=true]{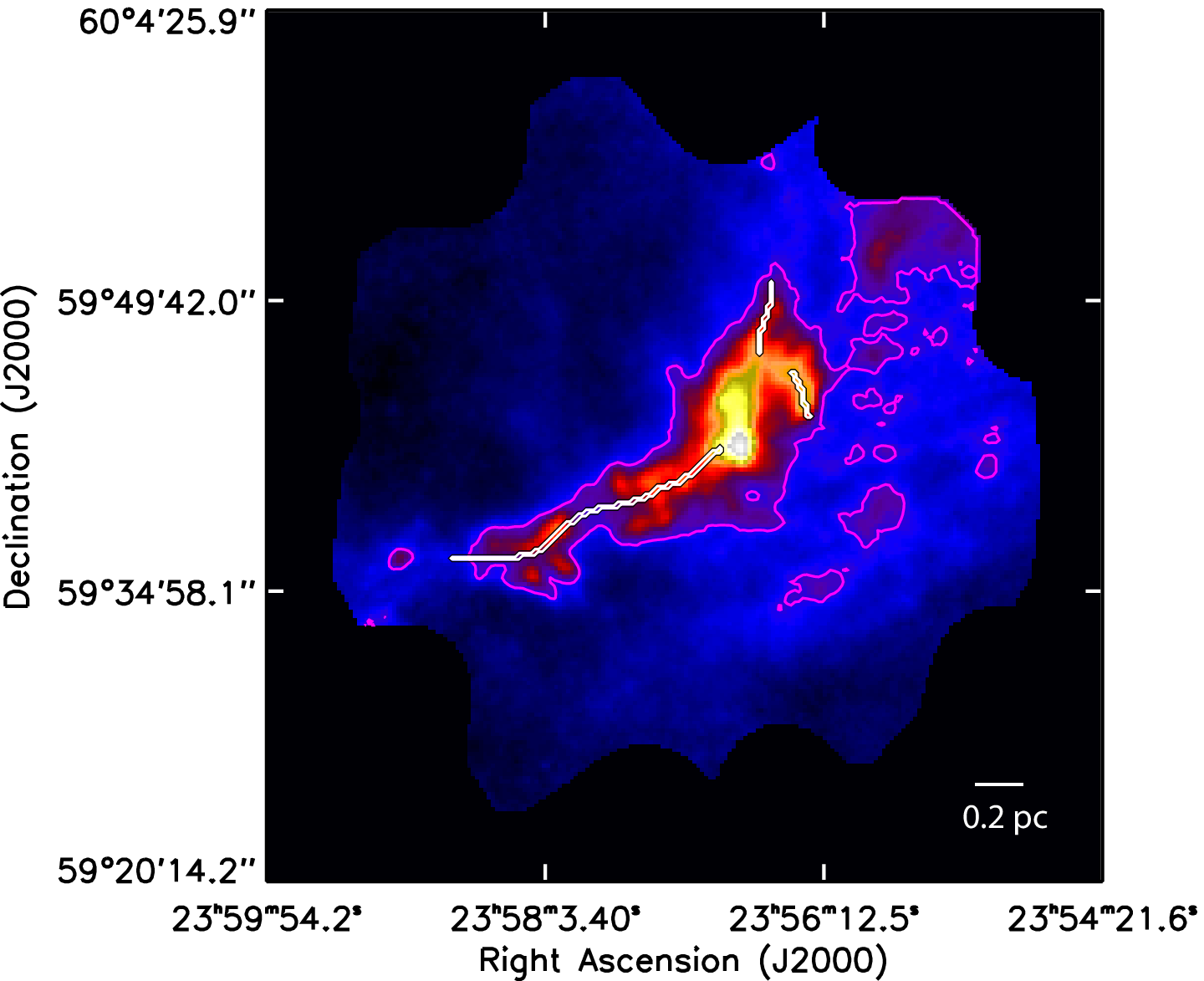}
}%
\subfigure[G126.63+24.55]{%
\label{fig:m5}
\includegraphics[scale=0.40,angle=0,trim=0cm 0cm 0cm 0cm,clip=true]{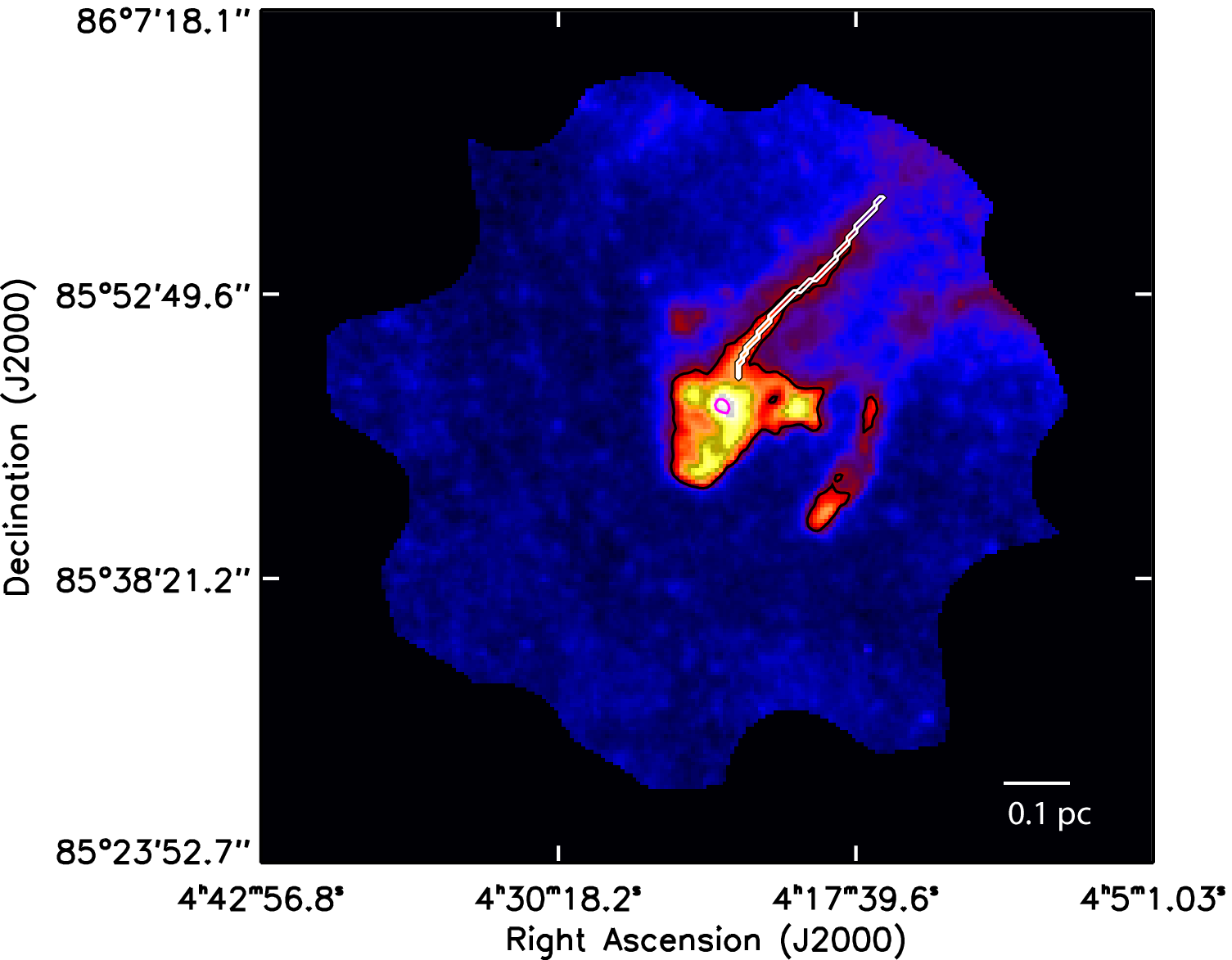}
}%
\subfigure[G150.47+3.93]{%
\label{fig:m6}
\includegraphics[scale=0.40,angle=0,trim=0cm 0cm 0cm 0cm,clip=true]{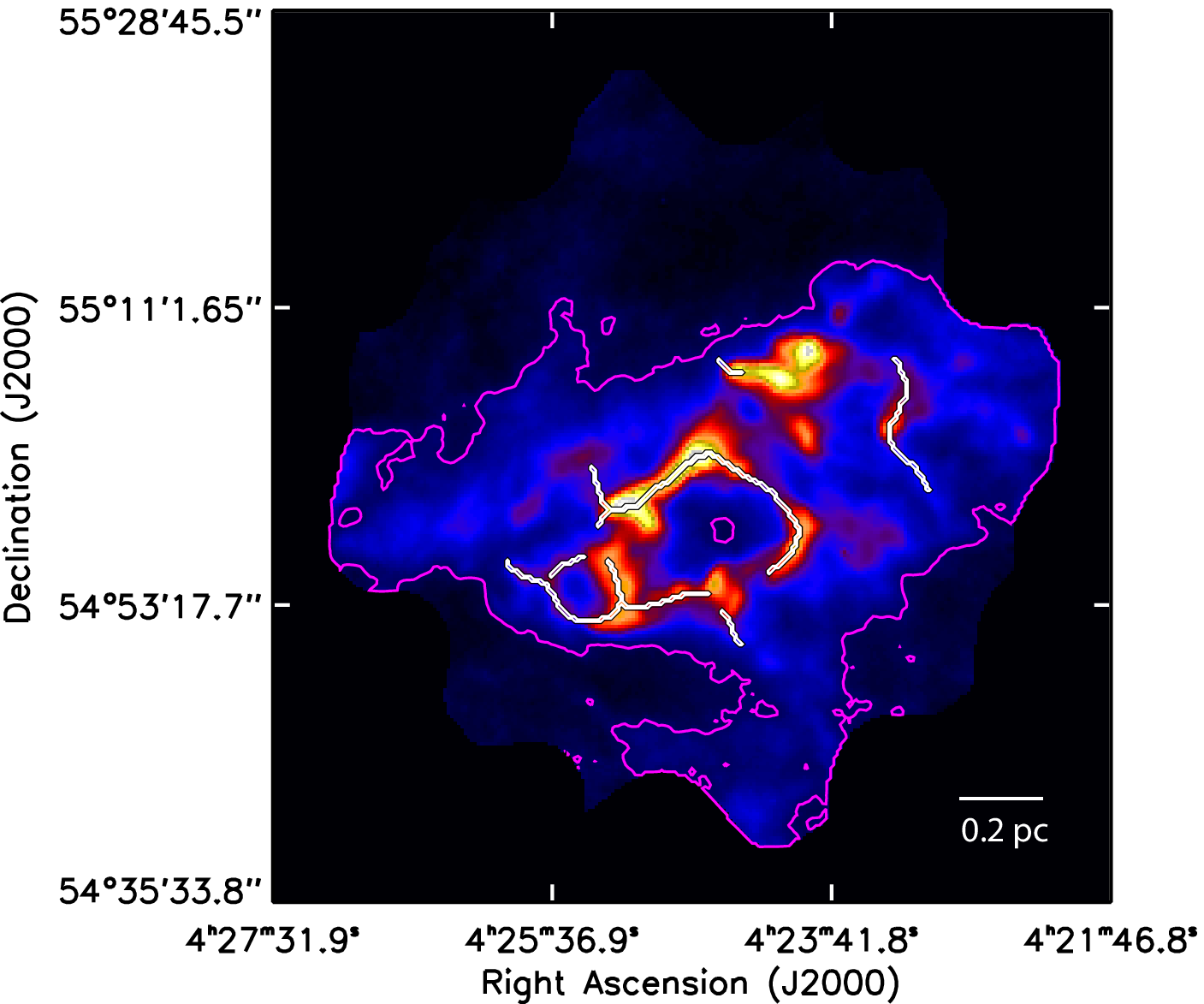}
}\\%
\label{fig:maps2}
\end{figure*}

\begin{figure*}[h!]
\centering
\subfigure[G159.23-34.51]{%
\label{fig:m7}
\includegraphics[scale=0.40,angle=0,trim=0cm 0cm 0cm 0cm,clip=true]{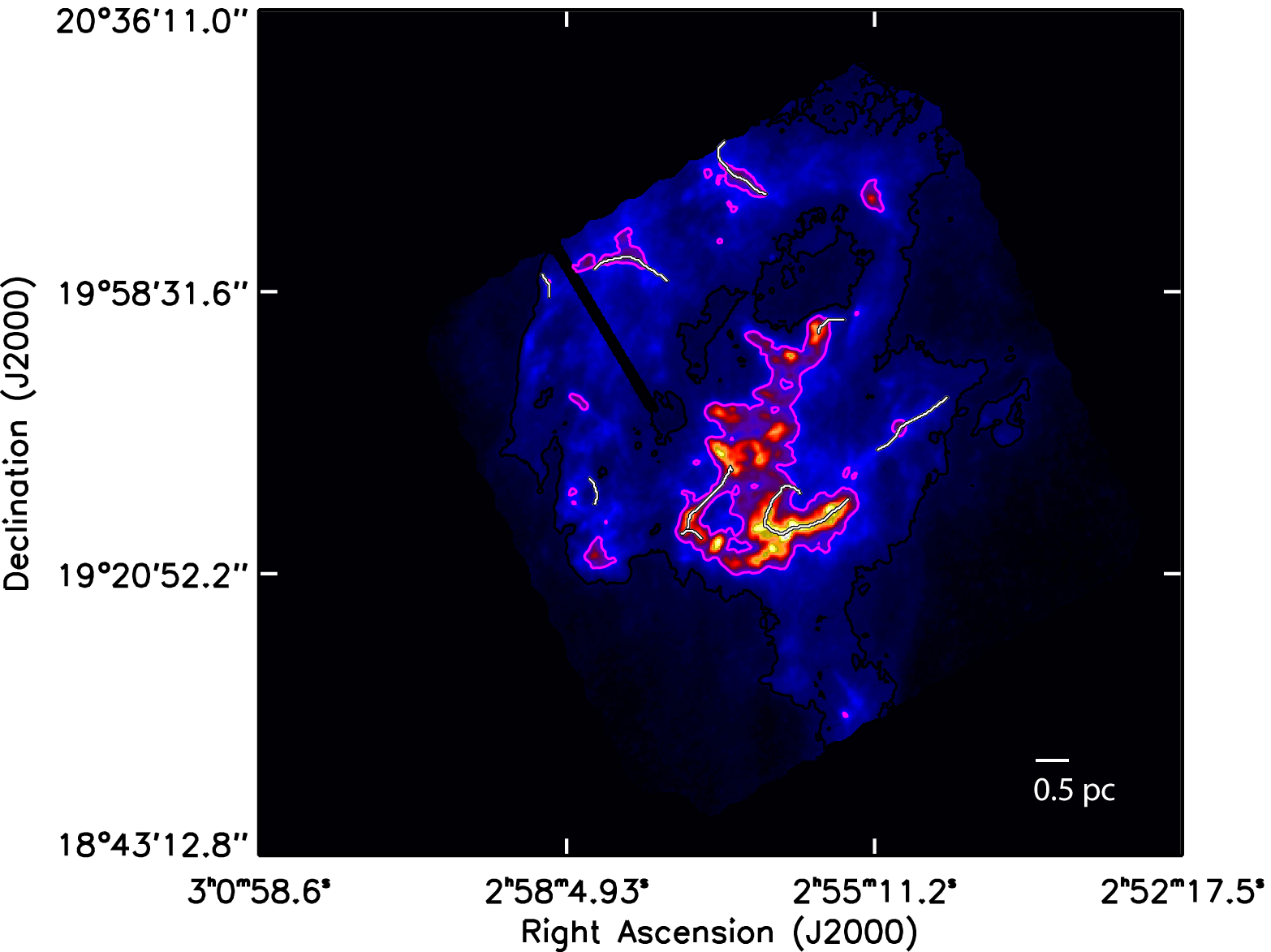}
}%
\subfigure[G173.43-5.44]{%
\label{fig:m8}
\includegraphics[scale=0.40,angle=0,trim=0cm 0cm 0cm 0cm,clip=true]{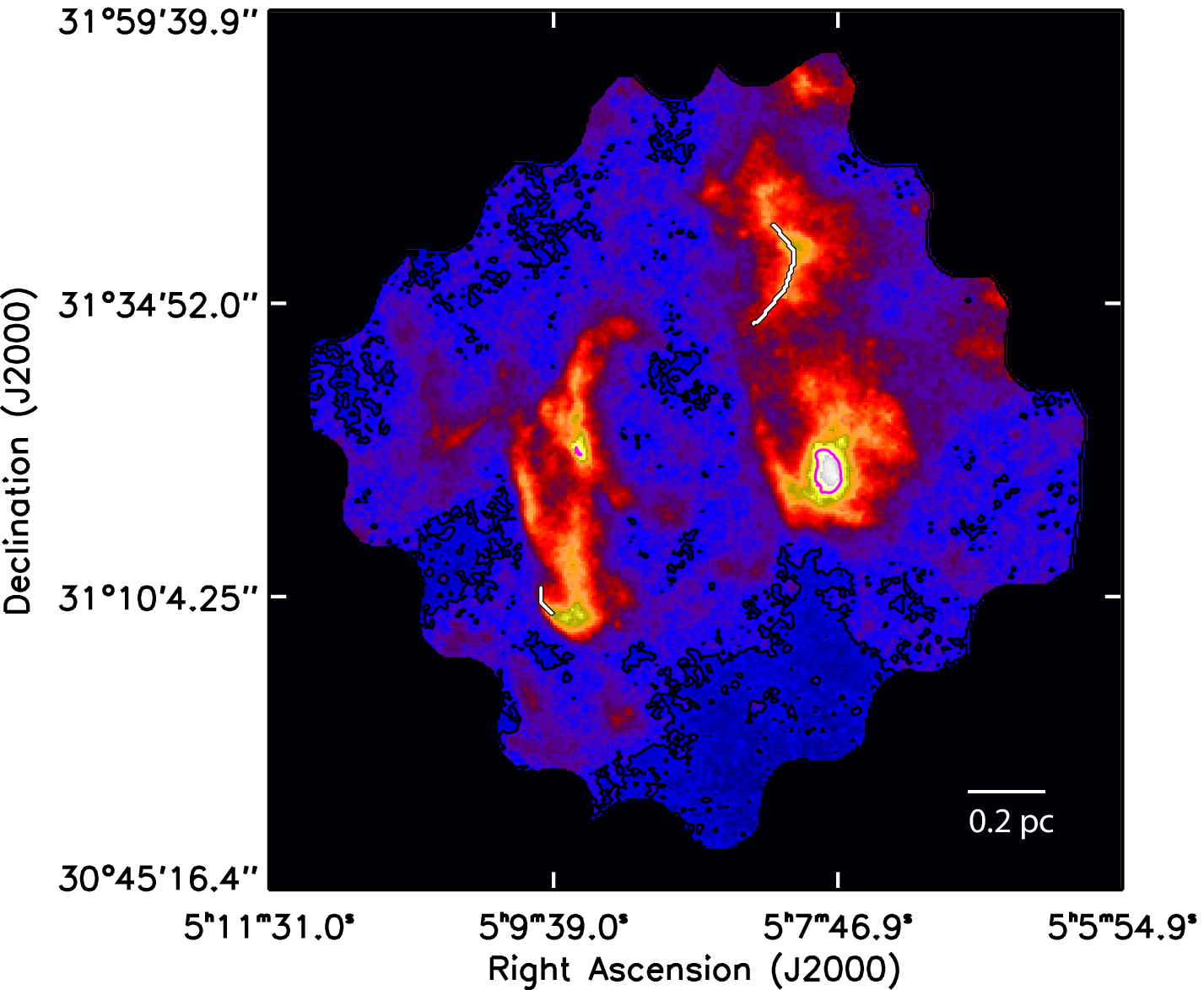}
}%
\subfigure[G206.33-25.94]{%
\label{fig:m9}
\includegraphics[scale=0.40,angle=0,trim=0cm 0cm 0cm 0cm,clip=true]{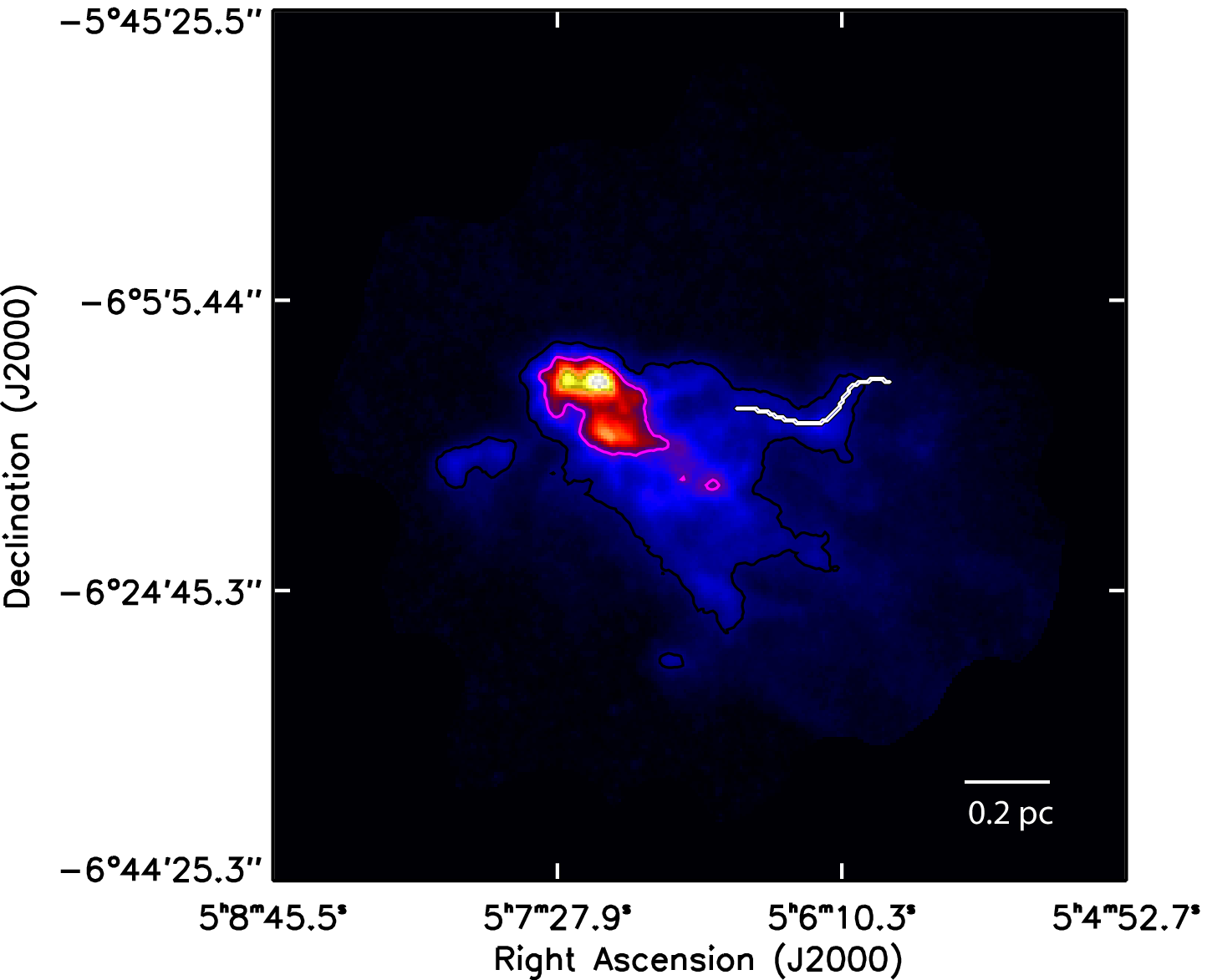}
}\\%
\label{fig:maps3}
\end{figure*}

\begin{figure*}[ht]
\centering
\subfigure[G210.90-36.55-1]{%
\label{fig:m10}
\includegraphics[scale=0.40,angle=0,trim=0cm 0cm 0cm 0cm,clip=true]{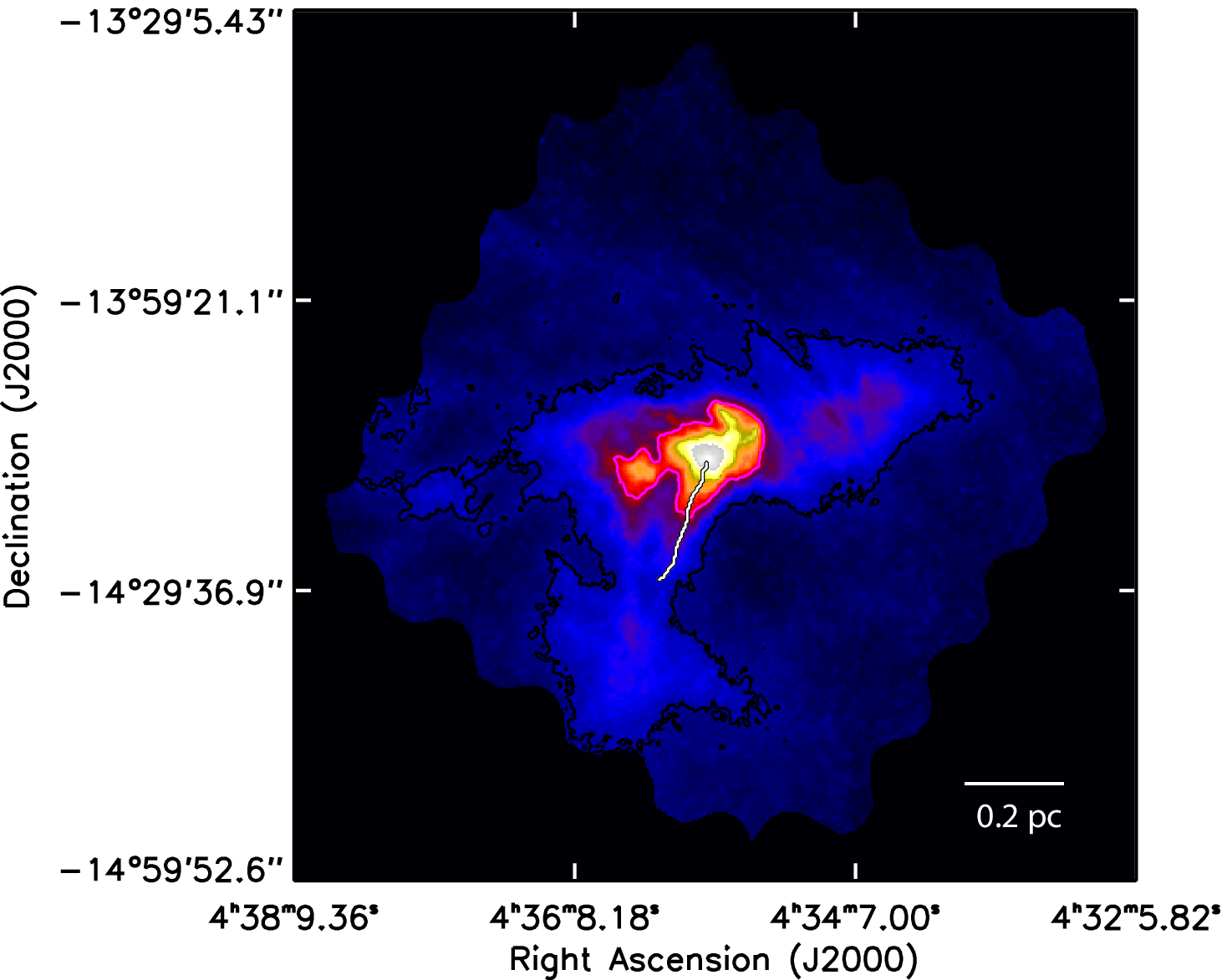}
}%
\subfigure[G300.61-3.13]{%
\label{fig:m11}
\includegraphics[scale=0.40,angle=0,trim=0cm 0cm 0cm 0cm,clip=true]{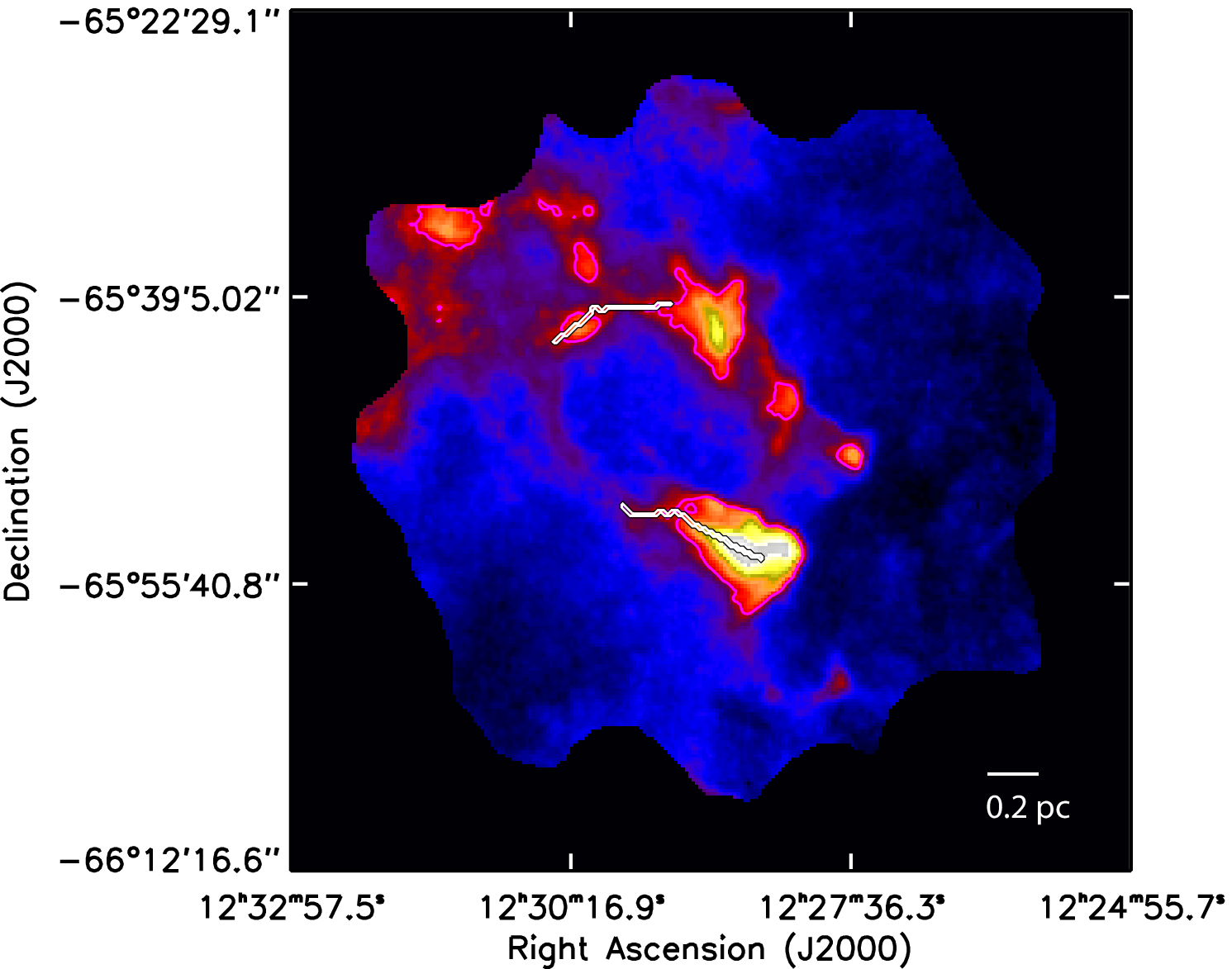}
}%
\subfigure[G300.86-9.00]{%
\label{fig:m12}
\includegraphics[scale=0.40,angle=0,trim=0cm 0cm 0cm 0cm,clip=true]{all_PCC550_10.eps}
}\\%
\label{fig:maps4}
\end{figure*}

\begin{figure*}[ht]
\centering
\subfigure[G315.88-21.44]{%
\label{fig:m13}
\includegraphics[scale=0.40,angle=0,trim=0cm 0cm 0cm 0cm,clip=true]{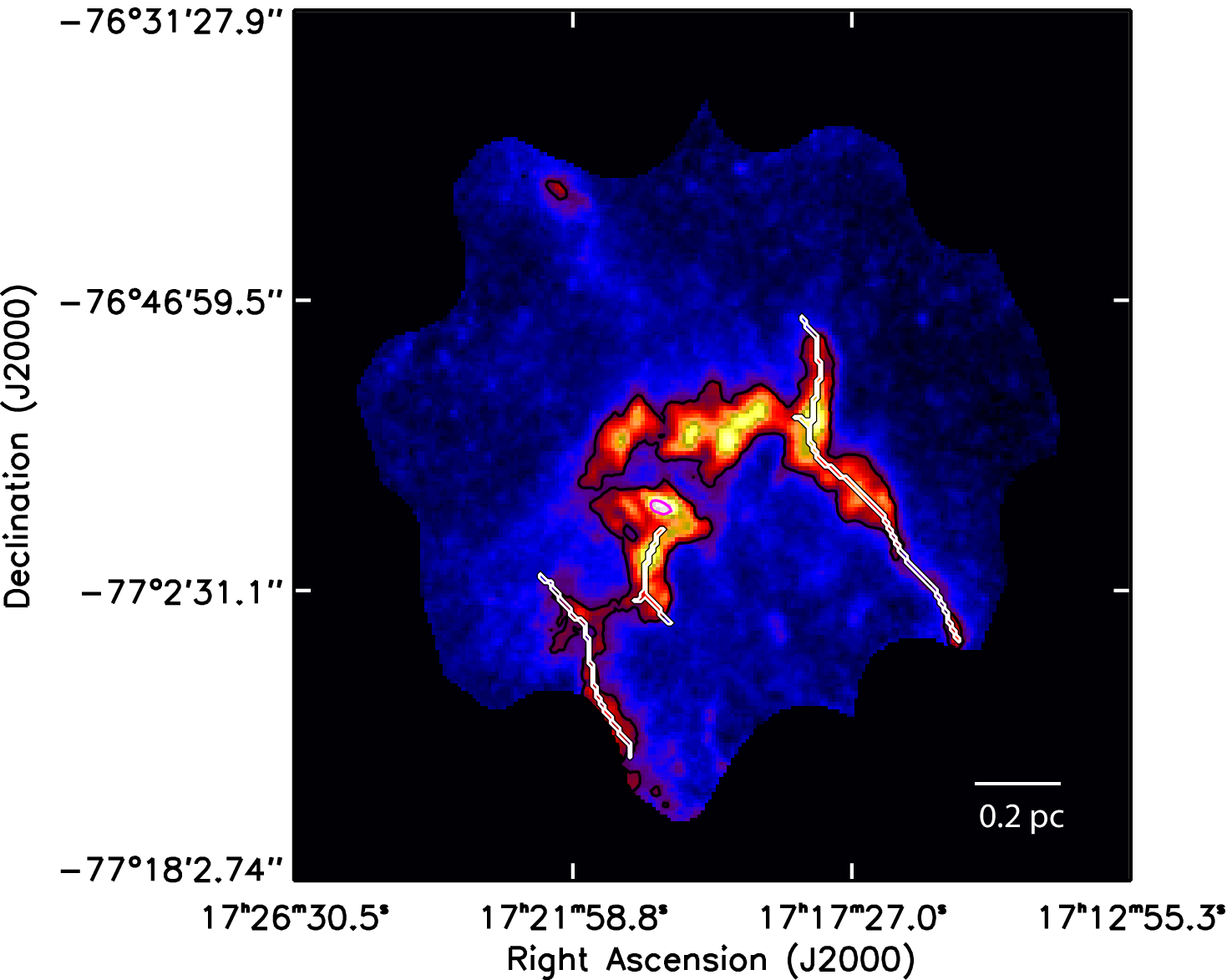}
}\%
\subfigure[G358.96+36.75]{%
\label{fig:m14}
\includegraphics[scale=0.40,angle=0,trim=0cm 0cm 0cm 0cm,clip=true]{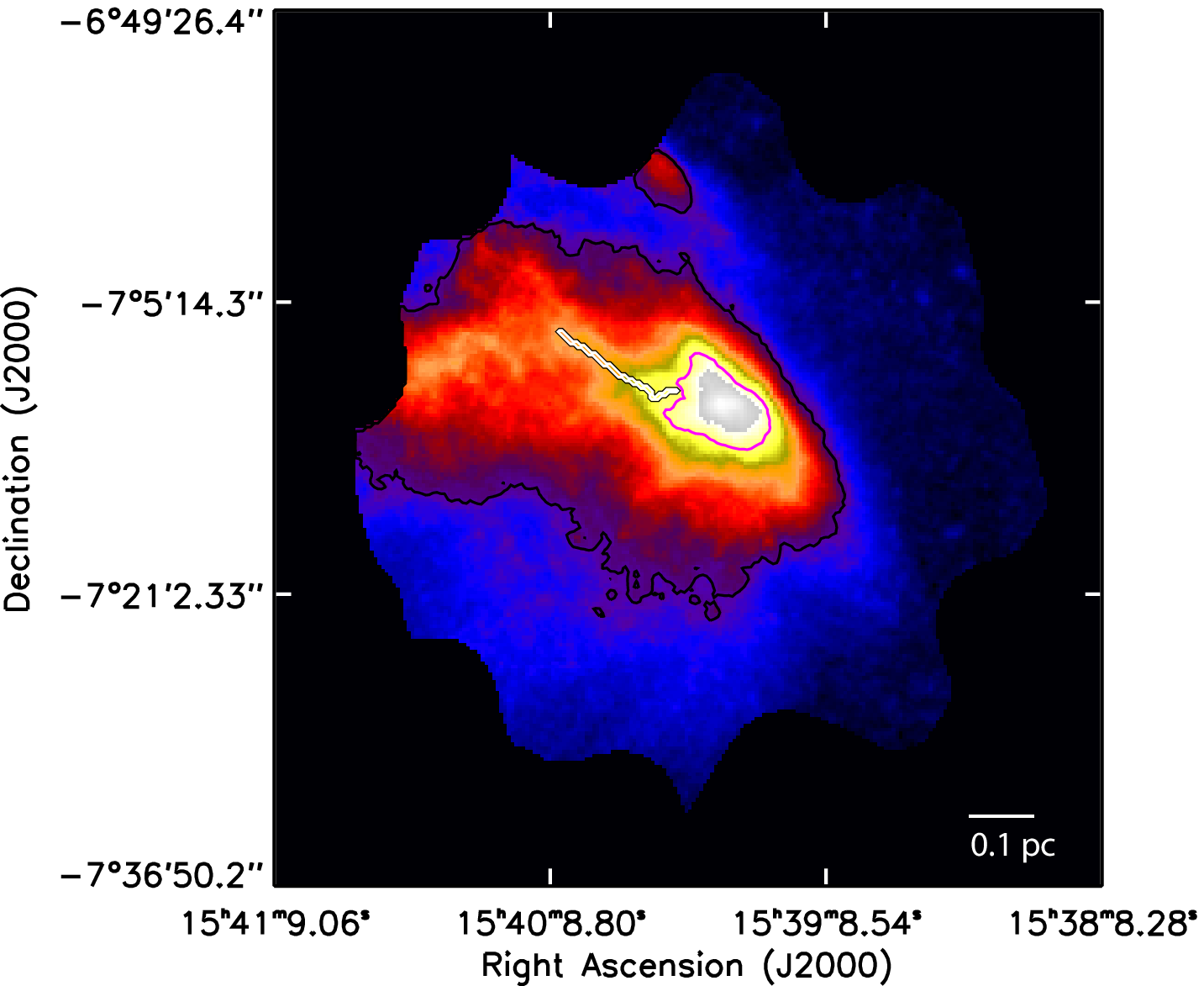}
}\\%
\label{fig:maps5}
\end{figure*}


\end{document}